%% file: main.tex
\documentclass[aps,nofootinbib,reprint,nobibnotes,showpacs,superscriptaddress]{revtex4-1}

\usepackage[english]{babel}
	\addto\captionsenglish{%
	}

\usepackage{graphicx}

\usepackage{amsmath}
\usepackage{amssymb}
\usepackage[version=3]{mhchem}	
\usepackage{bm}	
\usepackage{nicefrac}	
\usepackage{mathrsfs}

\usepackage[hidelinks,linktocpage]{hyperref}
\hypersetup{
	colorlinks,
	linkcolor={red!70!black},
	citecolor={green!70!black},
	urlcolor={blue!70!black}
}

\usepackage[usenames,dvipsnames]{xcolor}	
\usepackage{comment}	
\usepackage{upgreek}	

\newcommand{\bb}{$0\nu \beta \beta$} 
\newcommand{\bbvv}{$2\nu \beta \beta$} 

\newcommand{\mbb}{m_{\beta \beta}} 

\newcommand{\NH}{NH} 
\newcommand{\IH}{IH} 

\newcommand{\el}{\text{e}} 

\newcommand{\taubb}{t_{1/2}^{\,0\nu}}
\newcommand{\saubb}{S_{1/2}^{\,0\nu}}

\newcommand{\Cerenkov}{\u{C}erenkov} 

\newcommand{\gA}{g_{\mathrm{A}}} 
\newcommand{\gAn}{g_{\mathrm{A,nucl}}} 
\newcommand{\gAq}{g_{\mathrm{A,quark}}} 
\newcommand{\gAp}{g_{\mathrm{A,phen}}} 
\newcommand{\ggAn}{\mbox{\scalebox{.8}{$g_\mathrm{A,nucl}$}}} 
\newcommand{\ggAq}{\mbox{\scalebox{.8}{$g_\mathrm{A,quark}$}}} 
\newcommand{\ggAp}{\mbox{\scalebox{.8}{$g_\mathrm{A,phen}$}}} 

\newcommand{\ckty}{counts\,$\keV^{-1}\,\ton^{-1}\,\yr^{-1}$}
\newcommand{\ckky}{counts\,$\keV^{-1}\,\kg^{-1}\,\yr^{-1}$}

\newcommand{\uBqkg}{$10^{-6}\Bq\,\kg^{-1}$}

\newcommand{\K}{\text{K}}		\newcommand{\mK}{\text{mK}}

\newcommand{\mm}{\text{mm}}	\newcommand{\cm}{\text{cm}}	\newcommand{\m}{\text{m}}		

\newcommand{\yr}{\text{yr}}	\newcommand{\g}{\text{g}}		\newcommand{\kg}{\text{kg}}	\newcommand{\ton}{\text{t}}

		\newcommand{\cmsqs}{\text{cm}^{-2}\,\text{s}^{-1}}

\newcommand{\meV}{\text{meV}}	\newcommand{\eV}{\text{eV}}	\newcommand{\keV}{\text{keV}}		
\newcommand{\MeV}{\text{MeV}}	

\newcommand{\Bq}{\text{Bq}}

\newcommand{\nota}{$^{\mbox{\scriptsize $\,\dagger$}}$}
\newcommand{\notaB}{$^{\mbox{\scriptsize $*$}}$}
\newcommand{\notaC}{$^{\mbox{\scriptsize \S}}$}

\newcommand{\SB}{\hspace{-9pt}}
\newcommand{\SC}{\hspace{-7pt}}

\begin{document}

	\title{The saga of neutrinoless double beta decay search with TeO$_2$ thermal detectors}

	\author{Chiara Brofferio}
		\email{chiara.brofferio@mib.infn.it}
		\affiliation{Dipartimento di Fisica, Universit\`a di Milano-Bicocca, 20126 Milano, Italy \\}%
		\affiliation{INFN, Sezione di Milano-Bicocca, 20126 Milano, Italy \\}%
	\author{Stefano Dell'Oro}
		\email{sdelloro@vt.edu}
		\affiliation{Center for Neutrino Physics, Virginia Polytechnic Institute and State University, 
			Blacksburg, VA 24061, USA \\}%
		\affiliation{Gran Sasso Science Institute, 67100 L'Aquila, Italy \\}%

\date{\today}
	
	\begin{abstract}
		Neutrinoless double beta decay (\bb) is a direct probe of physics beyond the Standard Model.
		Its discovery would demonstrate that the lepton number is not a symmetry of nature and would provide us with unique
		information on the nature and mass of the neutrinos.
		Among the experimental techniques employed in the investigation of this rare process, thermal detectors 
		fulfill the requirements for a powerful search, showing an excellent energy resolution and the possibility
		of scaling to very large masses.
		In this work, we review the long chain of bolometric experiments based on \ce{TeO_2} crystals
		that were and continue to be carried out at the Laboratori Nazionali del Gran Sasso (Italy),
		searching for the \bb~of \ce{^{130}Te}.
		We illustrate the progress and improvements achieved in almost thirty years of measurements and compare 
		the various performance and results.
		We describe the several steps that led to the CUORE detector, the latest of this series and presently in data taking,
		and we highlight the challenges that a next bolometric experiment will face in order to further improve the 
		sensitivity, especially concerning the background abatement.
		Finally, we emphasize the advantages of \ce{^{130}Te} in the search for \bb~with a further future experiment. 
		\\[+9pt]
		Published on: \href{https://aip.scitation.org/doi/10.1063/1.5031485}{Rev.\ Sci.\ Instrum.\ {\bf 89}, 121502 (2018)}
	\end{abstract}


	\maketitle

	\tableofcontents

	\input{1_Introduction}

	\input{2_BolometricTechnique}

	\input{3_LNGSExperiments}

	\input{4_DBDLimits}

	\input{5_Challenges_Te}

	\input{6_Future}

	\input{7_Summary}

	\input{acknowledgments}

	\bibliography{ref}

\end{document}

%% file: 1_Introduction.tex
\section{Introduction}

	Neutrinoless double beta decay (\bb) \cite{Furry:1939qr} 
	is an extremely rare nuclear process not predicted by the Standard Model in which two neutrons inside a nucleus simultaneously 
	transform into two protons, with the emission of a pair of electrons:
	\begin{equation}
	\label{eq:DBD}
		(A,Z) \rightarrow (A,Z+2) + 2\el^-.
	\end{equation}

	The most evident feature of the \bb~transition is the \emph{violation of the lepton number of leptons}: its observation would thus demonstrate 
	that the lepton number is not a symmetry of nature.%
	\footnote{That in Eq.~\eqref{eq:DBD} is actually one of the forms that \bb~can take. 
		Depending on the relative numbers of protons and neutrons, it is also possible to have electron capture or the emission of \ce{e^+} instead of \ce{e^-}.
		The violation of the lepton number is present is all the processes, which are thus equally important in pointing at new Physics.}
	At the same time, \bb~is a key tool to study neutrinos, probing whether their nature is that of Majorana particles and providing us 
	with precious information on the neutrino absolute mass scale and ordering.

	The huge impact on Particle Physics has motivated and continues to motivate a strong experimental effort to search for \bb.

\subsection{Experimental search for \bb}
\label{sec:0nbb_exp_search}

	The experimental search for \bb~relies on the detection of the two emitted electrons. 
	In particular, since the energy of the recoiling nucleus is negligible, the sum of their kinetic energy is equal 
	to the Q-value of the transition, $Q_{\beta\beta}$, where we expect to observe a monochromatic peak.

	The observable probed by the experiments searching for \bb~is the half-life of the 
	decay of the isotope of interest, $\taubb$.
	In the fortunate event of a \bb~peak showing up in the energy spectrum, this parameter can be extracted starting from 
	the law of radioactive decay:
	\begin{equation}
		\taubb = \ln 2 ~ T ~ \varepsilon ~ \frac{N_{\beta \beta}}{N_\mathrm{peak}}
		\label{eq:exp_tau}
	\end{equation}
	where $T$ is the measuring time, $\varepsilon$ is the detection efficiency, $N_{\beta \beta}$ is the number of 
	$\beta \beta$ decaying nuclei under observation, and $N_\mathrm{peak}$ is the number of observed 
	decays in the peak. 

	If no peak is detected, the sensitivity of a given \bb~experiment, $\saubb$, is usually defined as the process half-life 
	corresponding to the maximum signal that can be hidden by the background fluctuations $n_B$. 
	In the assumption that the background counts scale linearly with the mass of the detector,%
	\footnote{This is true if impurities are uniform inside the detector.
		However, this might not be always the case. For example, if the main component of the background is superficial,
		it is the surface over volume ratio that matters.}
	from Eq.~(\ref{eq:exp_tau}) it is easy to obtain the expression~\cite{Cremonesi:2013vla}:
	\begin{equation}
	\label{eq:fiorini}
		\saubb = \ln 2 ~ \varepsilon ~ \frac{1}{n_\sigma} ~ \frac{x\,\eta\,N_A}{\mathscr{M}_A} ~
		\sqrt{\frac{M ~ T}{B  ~ \Delta}}
	\end{equation}
	where $B$ is the background level per unit mass, energy, and time, $M$ is the detector mass, $\Delta$ is the 
	energy resolution, $x$ is the stoichiometric multiplicity of the element containing 
	the $\beta \beta$ candidate, $\eta$ is the $\beta \beta$ candidate isotopic abundance, $N_A$ is the Avogadro number and $\mathscr{M}_A$ is the 
	compound molecular mass.
	Despite its simplicity, Eq.~(\ref{eq:fiorini}) is able to emphasize the role of the essential experimental parameters.
	Improving one or more of these quantities in order to enhance the sensitivity on $\taubb$ is what drives the experimental search for \bb.

	The previous expression is no longer valid when the background level $B$ is so low that the expected number of background events in the ROI along the 
	experiment life is close to zero.
	It is the so called ``zero background'' experimental condition.
	The transition between the two regimes can be identified with the intermediate situation in which the expected 
	number of counts is of the order of unity:
	\begin{equation}
	\label{eq:zb}
		M ~ T ~ B ~ \Delta = \mathcal O(1).
	\end{equation}
	In this case, $n_B$ is a constant, i.\,e.\ it is the maximum number of counts compatible with no observed counts at a given C.\,L., 
	and the expression for the sensitivity becomes:
	\begin{multline}
	\label{eq:0bkg_sens}
		S_{1/2,\,0\mathrm{bkg}}^{\,0\nu} = \ln 2 ~ T ~ \varepsilon ~ \frac{N_{\beta \beta}}{n_B} = \\
		\ln 2 ~ \varepsilon ~ \frac{x\,\eta\,N_A}{\mathscr{M}_A} ~ \frac{M\,T}{n_B}.
	\end{multline}
	
	Reaching the zero background condition is the goal of the future experimental searches (see the discussion in Secs.~\ref{sec:challenges} and \ref{sec:future}).
	In fact, in this case, the sensitivity grows much more rapidly, i.\,e.\ linearly with the exposure
	(the product detector mass times live-time),
	as it can be seen by comparing Eqs.~(\ref{eq:fiorini}) and (\ref{eq:0bkg_sens}). 

\label{sec:mbb_limit}

	\medskip
	From the theoretical point of view, there can be different mechanisms originating \bb. Still, the general 
	interest remains mostly focused on the \emph{neutrino mass mechanism}~\cite{Dell'Oro:2016dbc}.
	Within this scenario, it is useful to define the so called Majorana effective mass:
	\begin{multline}
	\label{eq:mbb}
		\mbb
		\equiv \biggl| \el^{i\alpha_1}|U_{\el1}^2|m_1 + \el^{i\alpha_2}|U_{\el2}^2|m_2 + |U_{\el3}^2|m_3 \biggr|
	\end{multline}
	where $m_i$ are the masses of the individual neutrinos $\nu_i$, $\alpha_{1,2}$ are 
	the Majorana phases and $U_{\el i}$ are 
	the elements of the mixing matrix that defines the composition of the electron neutrino: 
	$|\nu_\el \rangle~=~\sum_{i=1}^3 U_{\el i}^* |\nu_i \rangle$. 

	The Majorana effective mass carries the information on the neutrino masses that can be extracted
	from the \bb~searches. 
	An experimental measurement of $\taubb$ can be translated into a value of $\mbb$,
	while a lower limit on $\taubb$ becomes an upper bound on $\mbb$. 

	In order to operate this conversion,
	it is necessary to pass through quantities obtained by theoretical calculations of atomic and nuclear physics. 
	Within the hypothesis of ``ordinary'' neutrinos as mediators of the \bb~transition, a convenient parametrization 
	for $\taubb$ can be:
	\begin{equation}
		\left[\taubb \right]^{-1}=\frac{\mbb^2}{m^2_\el} \, G_{0\nu} \, \gA^4 \, \left|{\mathcal{M}}_{0\nu}\right|^2
		\label{eq:tau}
	\end{equation}
	where $G_{0\nu}$ is the Phase Space Factor (PSF), $\mathcal{M}_{0\nu}$ is the Nuclear Matrix Element (NME),  
	$\gA$ is the axial coupling constant and $m_\el$ is the electron mass, conventionally taken as a reference.

	By inverting Eq.~(\ref{eq:tau}) and by choosing proper values for $\gA$ and for the PSF~\cite{Kotila:2012zza} and 
	NME~\cite{Barea:2015kwa} for \ce{^{130}Te}, it is thus possible to obtain the limits on $\mbb$ 
	starting from the experimental sensitivities.
	
	To be fair, one should consider the uncertainties in the theoretical calculations while passing from $\taubb$ to $\mbb$.
	The PSFs are known with accurate precision for the nuclei of interest. The present uncertainty
	is about $7\%$~\cite{Kotila:2012zza}.
	For the NMEs, the situation is more complicated. 
	A relatively small intrinsic error of $\lesssim 20\%$~\cite{Simkovic:2013qiy,Barea:2015kwa} is presently assessed by the most recent calculations. 
	However, the disagreement between the results from different models is actually larger, 
	up to a factor $\sim 3$. Moreover, when processes ``similar'' to the \bb~are considered 
	(single $\beta$ decay, two-neutrino double beta decay (\bbvv), electron capture) and the calculations from the same models are compared to the measured rates, 
	the actual differences are much larger than 20\%.
	Finally, the value of $\gA$ remains an open issue. The measured value in the weak interactions and decays of nucleons
	($\gAn~\simeq 1.27$) could be indeed ``renormalized'' inside the nuclear medium toward the value appropriate for quarks 
	($\gAq~=1$) or, even, the possibility of a further reduction (quenching) has been argued, based on the systematic 
	over-prediction of the $\beta$ and \bbvv~NMEs (worst scenario: $\gAp\simeq \gAn\cdot A^{-0.18}$, where $A$ is the mass number 
	\cite{Faessler:2007hu,Barea:2013bz}).%
	\footnote{If the latter was the case, the implications for the \bb~searches could be tremendous, 
		given the strong dependence on $\gA$ of the half-life~\cite{Robertson:2013cy,Dell'Oro:2014yca}. 
		However, the answer depends on the source of the quenching, which is still unknown 
		(see Ref.~\cite{Engel:2016xgb} for an extensive review on the topic).
		Also, some theoretical models find a dependence of $\taubb$ on $\gA$ milder than quartic, 
		so that the effects of a suppression of this parameter are reduced~\cite{Lisi:2015yma,Suhonen:2017rjf}.}
	
	Therefore, an experimental limit on $\taubb$ actually translates into a range of values for $\mbb$ (see e.\,g.\ Fig.~\ref{fig:mbb_ml}).
	The broadness of the band depends on the adopted approach in discussing these theoretical uncertainties.

\subsection{The case for $^{130}$Te}

	The great importance of \bb~motivates a strong experimental effort to search for this process in different isotopes.
	In practice, the choice among all the possible \bb~candidate emitters is narrowed down by a series of important requests:
	a high transition Q-value ($Q_{\beta\beta}$), since this directly reduces the natural radioactive background
	and increases the decay probability due to the higher PSF (see Eq.~\eqref{eq:tau});
	a large isotopic abundance for either the natural or the enriched material; the compatibility with a suitable detection technique. 
	The group of ``commonly'' studied isotopes includes \cite{Giuliani:2012zu,Das:2015yba}: 
	\ce{^{48}Ca}, \ce{^{76}Ge}, \ce{^{82}Se}, \ce{^{96}Zr}, \ce{^{100}Mo}, \ce{^{116}Cd}, \ce{^{124}Sn}, \ce{^{130}Te}, 
	\ce{^{136}Xe} and \ce{^{150}Nd}.	
	Among these, \ce{^{130}Te} presents important advantages (Table~\ref{tab:te}) and it is therefore a very favorable choice. 

	The $Q_{\beta\beta}$ of \ce{^{130}Te} lies in between the Compton edge and the full-energy peak of the $2615\,\keV$ \ce{^{208}Tl} line
	and therefore in an acceptable natural background region for a detector with good energy resolution,
	while the long \bbvv~half-life time reduces the unavoidable background from this process.
	At the same time, the isotopic abundance of \ce{^{130}Te} is over 30\% in natural \ce{Te} -- it is the 
	largest value among those of the above-mentioned isotopes -- and a further enrichment is a viable option.
	
	Another fundamental characteristic is that \ce{^{130}Te} is compatible with more than one detection technique. 
	The first limits on the \ce{^{130}Te} half-life were placed by using scintillators~\cite{Zdesenko:1980} 
	and \ce{CdTe} solid state detectors~\cite{Mitchell:1988zz},
	while the combination of a tracking device and a calorimeter~\cite{Arnold:2011gq} allowed to further improve the sensitivity.
	However, all these detectors show a limitation in terms of scalability of the amount of isotope mass, not extendible to more than some tens of kg.
	At present, two detector technologies seem to be the most promising choices: metal-loaded liquid scintillator 
	detectors~\cite{Biller:2013wua}
	and thermal detectors.

	In this work, we will discuss the use of thermal detectors for the search of \bb~of \ce{^{130}Te}.

	\begin{center}
		\begin{table}[tb]
		\caption{Relevant features of \ce{^{130}Te} for the search of \bb.}
			\begin{ruledtabular}
				\begin{tabular}{l r}

					\ce{^{130}Te} properties		&	\\
					\cline{1-1}

					$Q_{\beta\beta}$ [keV],~\cite{Rahaman:2011zz,Redshaw:2009zz,Scielzo:2009nh}	&$2527.515 \pm 0.013$\nota	\\

					\bbvv~half-life [$10^{20}$\,yr],~\cite{Alduino:2016vtd}		
							&$8.2 \pm 0.2\,\mbox{\scriptsize (stat.)} \pm 0.6\,{\mbox{\scriptsize (syst.)}}$		\\

					isotopic abundance (\%),~\cite{Fehr200483}			&$34.167 \pm 0.002$											\\[+3pt]	

					affordability		&	\\
					\cline{1-1} \\[-11pt]

					\ce{^{nat}Te} [$\,\$/\kg$],~\cite{Te-price}							&$45$						\\

					\ce{^{enr}Te} ($95\%$ \ce{^{130}Te}) [$\,\$/\kg$],~\cite{Avignone_chat}		&$20,000$		\\
				\end{tabular}
			\end{ruledtabular}
			\begin{flushleft}
				\nota {\scriptsize Combined limit from the weighted mean of the individual results.}
			\end{flushleft}
		\label{tab:te}
		\end{table}
		\vspace{-10pt}
	\end{center}

\subsection{$\text{TeO}${$_2$} bolometers and \bb}
\label{sec:teo2_bol}

	The concept of calorimetric detection of energetic particles is very old, since it appears as a pure application 
	of the first law of thermodynamics. 
	The first thermal detection of radiation was performed by S.\,P.\ Langley and dates back to the end of the XIX 
	century%
	\footnote{Langley calls his new instrument \emph{bolometer} 
		(from the greek $\beta o \lambda \acute{\eta}$ + $\mu \acute{\epsilon} \tau \rho o \nu$ = ray + meter).}
	\cite{Langley:1881}, while the observation by P.\ Curie and A.\ Laborde of radioactive particles via their heat production 
	is only a few years younger~\cite{Curie_Laborde:1903}.

	With the birth and evolution of Particle Physics, the same idea was also applied to the detection of individual 
	particles and in the late 1940s Andrews and collaborators were able to identify single $\alpha$ particles
	thanks to a superconducting calorimeter~\cite{Andrews:1949}.%
	\footnote{In the 1930s, F.\ Simon showed that operating calorimetric detectors at cryogenic temperatures could 
	significantly improve the sensitivity~\cite{Simon:1935}.}
	However, it was only in the 1980s that new generations of low temperature detectors could be proposed 
	as competitors in several important applications in neutrino physics, nuclear physics and 
	astrophysics~\cite{Mitsel'makher:1982bb,Fiorini:1983yj,Drukier:1983gj,Moseley:1984}. Among these, the study of \bb.%
	\footnote{Actually, one should distinguish between radiation and single particle detectors. 
		The word bolometer refers to the former family of devices, the latter being indicated as 
		macro/micro-low-temperature-calorimeters.
		This separation is especially remarked within the astrophycs community. 
		However, within the \bb~community, only macro-low-temperature calorimeters are considered, and these thermal detectors 
		are simply called bolometers.
		Since this is the custom in the field of \bb~search, the same choice has been adopted for this work.}

	In particular, the group of E.\ Fiorini and collaborators began to develop bolometers made of materials containing 
	$\beta\beta$-emitters, focusing on the search for the \bb~of \ce{^{130}Te} using \ce{TeO_2} crystals.
	The series of measurements performed by this group covers almost thirty years, 
	proving \ce{TeO_2} bolometers to be competitive players in the search for \bb.

\begin{figure}[t]
		\centering
		\includegraphics[width=.8\columnwidth]{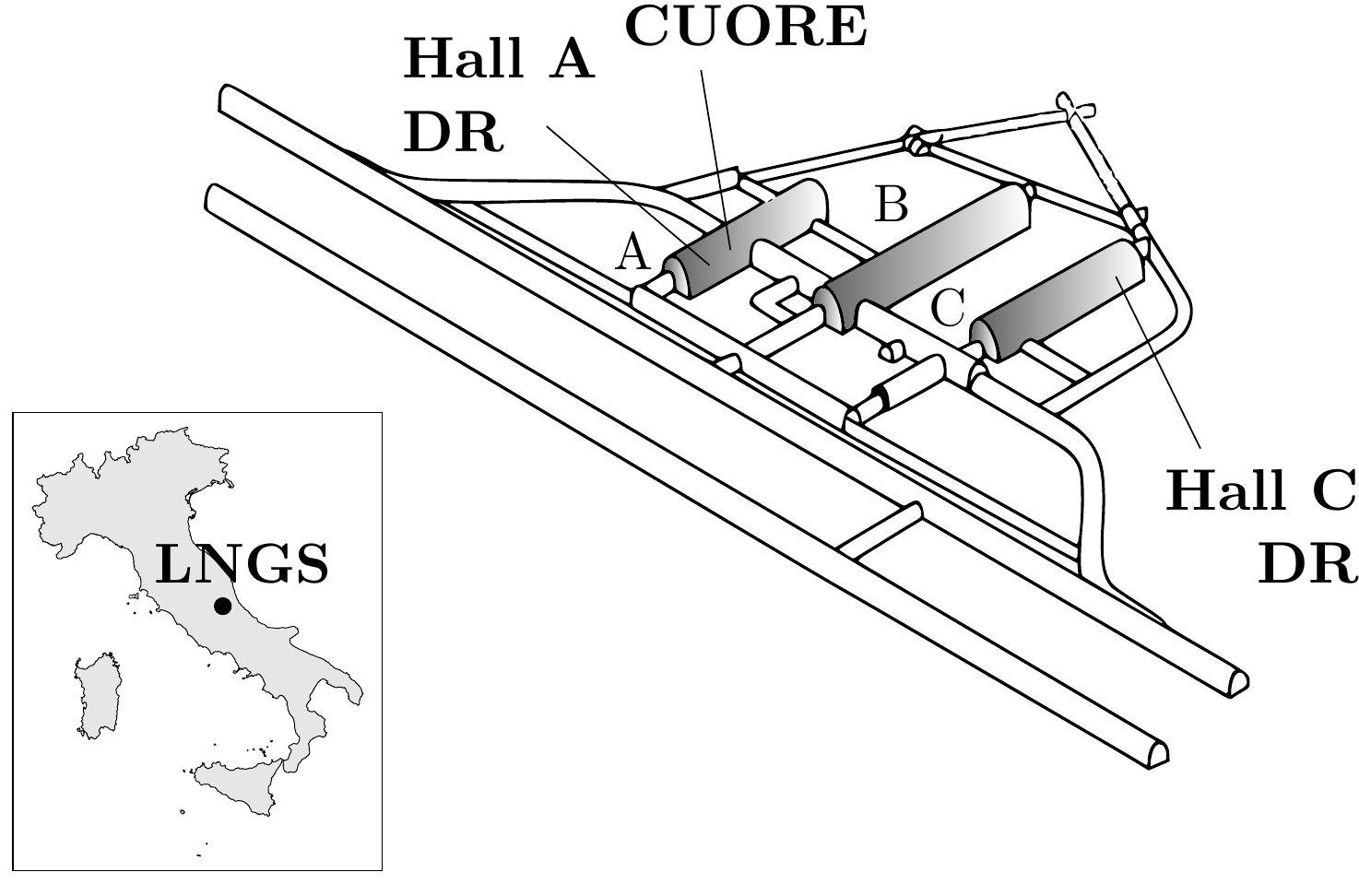}
		\caption{The LNGS underground laboratories in Italy. 
			The position inside the experimental halls of the infrastructures used for the measurements with \ce{TeO_2} 
			bolometers (see the text for details) is indicated.}
		\label{fig:LNGS}
	\end{figure}

	The first tests on these crystals were performed in the Laboratorio Acceleratori e Superconduttivit\`a Applicata of 
	I.\,N.\,F.\,N. (LASA, Segrate (MI), Italy) in 1990.
	These allowed the performance of a series of initial measurements. However, an underground facility represents a fundamental 
	requirement in order to run a rare event physics experiment.
	The Laboratori Nazionali del Gran Sasso of I.\,N.\,F.\,N. (LNGS~\cite{LNGS_site}, Assergi (AQ), Italy, Fig.~\ref{fig:LNGS}), 
	whose construction had been recently completed at the time, offered a suitable environment for the search for \bb.

	With their average coverage of about $3600$\,m.\,w.\,e.\ (meters water equivalent) and the mostly calcareous rock composition~\cite{Catalano:1986M},
	LNGS guarantee very low muon and neutron fluxes of about $3\cdot10^{-8}\,\cmsqs$~\cite{Ambrosio:1995cx} and 
	$4\cdot10^{-6}\,\cmsqs$~\cite{Best:2015yma}, respectively.
	Therefore, starting from 1991, these laboratories became the favorable location of a rich activity of experiments and R\&D
	searching for the \bb~of \ce{^{130}Te} with \ce{TeO_2} bolometers.
	Since then, many progresses and improvements have been achieved in almost thirty years of measurements
	and today this technique, that began with crystals of few grams, is now active at the tonne-scale.
	The growth in complexity of the subsequent experiments was accompanied by the expansion of the original group,
	from the initial number of slightly fewer than ten persons to an international collaboration of about 150 
	people~\cite{CUORE_site}.

%% file: 2_BolometricTechnique.tex
\section{Bolometric technique}

	Bolometers are calorimeters in which the energy released in the absorber by an interacting particle is converted 
	into phonons and measured via temperature variation. 
	These detectors can be operated only at cryogenic temperatures of about 10 or few tens of mK.
	In fact, the elementary excitation energy is of the order of $10\,\meV$, with a typical corresponding temperature 
	increase of about a few tens/hundreds of $\upmu$K per MeV. 
	
	Unlike most of the conventional spectroscopic techniques, which are based on the detection of the energy 
	released in the form of ionization and/or excitation of the detector's molecules,
	the bolometric technique measures the phonon component.
	This is a great advantage since a considerable fraction, if not all, the released energy is indeed converted into phonon 
	excitations inside the detector.
	The resulting excellent energy resolution makes these detectors very suitable for the use in rare event physics,
	such as the search for \bb.
	
\subsection{Simplified thermal model}
\label{sec:thermal_model}

	A bolometer consists essentially of two elements: an energy absorber, in which the energy from the interacting particle 
	is deposited, and a phonon sensor that converts this energy (i.\,e.\ the phonons) into a measurable signal.
	
\begin{figure}[tb]
		\centering
		\includegraphics[width=.6\columnwidth]{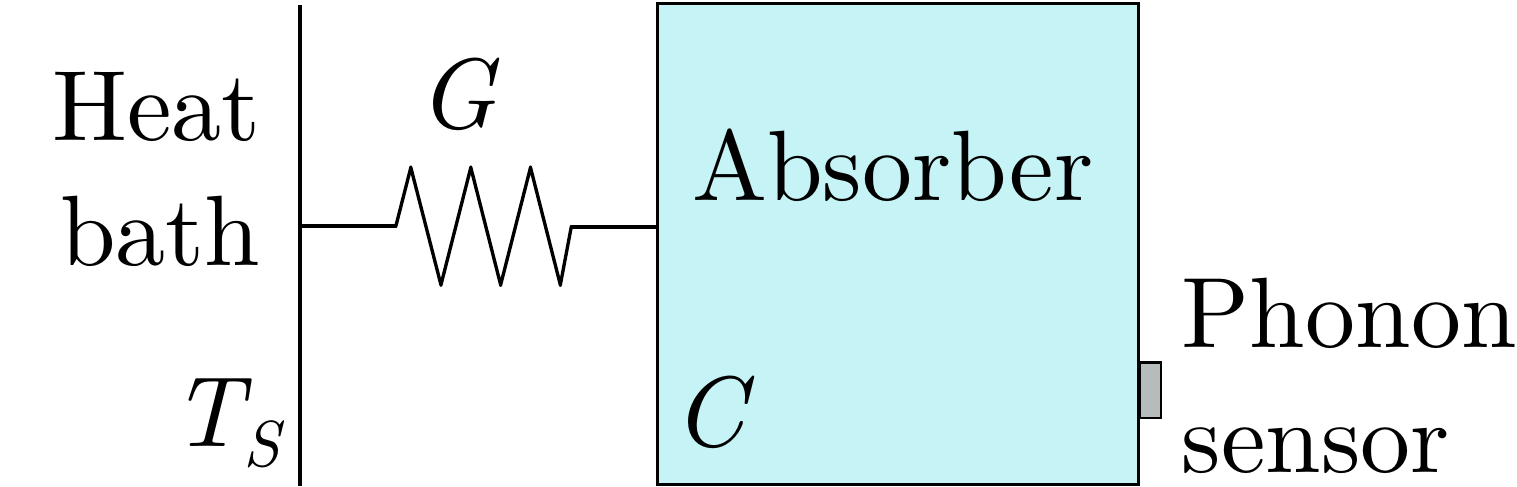}
		\caption{Simplified bolometer thermal model. The detector is modeled as a single object with heat capacity $C$ 
			coupled to the heat sink (with constant temperature $T_S$) through the thermal conductance $G$.
			The phonon sensor plays no active role.}
		\label{fig:thermal_model}
	\end{figure}

	In a very simplified model, a bolometer can be represented as a calorimeter with heat capacity $C$ 
	connected to a heat bath with constant temperature $T_S$ through a thermal conductance $G$ (Fig.~\ref{fig:thermal_model}).
	If a certain energy $E$ is released in the absorber, this will produce a change in temperature $\Delta T$ equal to the 
	ratio between $E$ and $C$. Let $T(t)$ be the absorber temperature as a function of the time $t$ and let us assume that
	\begin{equation}
		\Delta T \equiv |T(t) - T_S| \ll T_S \quad \forall t,
	\end{equation}
	so that $C$ and $G$ can be considered constant quantities.
	Then, the temperature variation can be described by the time evolution
	\begin{equation}
		\Delta T(t) = \frac{E}{C}\,\el^{-t/\tau} \quad \mbox{where} \quad \tau \equiv \frac{C}{G}.
	\label{eq:time_evolution}
	\end{equation}
	The characteristic time of the thermal pulse is very long, up to the order of few seconds, depending on the 
	values of $C$ and $G$. 
	From Eq.~(\ref{eq:time_evolution}), the crucial importance of the heat capacity appears evident: to get a large 
	signal amplitude, $C$ has to be as small as possible.

	In order to minimize $C$, very low temperatures are needed and a suitable material for the absorber has to be selected. 
	Dielectric and diamagnetic crystals or superconductors below the transition phase are preferred, 
	since in these cases the main contribution to $C$ is the one from the crystal lattice
	and the Debye law for the specific heat at low temperatures can be applied. We thus get:
	\begin{equation}
		C \propto \left( \frac{T}{\Theta_\mathrm{D}} \right)^3
	\end{equation}
	where the $\Theta_\mathrm{D}$ is the Debye temperature and is a property of the material.
	Since the latter parameter depends on the mass number $A$ and on the material density $\rho$ as 
	$\Theta_\mathrm{D} \propto A^{-1/3}\,\rho^{-1/6}$~\cite{EnssSiegfried:2005}, low atomic mass and low density should be 
	preferred for their high $\Theta_\mathrm{D}$. 
	On the other hand,
	one has to take into account the requirements from particle physics, too. In particular, a high $Z$ guarantees a larger detection efficiency for $\gamma$ radiation, 
	while a high $Z/A$ ratio a larger detection efficiency for $\beta$s.
	
	Of course, a real bolometer is much more complicated than the ``monolithic'' detector here represented. 
	For example, the parameters $C$ and $G$ cannot be considered as global quantities, but rather as the sum of 
	many contributions, each with different behaviour. Also the role of the phonon sensor is not negligible in the
	development of the signal.
	However, this simple description gives a qualitative idea of a bolometric detector.
	
\subsection{Phonon sensor}
\label{sec:phonon_sensor}

	A phonon sensor collects the phonons produced in the absorber and generates an electric signal.
	It is possible to identify many classes of such devices, sensitive to phonons at different phases 
	of the thermalization process (see e.\,g.\ Ref.~\cite{Stahl:2005} for a review).
	Anyway, since the series of experiments considered in this 
	work used and currently use semiconductor thermistors, we focus our attention only on this kind of sensors.
	
	Semiconductor thermistors are glued on the crystal absorbers and are intrinsically slow devices.
	They are mainly sensitive to thermal phonons	and act as temperature sensors, giving information about the system in thermal equilibrium.
	In general, semiconductor thermistors consist of small \ce{Ge} (or \ce{Si}) crystals with a doped region.
	Their resistance presents a steep slope as a function of the temperature, so that a measurable variation of the former is associated
	to an even very small variation of the latter.

	A very uniform and large dopant distribution and a very accurate net dopant concentration are usually obtained through 
	the Neutron Transmutation Doping (NTD)~\cite{Larrabee:1984}: the semiconductor sample is bombarded with neutrons, which induce nuclear reactions on the 
	various target stable isotopes leading to the formation of n- and p-dopants.%
	\footnote{The thermistors that undergo this process are in turn referred to as NTDs.}
	The presence of such ``impurities'' is crucial for the operation at low T.

	In fact, for intrinsic semiconductors (i.\,e.\ without impurities) the conduction can happen only
	with an activation energy equal or larger than the energy gap. This in turn requires high temperatures 
	since the concentration of free charges $n$ follows an exponential law
	with $T$.
	As an example, $n \simeq 2.4\cdot 10^{13}\,\cm^{-3}$ for \ce{Ge} at room temperature
	but close to zero already at $100\,\K$.

	Instead, if impurities are present in the semiconductor lattice, 
	it is possible for the electronic conductance to take place also at lower $T$. 
	The mechanisms of ``banding'' and ``hopping'' occur in semiconductors which are heavily doped and 
	compensated (equal number of donors and acceptors), respectively.
	In these cases, the charge carriers move from one impurity to the next without reaching the device conduction band.

	The banding mechanism requires an impurity concentration high enough to lead to the substantial overlap between 
	neighboring wave functions. The individual impurity states form an ``impurity band''~\cite{Erginsoy:1950}.
	Depending on the number of doped atoms, therefore, the semiconductor can behave either as an insulator or as a metal, 
	even at low $T$. In general, there exists a critical concentration $N_c$ characterizing the (abrupt) transition between 
	the two opposite situations, which is called the metal-insulator transition (MIT)~\cite{Mott:1969nc}. 

	The hopping mechanism, instead, occurs when compensating (or minority) impurities create a number of majority 
	impurities which remain ionized down to $T=0\,$K.
	In this case, the charge carriers can ``hop'' from an occupied majority impurity site to an empty one
	via quantum mechanical tunneling through the potential barrier which separates the two dopant sites.
	The conduction is activated by phonon mediation.
	The tunneling probability is exponentially dependent on the inter-impurity distance and this is why an extreme 
	homogeneity of the dopant distribution is fundamental.

	In particular, if the dopant concentration is slightly lower than $N_c$, then the resistivity is strongly dependent
	on the temperature. This is why it is usually chosen to operate semiconductor thermistors just below the MIT 
	region. The conduction mechanism in these conditions takes the name of 
	``variable range hopping'', since the carriers can also migrate to far sites if their energy levels are 
	localized around the Fermi energy.
	The resistivity as a function of the temperature is described by the following law:
	\begin{equation}
		\rho = \rho_0 \, \exp \left(\frac{T_0}{T} \right)^\gamma
	\label{eq:NTD_general}
	\end{equation}
	where $\rho_0$ and $T_0$ are parameters depending on the doping level and compensation.
	The exponential $\gamma$ in the Mott model for a 3-dimensional crystal is equal to $1/4$ for low compensation 
	values~\cite{Mott:1969nc}. It becomes $1/2$ for large values of compensation, where the Coulomb repulsion 
	among the electrons leads to the formation of a gap in the electron state density near the 
	Fermi energy~\cite{Shklovskii&Efros:1984}.

\subsection{Energy resolution}
\label{sec:resolution}

	The intrinsic energy resolution of a detector is primarily determined by the statistical fluctuation in the number 
	of elementary events contributing to the signal. 
	In general, more channels are available for the energy transfer, while usually only one (e.\,g.\ scintillation or ionization) is used for the detection.
	The energy resolution thus depends on the fluctuation in the fraction of these useful events.

	Instead, in the case of a sensor sensitive to thermal phonons, one expects no such fluctuation, 
	since all the released energy is sooner or later converted into phonon excitations.
	The effective number of phonon modes in the detector at thermal equilibrium is $N = C(T)/k_\mathrm{B}$, 
	each with quantum occupation number 1, RMS fluctuation of 1 phonon, and mean energy $\varepsilon = k_\mathrm{B}\,T$~\cite{Moseley:1984}.
	As an example, for a $750\,\g$ \ce{TeO_2} crystal at $10\,\mK$ ($C \simeq 2.2\,\text{nJ\,K}^{-1}$~\cite{Baruccietal:2001}),
	this gives $N \simeq 1.6\cdot10^{14}$.
	
	The number of these elementary events follows a Poisson distribution with $\Delta N = \sqrt{N}$
	due to the continuous phonon exchange between the absorber and the heat sink.
	Therefore, the intrinsic  energy resolution $\Delta E$ is proportional to the mean energy of the elementary excitation:
	\begin{equation}
	\label{eq:en_res}
		\Delta E \propto \varepsilon \, \Delta N = \xi \sqrt{k_\mathrm{B}\,C(T)\,T^2}.
	\end{equation}
	The factor $\xi$ is a dimensionless factor that depends on the details of the real detector. 
	It can become of the order of unity with a proper optimization work~\cite{Moseley:1984}.
	It is worthy of note that the expression obtained in Eq.~(\ref{eq:en_res}) is independent of energy.
	To give an idea of the potentiality of these detectors, the intrinsic resolution 
	for the considered $750\,\g$ \ce{TeO_2} crystal lays within the range $(20-100)\,\eV$.

	However, most of the time the effective energy resolution of these kind of detectors is dominated by the ``extrinsic'' noise generated by a variety of 
	uncorrelated sources that are almost impossible to cancel completely~\cite{Mather:1982,Mather:1984}.
	These include Johnson noise, thermal noise (i.\,e.\ noise due to temperature instabilities of the heat sink), 
	noise from the electronics readout and the cryogenic system, mechanical vibrations, \dots\
	Eventually, the combination of the multiple factors often results in an energy resolution of the order of a few keV at $1\,\MeV$ (see Sec.~\ref{sec:LNGS_exp}), 
	a value anyhow comparable with that of the best performing detectors.

\subsection{Detector operation}
\label{sec:det_operation}

	\begin{figure}[t]
		\centering
		\includegraphics[width=.8\columnwidth]{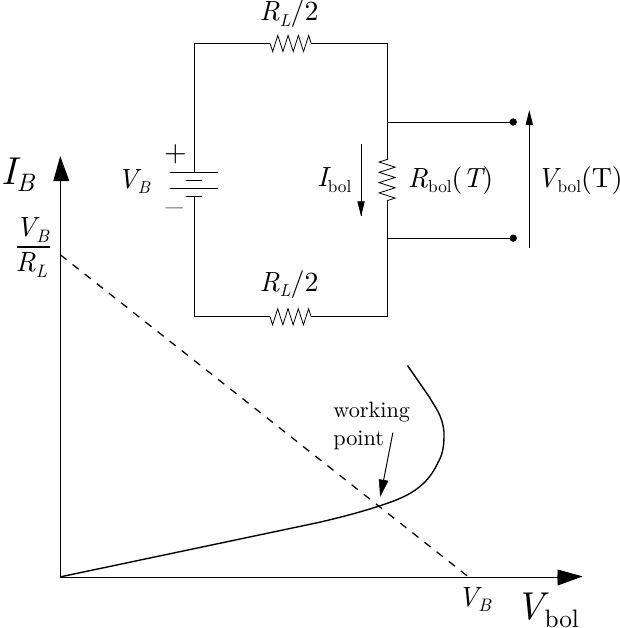}
		\caption{(Top) Electric scheme of the biasing system used for the NTD readout. 
			(Bottom) Typical load curve of a semiconductor thermistor. The working point is set as the intersection between 
			the $V$-$I$ characteristic curve and the load line $V_\mathrm{bol} = V_B - I_B \, R_L$.}
		\label{fig:VI_curve}
	\end{figure}
	
	In order to obtain a voltage signal, a steady bias current $I_B$ is sent through the thermistor by means of the bias circuit shown in Fig.~\ref{fig:VI_curve}.
	The voltage generator is closed on 
	a load resistor $R_L$ in series with the thermistor $R_\mathrm{bol}$, whose resistance is negligible in comparison to 
	$R_L$ ($R_L \gg R_\mathrm{bol}$ therefore $I_B$ is independent from $R_\mathrm{bol}$).
	
	Once a working point is set by forcing a bias current, a voltage drop 
	$V_\mathrm{bol}(T) = I_B\,R_\mathrm{bol}(T)$ is established across the thermistor. 
	The consequent power dissipation $P = I_B \, V_\mathrm{bol}$ produces a temperature rise and acts back on the resistance 
	$R_\mathrm{bol}(T)$ until an equilibrium is reached:
	\begin{equation}
		T_\mathrm{bol} = T_S + \frac{P}{G}
	\end{equation} 
	where $G$ the thermal conductance between the detector and the heat sink ($T_S$).
	
	This phenomenon makes the $V$-$I$ relation deviate from linearity and leads to a non-ohmic behavior.
	It is often referred to as ``electrothermal feedback''.
	For a given $I_B$, the ``static'' resistance is simply the ratio $V_\mathrm{bol}/I_B$ while the 
	``dynamic'' resistance is the inverse of the tangent to the $V$-$I$ curve, usually called ``load curve''.
	By further increasing $I_B$, the dynamic resistance crosses the so called inversion point (where it vanishes) and 
	becomes negative.
	The intersection of the load curve, with the load line imposed by the biasing 
	system determines the working point of the sensor (plot in Fig.~\ref{fig:VI_curve}). 
	This is typically chosen before the inversion point in order to maximize the signal amplitude and the signal-to-noise 
	ratio. By a combined fit of load curves measured at different base temperatures, it is possible to evaluate 
	the thermistor intrinsic parameters $\rho_0$, $T_0$ and $\gamma$.
	
	In first approximation, the thermal pulse produced by an energy release in the absorber is characterized by a 
	very fast rise time (instantaneous if we assume the model described in Sec.~\ref{sec:thermal_model} and a 
	negligible thermalization time). The fall time, instead, follows an exponential decay whose time constant depends 
	on the physical characteristics of the detector (recall Eq.~(\ref{eq:time_evolution})).
	The relationship between the electrical pulse height $\Delta V_\mathrm{bol}$ and the energy deposition $E$
	can be obtained by resolving the circuit of Fig.~\ref{fig:VI_curve}:
	\begin{equation}
		\Delta V_\mathrm{bol} = \text{const} \cdot \gamma \left( \frac{T_0}{T_\mathrm{bol}} \right)^\gamma \cdot 
		\frac{E}{C \cdot T_\mathrm{bol}} \cdot \sqrt{P \cdot R_\mathrm{bol}}
	\end{equation}
	where $P$ is the power dissipated in the thermistor $R_\mathrm{bol}$ by Joule effect.
	In particular, this expression vanishes in the limits $P \rightarrow \infty$ and $C \rightarrow \infty$. 
	To make an example, an energy deposition of $1\,\MeV$ in a $750$\,g \ce{TeO_2} absorber at 10\,mK (see Sec.~\ref{sec:teo2_bol}) 
	typically produces a temperature increase $\Delta T \sim 100\,\upmu$K, with a related voltage drop 
	$\Delta V_\mathrm{bol} \sim 100\,\upmu$V.
	
\label{sec:heaters}

	A critical issue when operating bolometric detectors over long periods of time, months or even years, consists in maintaining a stable response,
	despite the unavoidable temperature fluctuations due to the cryogenic setup.
	
	An effective approach to this issue can be found in the use of a pulser able to periodically deliver a fixed 
	(and extremely precise) amount 
	of energy to the detector and to generate a pulse as similar as possible to the signal corresponding to a real 
	event~\cite{Alessandrello:1998bf,Arnaboldi:2011zza}. 
	In this way, the study of the variation of the detector response to the same energy deposition can be used to correct 
	(off-line) the effects of the cryogenic instabilities.
	
	A particle-based stabilization would present the advantage of getting an identical detector response to the pulser and 
	to the events of interest. However, this method would present some important drawbacks: a series of undesired cascade
	peaks would be generated by the same source and the Poisson time distribution of the events would limit the calibration rate. 
	Furthermore, the calibration pulses could not be ``flagged'' and their off-line identification would be based only on their amplitude.

	Instead, a resistive element thermally coupled to the crystal, can be used to inject calibrated amounts of energy 
	via Joule heating. This solution offers the advantage of a complete control of the calibration mechanism: 
	the pulses can be equally spaced in time and their rate and amplitudes can be easily tuned to the necessity of 
	the experiment. 
	Anyhow, it has to be noted that such a heating element needs to satisfy some (non-obvious) requirements.
	In particular, its resistance must be reasonably independent of temperature and applied voltage, while 
	its heat capacity must be negligible with respect to the detector's one.
	In order to provide an almost instantaneous energy release, the relaxation time to the crystal of the developed heat 
	must be much shorter than all the typical thermal time constants.
	Finally, the mechanism of signal formation for particle interactions and for Joule heating must be similar enough to 
	assure that the pulse amplitude dependence on time, baseline level and other operation conditions are the same for the 
	two processes.
	
	Steady resistances useful for this purpose can be realized through a heavily doped semiconductor, 
	well above the MIT region, 
	so that a low-mobility metallic behaviour is exhibited~\cite{Andreotti:2012zz}. 
	In particular, silicon heaters have been used for the detector response stabilization in the 
	experiments described in Sec.~\ref{sec:LNGS_exp}.

\subsection{Properties of TeO$_2$ bolometers}
\label{sec:teo2_bol}

	Tellurium dioxide, \ce{TeO_2}, is a particularly suitable material to be employed
	in cryogenic particle detectors. \ce{TeO_2} is the most stable oxide of \ce{Te}~\cite{Dutton:1966a} 
	and presents favorable thermodynamic characteristics. 
	\ce{TeO_2} crystals are both dielectric and diamagnetic 
	with a relatively high value of the Debye temperature ($\Theta_\mathrm{D} = (232 \pm 7)$\,K~\cite{Baruccietal:2001}).
	The very low heat capacity at cryogenic temperatures leads to large temperature variations from tiny energy 
	releases, which is the basis for a high energy resolution bolometer.
	Moreover, the fact that the thermal expansion of \ce{TeO_2} crystals is very close to that of 
	copper~\cite{White:1990,Kroeger:1977}, allows the use of the latter for the detector mechanical support structure without 
	placing too much strain on the crystals during the system cool down.
	
	\ce{TeO_2} is present in nature in two mineral forms: orthorhombic tellurite ($\beta$-\ce{TeO_2}) and paratellurite 
	($\alpha$-\ce{TeO_2}). 
	In particular, paratellurite is commercially produced at the industrial scale and employed in ultrasonic light deflectors and in laser 
	light modulators~\cite{El-Mallawany:2002}, thanks to its acoustic and optical properties. 
	The crystals have mechanical and optical 
	characteristics fully compliant with the requirements for the use as thermal detectors. 
	Moreover, the numerous improvements in preparing the \ce{TeO_2} powder and in growing the crystals allow the 
	production of almost perfect large specimens (bubble-free, crack-free and twin-free%
	\footnote{Crystal twinning is the intergrown of two or more separate crystals which share some of the crystal lattice points.}%
	) from which it is possible to construct bolometers with masses of the order of 
	$750$\,g~\cite{Chu2006158,Arnaboldi:2010fj} or more~\cite{Arnaboldi:2005yf,Cardani:2011vg}. 

	It is fundamental to notice that the use of a material inside a \bb~detector imposes very 
	stringent constraints in terms of radiopurity. Therefore, tight limits must be set for the crystal impurity content
	in order to get an acceptable background rate.
	\ce{TeO_2} also satisfies these requirements.
	Thanks to dedicated production lines for the raw powder synthesis, the crystal growth and the 
	surface processing, crystals with bulk contaminations of the 
	order of $1$\,\uBqkg for \ce{^{238}U} and \ce{^{232}Th} are 
	available~\cite{Arnaboldi:2010fj,Alessandria:2011vj}.
	At the same time, dedicated surface treatments translate in surface contamination levels
	of a few $10^{-9}\,\text{Bq\,cm}^{-2}$ for both \ce{^{238}U} and \ce{^{232}Th}~\cite{Alessandria:2011vj,Alessandria:2012zp}.


%% file: 3_LNGSExperiments.tex
\section{A long chain of experiments}
\label{sec:LNGS_exp}

	\begin{figure}[t]
		\centering
		\includegraphics[width=\columnwidth]{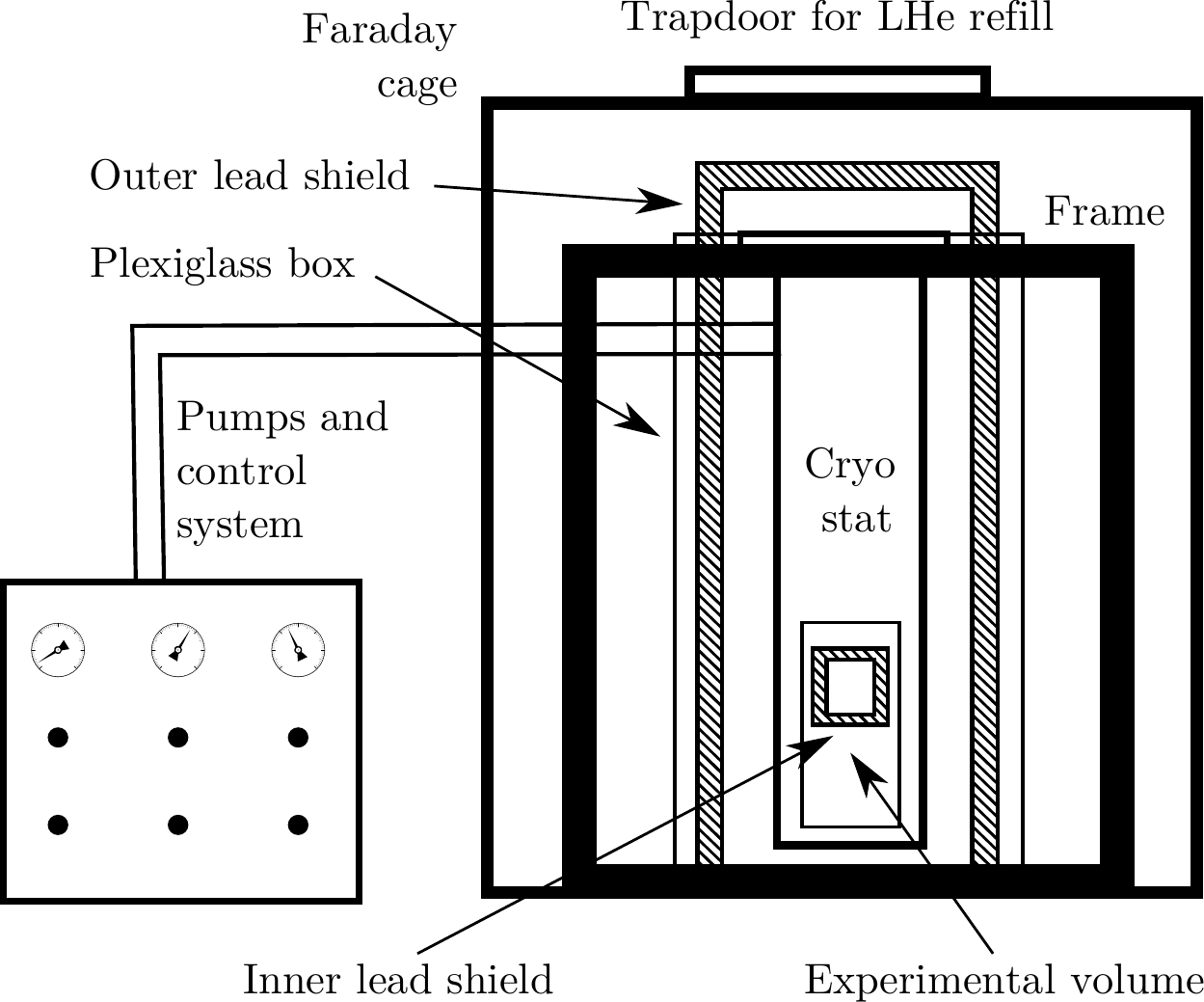}
		\caption{Schematic of the cryogenic system in the Hall A of the Laboratori Nazionali del Gran Sasso.
			Figure from Ref.~\cite{Alessandrello:1992jc}.}
		\label{fig:LNGS_DR}
	\end{figure}

	Early measurements of $\gamma$-ray spectroscopy with bolometers were carried out at LASA,
	where a Dilution Refrigerator (DR) from Oxford Instruments was installed in 1987~\cite{Alessandrello:1988dv}.
	Initially, pure \ce{Te} crystals ($2.1$\,g~\cite{Alessandrello:1990wd} and $28$\,g~\cite{Alessandrello:1990te})
	were tested, but the metal was found to be too brittle at low temperatures.
	A positive outcome was instead obtained by the use of \ce{TeO_2} crystals 
	($6\,\g$~\cite{Alessandrello:1992aa} and $21\,\g$~\cite{Giuliani:1991ze}).
	Given the promising results, the same crystals were operated also at LNGS.

	A DR (also by Oxford Instruments) was installed in the Hall A of LNGS in 1989~\cite{Alessandrello:1989bu,Alessandrello:1991fig}.
	A schematic of the cryogenic system is shown in Fig.~\ref{fig:LNGS_DR}.%
	\footnote{The setup actually evolved over the years, as it will become 
	clear from the following sections.}
	The detector was contained in an Oxygen Free High Conductivity (OFHC) copper frame.
	The DR was constructed with specially chosen low radioactivity components. 
	All materials inside the refrigerator had been previously tested for radioactivity with a germanium spectrometer.
	The system was built in order to mechanically support loads for tens of kg at low temperature, so that it would have become possible to 
	install radioactive shields inside the cryostat and run large bolometric arrays.
	Indeed, starting after the first few runs, different lead shields were added and changed 
	depending on the geometry of the current detector.

	Outside the DR, a $10$\,cm minimum thickness lead layer protected the system against external radioactivity.
	Later, another $10$\,cm layer made of low radioactivity lead was added (internally to the previous one)
	and the shield was enclosed in a Plexiglas box with anti-radon purpose, constantly flushed with nitrogen gas from a 
	\ce{LN_2} evaporator.
	The cryogenic system, together with the readout electronics, were in turn contained inside a Faraday cage in order to 
	minimize electromagnetic interferences.
	
	This low-radioactivity cryogenic infrastructure inside an underground site
	represented the ideal facility for the study of rare events, such as \bb.
	
\subsection{Measurements with single crystals}
\label{sec:single_crystals}

	Given the favorable environment offered by the Hall A DR, in order to perform an experimental search for \bb, 
	it was now crucial to construct a suitable low-background bolometric detector.
	In particular, a critical point consisted in the design of the crystal holder, since this element 
	would have strongly affected both the detector performance and the background rate. 

\begin{figure}[tb]
		\centering
		\includegraphics[width=.75\columnwidth]{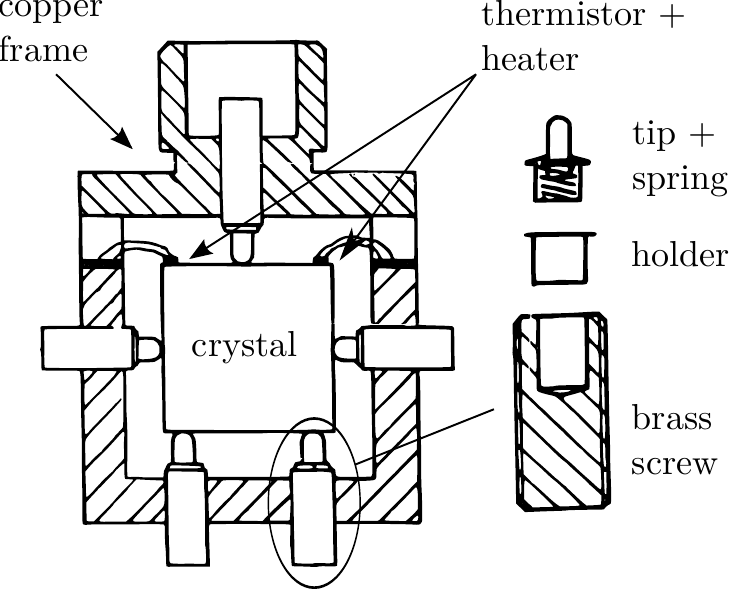}
		\caption{Cross section of the frame used to hold the $6\,\g$, $21\,\g$ and $34\,\g$ crystal bolometers. 
			Figure from Ref.~\cite{Alessandrello:1992aa}.}
		\label{fig:6g_holder}
	\end{figure}

	The holder had to provide mechanical support to the detector in place while, at the same time,
	it had not to constitute a (strong) thermal link to the heat sink.
	In this way, the heat would have been conducted away from the absorber only via the thermistor (and hence measured) 
	and the crystal would have been prevented from oscillating around its equilibrium position, due to the vibrations induced by the refrigerator cooling unit.%
	\footnote{Vibrations cause an increase of the operating temperature since they dissipate power by means of mechanical 
		friction, in turn increasing the detector heat capacity (recall Eq.~(\ref{eq:time_evolution})).
		This mechanism generates noise due to the random vibrational heating and microphonic noise due to the 
		modification of the parasitic electrical capacitance shunting the current across the thermistor.}
	A compromise between a good mechanical support and a weak thermal contact had to be found in order 
	to minimize the effects of the loss in the signal amplitude and to prevent the generation of microphonic noise.
	Moving in this direction, the crystal was suspended with brass screws whose extremities had small spring loaded needles 
	which could exert an adjustable pressure (Fig.~\ref{fig:6g_holder}). 

	Therefore, the first measurements performed at LNGS had also the goal of investigating the background level of the whole 
	experimental apparatus and of testing the effectiveness of the adopted solutions.

\subsubsection{6\,g crystal}	
\label{sec:6g_crystal}

\begin{figure}[tb]
		\centering
		\includegraphics[width=1.\columnwidth]{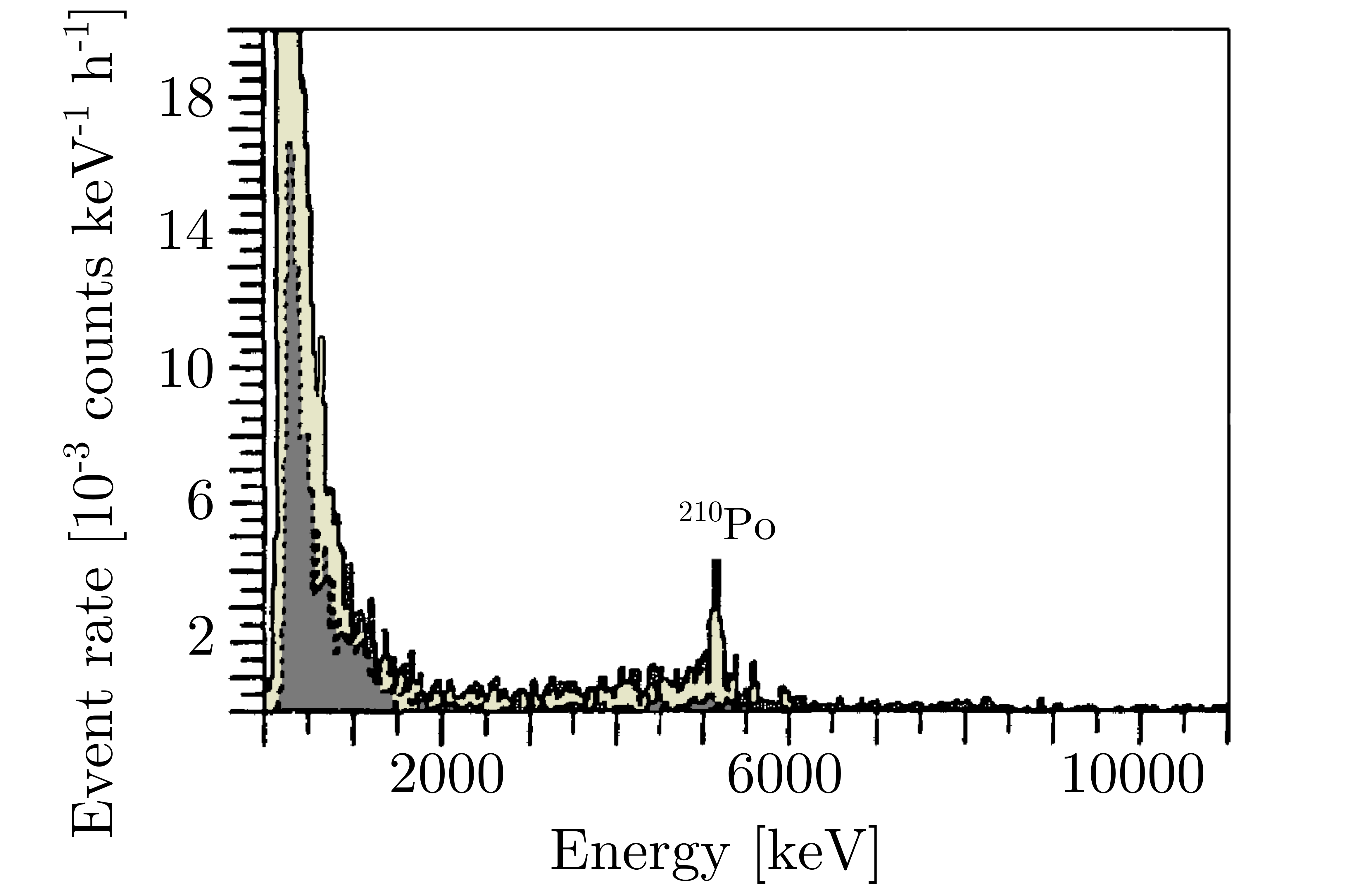}
		\caption{Background spectra of the 6\,g \ce{TeO_2} bolometric detector acquired during the first (light) 
			and second (dark) runs.
			The contribution to the $\alpha$ background coming from the solderer is clearly visible in the former spectrum.
			Figure from Ref.~\cite{Alessandrello:1992aa}.}
		\label{fig:6g_spectrum}
	\end{figure}

	The $6\,\g$ crystal~\cite{Alessandrello:1992aa} was operated in two following runs of $370$\,h and $260$\,h, respectively.
	The first run presented an excessively high $\alpha$ counting, with the \ce{^{210}Po} peak clearly visible in the 
	background spectrum (light spectrum in Fig.~\ref{fig:6g_spectrum}). 
	The source of the contamination was identified in the tin used for the soldering required to fix the needles, which was
	directly facing the detector, to the screws.
	Once the tin was eliminated (dark spectrum in Fig.~\ref{fig:6g_spectrum}), 
	the average background rate in the $(3-11)\,\MeV$ interval passed from 
	$(560 \pm 20)$\,\ckky~to $(115 \pm 10)$\,\ckky.
	In the Region of Interest (ROI) around the \ce{^{130}Te} $Q_{\beta\beta}$, the measured value was $123$\,\ckky. 
	The achieved energy resolution was $\sim50\,\keV$ FWHM, rather independent of the energy.
	
	The $6\,\g$ crystal allowed to set the first, already competitive, limit on the \ce{^{130}Te} \bb~half-life 
	with a \ce{TeO_2} bolometer (see Sec.~\ref{sec:limits}).
	This was one of the first physics results ever obtained with the bolometric technique.
	
\subsubsection{21\,g crystal}	

	The $21\,\g$ crystal~\cite{Giuliani:1991ze} was operated for $\sim 160$\,h, keeping the same configuration for the 
	detector holder (Fig.~\ref{fig:6g_holder}).
	Nevertheless, the new crystal exhibited a FWHM energy resolution of $\sim 20\,\keV$ at $2615\,\keV$, which represented an 
	improvement of more than a factor $2$.
	This was obtained thanks to a better thermistor and an optimized mounting. 
	The main limitations to the resolution remained the microphonic and the electronic noise and the instabilities in 
	the operating temperature.
	At the same time, the careful selection of the material facing the detector and a proper choice of the crystal itself allowed 
	to keep the background in the ROI, mainly due to $\alpha$ contamination, at the level of $88$\,\ckky. 
	
\subsubsection{34\,g crystal}	

	\begin{figure}[b]
		\centering
		\includegraphics[width=1.\columnwidth]{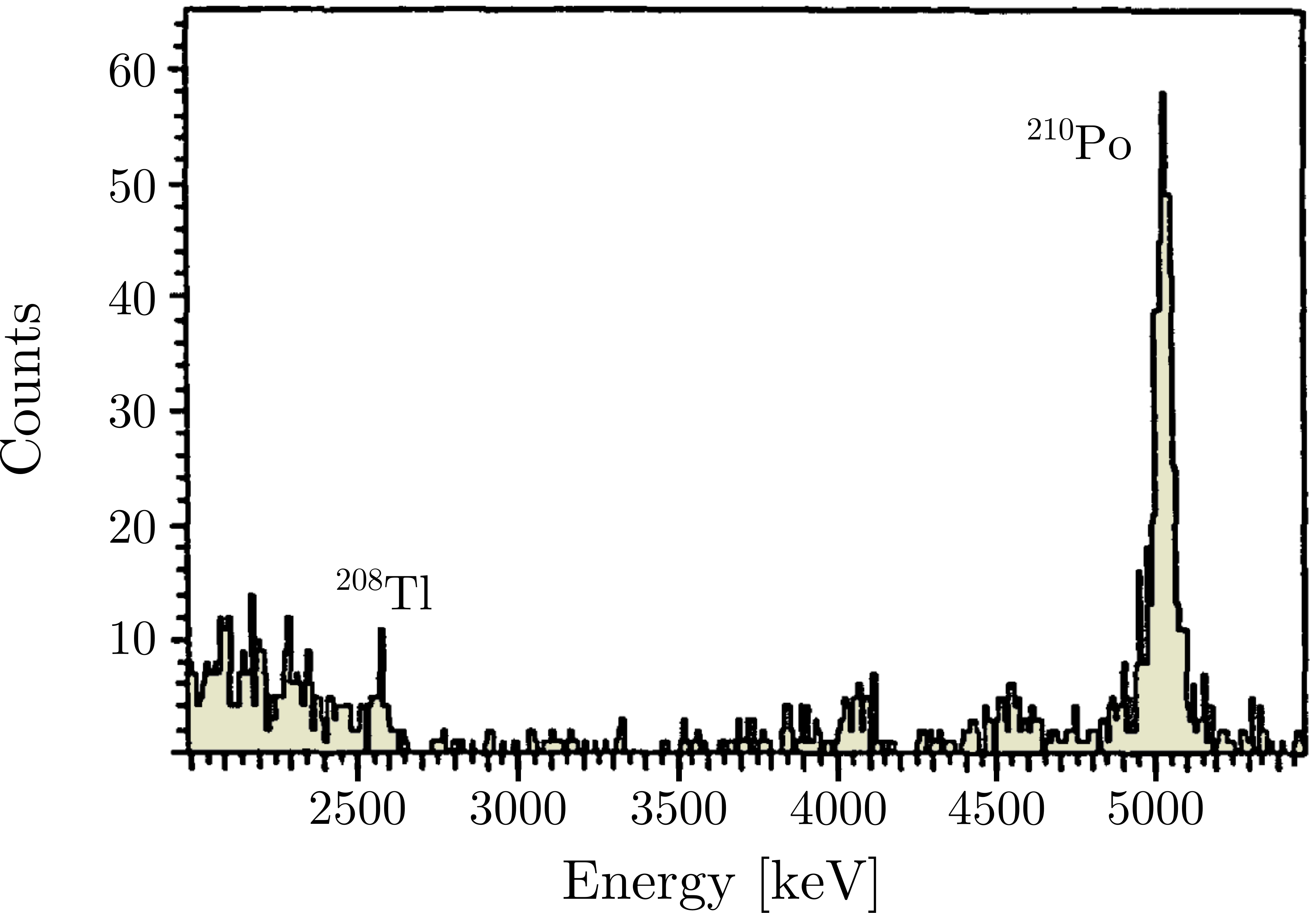}
		\caption{Background spectrum of the 34\,g \ce{TeO_2} bolometric detector obtained in $1051.4$\,h.
		 	The wrong position of the \ce{^{210}Po} peak ($5\,\MeV$ instead of $5.4\,\MeV$) is due to the linear fit used in the calibration, 
		 	while bolometers are naturally non-linear in the energy response (due to the electrothermal feedback).
			Figure from Ref.~\cite{Alessandrello:1992jc}.}
		\label{fig:34g_spectrum}
	\end{figure}

	A $34\,\g$ crystal~\cite{Alessandrello:1992jc} was operated shortly after the $21$\,g one in the usual setup.
	The live time of the experiment was over $1050$\,h, thus obtaining a $\sim 10$ times larger exposure.
	The total background spectrum is shown in Fig.~\ref{fig:34g_spectrum}.
	The background level was compatible with the one obtained in the previous run.
	The energy resolution worsened instead by a factor $2$ ($\sim 40\,\keV$ above $1\,\MeV$).
	In fact, in order to check a possible background contribution coming from the heater resistor, the chip was not glued on the crystal. 
	The detector response stability was routinely checked by means of \ce{^{60}Co} and \ce{^{232}Th} sources illuminating the detector 
	through a small window in the lead shield that could be opened.
	However, the effectiveness of this system cannot reach the same level of performance of a heater-based one (see Sec.~\ref{sec:heaters}).

	In the end, the hypothesis of a significant contribution to the background coming from the 
	heater itself was found to be wrong.
	
\subsubsection{73\,g crystal}	
\label{sec:73g}

	\begin{figure}[tb]
		\centering
		\includegraphics[width=.9\columnwidth]{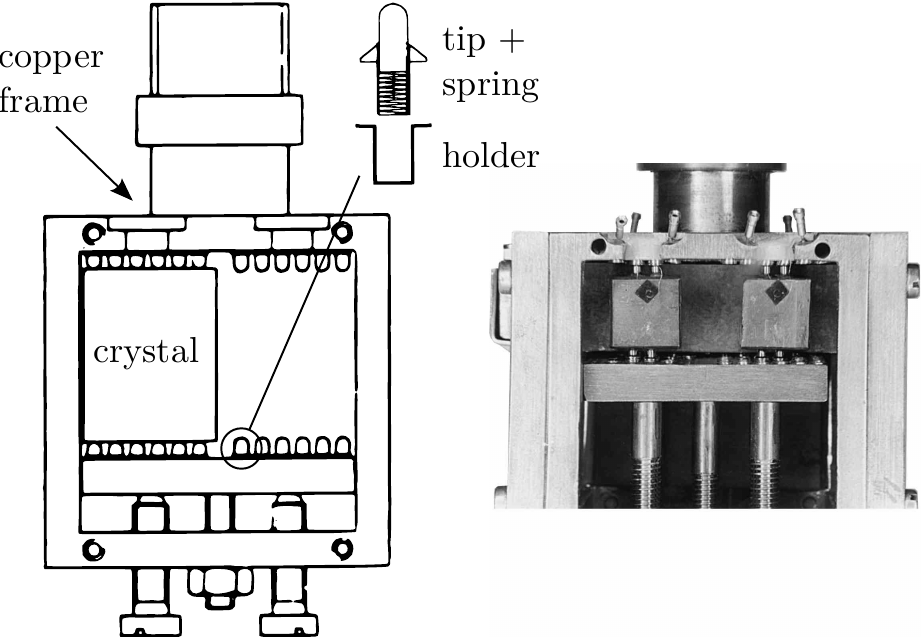}
		\caption{(Left) Cross section of the frame used to hold the $73\,\g$ and $334\,\g$ crystal bolometers. 
			(Right) The same frame during the radioactivity measurements of the Roman lead used for the DR internal shielding. 
			Figure from Ref.~\cite{roman_lead:1998}.}
		\label{fig:73g_holder}
	\end{figure}

\begin{figure}[tb]
		\centering
		\includegraphics[width=.8\columnwidth]{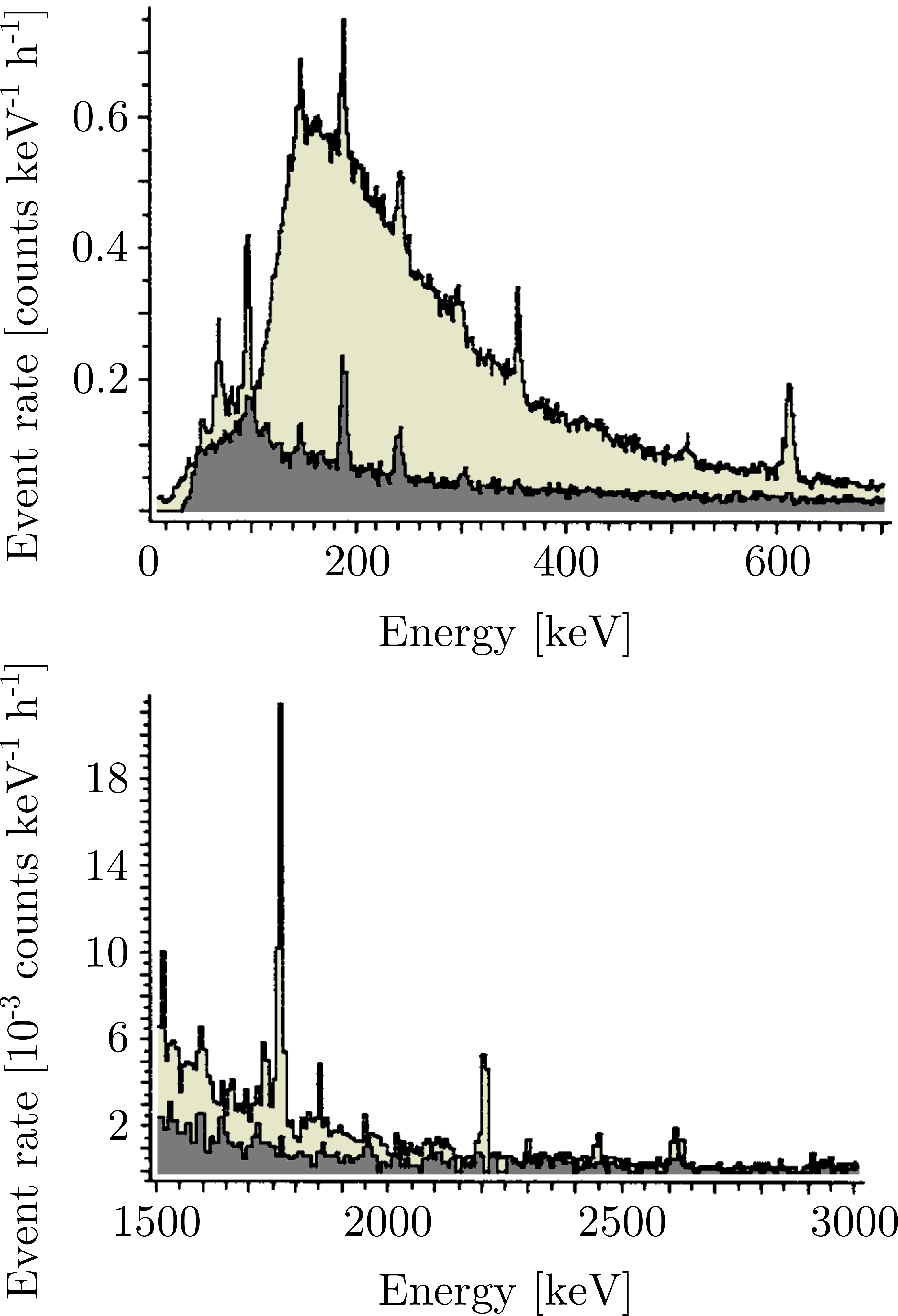}
		\caption{Background spectra of the 73\,g \ce{TeO_2} bolometric detector acquired during Run 1 (light) 
			and Run\,II (dark). The effect of the new lead shield on the background is clearly visible. 
			(Top) $(0-700)\,\keV$ energy region (Bottom) $(1500-3000)\,\keV$ energy region.
			Figure from Ref.~\cite{Alessandrello:1993cs}.}
		\label{fig:73g_spectra}
	\end{figure}

	The next measurement was carried out with a $73\,\g$ crystal~\cite{Alessandrello:1993cs,Alessandrello:1992fig}. 
	This time, an improved version for the crystal holder was used (Fig.~\ref{fig:73g_holder}).
	The detector was operated for more 
	than four months in two following runs of $1389$ and $1046$ hours of effective live-time, respectively. 
	In between, an inner ultra-low activity lead shield of $3.5$\,cm minimum thickness was inserted inside the cryostat 
	(Fig.~\ref{fig:LNGS_DR}). 
	This was completely surrounding the detector and allowed to reduce the background of a factor $\sim 2$ in the ROI (around $2.5\,\MeV$),
	bringing it to less than one quarter with respect to that of the $34$\,g crystal.
	The strong background suppression is clearly visible in the comparison between the spectra acquired in the two runs,
	as shown in Fig.~\ref{fig:73g_spectra}.
	The obtained energy resolution was extremely good: $7\,\keV$ FWHM in the ROI, not too different from that of a Ge diode.

\subsubsection{334\,g crystal}

	The successive measurement represented an important breakthrough for the bolometric technique. The new detector,
	a $3\times3\times6\,\cm^3$ crystal of $334$\,g~\cite{Alessandrello:1994tm}, was indeed the largest 
	thermal detector operated so far.
	Although the resolution was worse than that of the $73$\,g crystal, still mainly limited by the microphonic and electronic 
	noise, it remained similar to the ones previously obtained.
	On the contrary, the background in the ROI was reduced to about $3.4$\,\ckky. 
	In total, more than $10500$\,h of effective running time~\cite{Alessandrello:1994_XX} 
	were collected in about 18 months. This translated into a factor $50$ of increase in the exposure.

\subsection{Detector arrays}

	During the period of the $334$\,g crystal run (around 1994), the growth technique for \ce{TeO_2} did not allow the 
	production of specimens with mass larger than $\sim 0.5$\,kg without risks of defects and fragility~\cite{Brofferio:1995wx}.
	Therefore, in order to improve the sensitivity by further increasing the exposure, it became necessary 
	to begin to assemble arrays of bolometers.

	With the growth in complexity of the detectors and, especially, due to the very long data taking periods
	in the Hall A DR, the use of a dedicated test facility became fundamental in view of the new generation of
	experiments~\cite{Alessandrello:1995hq}.
	In 1992, a new DR was installed inside the Hall C (Fig.~\ref{fig:LNGS}).
	Similarly to the one in Hall A, the Hall C DR was built with low radioactive materials and,
	over the years, provided with an external shielding in 
	all directions ($15$\,cm lead + copper) and an anti-radon Plexiglas box.
	The internal shielding (against the contamination from the dilution unit materials) 
	was provided by a $5.5$\,cm lead layer just above the detector .
	This was made of ancient Roman lead (I century BCE, Hispanic origin) 
	with a \ce{^{210}Po} activity less than $4\,\cdot 10^{-3}\,\Bq\,\kg^{-1}$~\cite{roman_lead:1998}.
	Inside the cryostat, a new system of mechanical suspensions was implemented to reduce vibration and thermal 
	noise~\cite{Pirro:2000vib,Pirro:2006mu}.
	
\subsubsection{4 crystal array}

\begin{figure}[tb]
		\centering
		\includegraphics[width=.6\columnwidth]{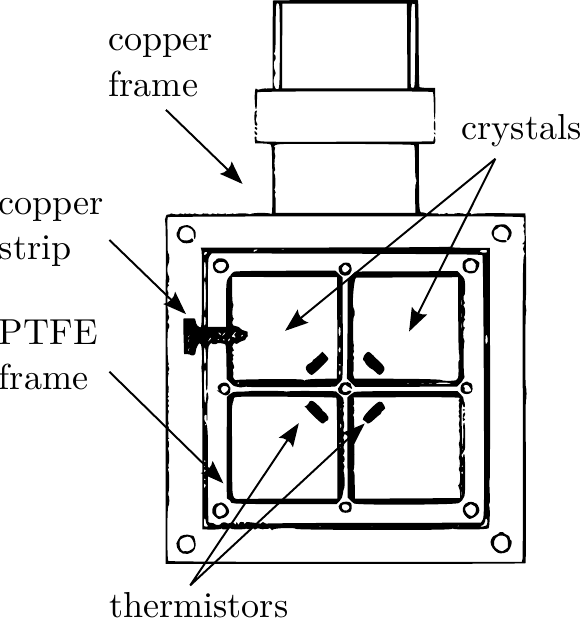}
		\caption{Cross section of the frame used to hold the 4 crystal array. 
			Figure from Ref.~\cite{Alessandrello:1995hq}.}
		\label{fig:4Xtal_array_holder}
	\end{figure}

	The first array of \ce{TeO_2} bolometers was constituted by 4 crystals very similar both in mass 
	and in dimensions to the last operated one~\cite{Alessandrello:1995hq}.
	The levels of the crystal internal contamination were satisfactory: 
	$4.8\cdot 10^{-6}\,\Bq\,\kg^{-1}$ in both \ce{^{238}U} and \ce{^{232}Th}~\cite{Brofferio:1995wx}.
	A new concept for the crystal holder was adopted
	and tested in the Hall C DR in 1994.
	The spring loaded tip system was substituted with polytetrafluoroethylene (PTFE) frames,
	which allowed to reduce the radioactive contamination (Fig.~\ref{fig:4Xtal_array_holder}).

	A physics run with the new detector was then performed in the Hall A cryostat~\cite{Alessandrello:1996dd}.
	The 4 crystal array constituted the first bolometric \bb~experiment with over $1$\,kg of mass.
	Unfortunately, one thermistor out of 4 turned out to be damaged.
	In 1995, $\sim 4$ months of active live-time could be collected~\cite{Brofferio:1995wx}.
	The installation of a re-liquefier (directly connected to the DR) required quite a time to reassess the system,
	decreasing the duty cycle of the experiment.%
	\footnote{This system avoided the need of frequent LHe refills and guaranteed a constant LHe level, 
		thus stabilizing the internal temperatures of the DR. 
		It was used until September 2004, during the run of Cuoricino (see Sec.~\ref{sec:Cuoricino}), 
		and then abandoned due to its complexity. In fact, the liquefier required very frequent maintenance, thus reducing the live-time of the experiment.}
	The final exposure was therefore lower than that obtained with the $334$\,g crystal alone. 
	Anyway, this measurement proved that the same good conditions of resolution and background rate 
	were achievable even with a complex system of bolometers.
	The 4 crystal array represented a prototype for a more complex detector to come.
	
\subsubsection{MiDBD}

	\begin{figure}[t]
		\centering
		\includegraphics[width=.9\columnwidth]{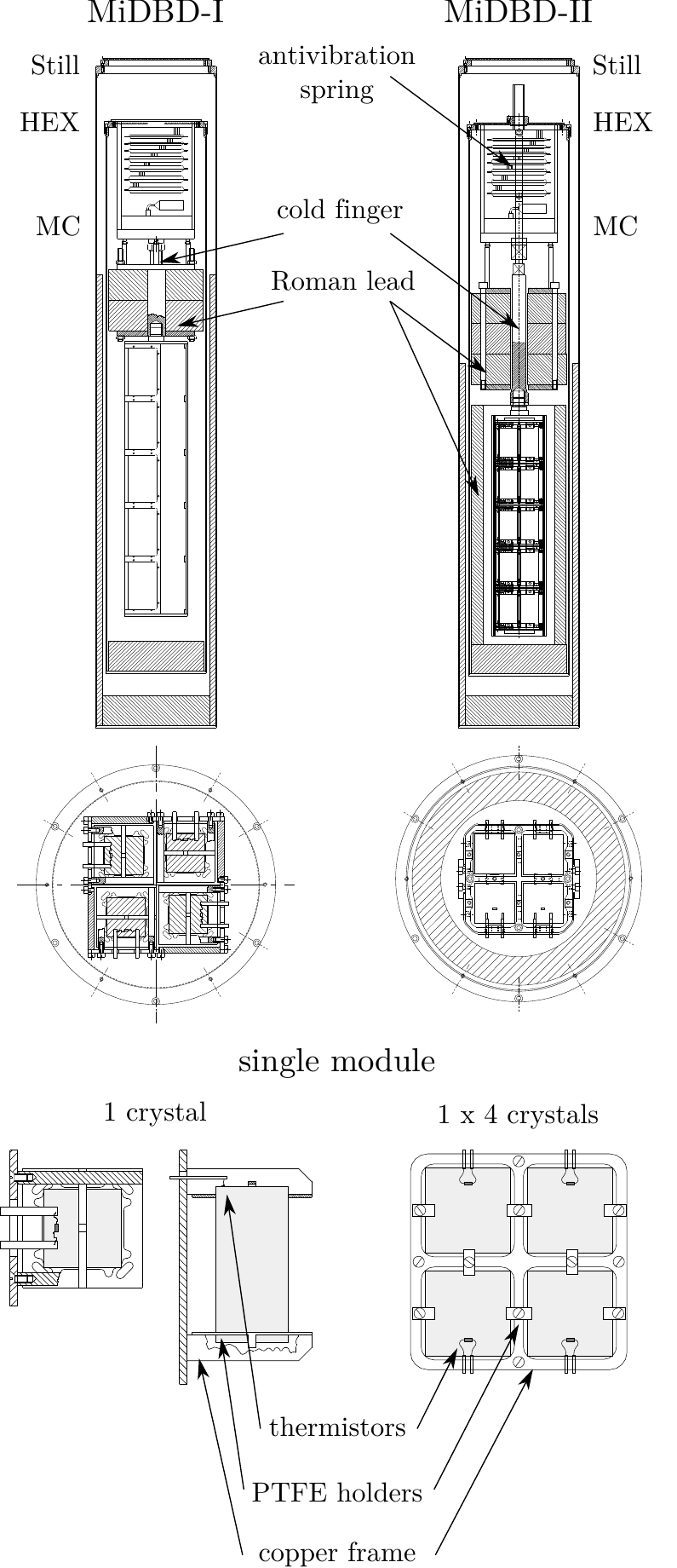}
		\caption{Schematic of the MiDBD experiment: comparison between the old and the new configuration with a detailed view of the single modules.
			Adapted from Ref.~\cite{Arnaboldi:2002te}.}
		\label{fig:MiDBD_scheme}
	\end{figure}

	In the summer of 1997, a tower made of 20 bolometers (5 floors of 4 crystals each) was assembled and installed inside 
	the Hall A DR~\cite{Alessandrello:2000kt}. This experiment was later named MiDBD (Milan Double Beta Decay).
	The single module absorber consisted of a ``standard'' $3\times3\times6\,\cm^3$ crystal, for a total \ce{TeO_2} mass of about $6.8$\,kg. 
	MiDBD was thus the new largest operating cryogenic mass.
	In March 1999, four natural crystals were replaced with four isotopically enriched crystals,
	two in \ce{^{128}Te} ($82.3\%$) and two in \ce{^{130}Te} ($75.0\%$).
	The enriched crystals allowed to study the \bbvv~by analyzing the different spectra~\cite{Arnaboldi:2002te}.

	The tower frame was made of OFHC copper connected via an OFHC copper cold finger to the MC.
	As for the 4 crystal array, the individual bolometers were fastened to this structure by means of PTFE and copper supports 
	(left side in Fig.~\ref{fig:MiDBD_scheme}).
	The detector was laterally surrounded by a $1\,\cm$ Roman lead shield and enclosed between two $10\,\cm$ 
	thick disks, also made of Roman lead.
	
	A new dedicated front-end system for the readout of the large array was developed, which allowed all the necessary parameters for each detector
	to be set remotely~\cite{Alessandrello:2000fs,Pessina:2000fu}.
	In a preliminary run, intended to test the overall performance of the array, only a part of the readout electronics 
	was ready. As a consequence only eight channels could be read out simultaneously~\cite{Alessandrello:1998ey}.
	Then, the system became fully operative and data for a total of $3.60$\,kg\,yr of exposure were collected
	(MiDBD-I). The major improvement with respect to the previous measurement consisted in the 
	drastic (almost a factor 10) suppression of the background rate, now at the level of ($0.59 \pm 0.06$)\,\ckky. 
	The results from MiDBD-I showed that, to gain in terms of overall detector performance, it was necessary to reduce 
	the spread among the single bolometer performance.
	
	\begin{figure}[b]
		\centering
		\includegraphics[width=.9\columnwidth]{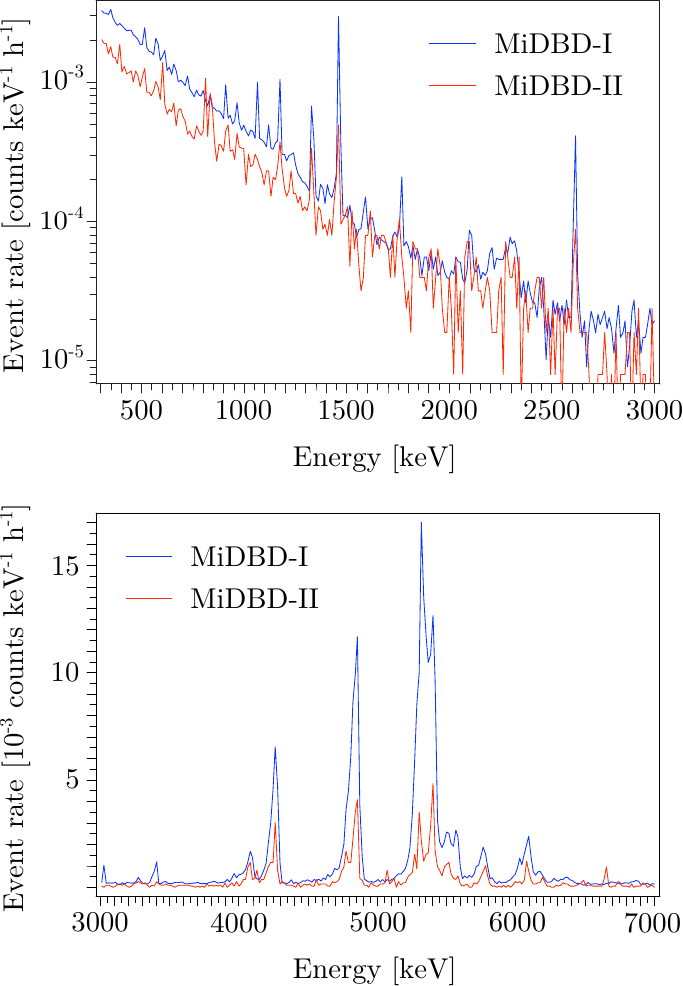}
		\caption{Background comparison between MiDBD-I (blue) and MiDBD-II (red).
			(Top) $(300-3000)\,\keV$ energy region (Bottom) $(3000-7000)\,\keV$ energy region.
			Figure from Ref.~\cite{Capelli_PhD-thesis:2005}.}
		\label{fig:MiDBD_spectra}
	\end{figure}

	At the beginning of 2001, the MiDBD detector was therefore completely dismounted and remounted in a new configuration~\cite{Arnaboldi:2002te}.
	A comparison between the old and the new configurations is shown in Fig.~\ref{fig:MiDBD_scheme}.
	it can be seen that the differences between MiDBD-I and \mbox{MiDBD-II} were very significant.
	A more compact assembly of the crystals was adopted. The fundamental module was no more constituted 
	by a single crystal, but by each plane of the tower (bottom panel in the figure). 
	The new design allowed both for a better anti-coincidence analysis and for the reinforcement of the Roman lead shield. 
	The thickness of the lead layer between the MC and 
	the detector was increased by 5\,cm and a new lateral shield of 2\,cm minimum thickness was added.
	Moreover, the PTFE masks were substituted by small PTFE pieces, with a consequent reduction of the material 
	facing the crystals.
	New electric connections with copper pins avoided the use of the soldering indium (rich in the contaminant \ce{^{115}In}),
	since the gold wires from the thermistors could now be crimped to the pins, 
	and an anti-vibration damping method was implemented to reduce mechanical noise.
	Finally, both the crystals and the copper were cleaned in order to reduce their own radioactive contamination and that 
	coming from the production process.

	The detector was operated in the new configuration for almost a year ($0.65$\,kg\,yr of exposure).
	The individual FWHM energy resolutions at the $2615\,\keV$ line ranged from 5 to 15 keV.
	As shown in the spectra of Fig.~\ref{fig:MiDBD_spectra},
	the introduced changes translated in a clear reduction both in the continuum and in the peak intensities 
	with respect to MiDBD-I: a factor $\sim 2$ for the \ce{^{238}U} and \ce{^{232}Th} $\alpha/\gamma$ peaks and for the 
	$(3-4)\,\MeV$ continuum and a factor $4$ for the 5.3 and 5.4\,MeV peaks of \ce{^{210}Po}.
	A reduction of a factor $\sim 1.5$ was observed in the gamma continuum between 1 and 3\,MeV, 
	with the background level in the \bb~energy region of ($0.33 \pm 0.11$)\,\ckky (see the discussion on the 
	background in Sec.~\ref{sec:bkg_study}).

\subsubsection{Cuoricino}
\label{sec:Cuoricino}

	\begin{figure}[p]
		\centering
		\includegraphics[width=.9\columnwidth]{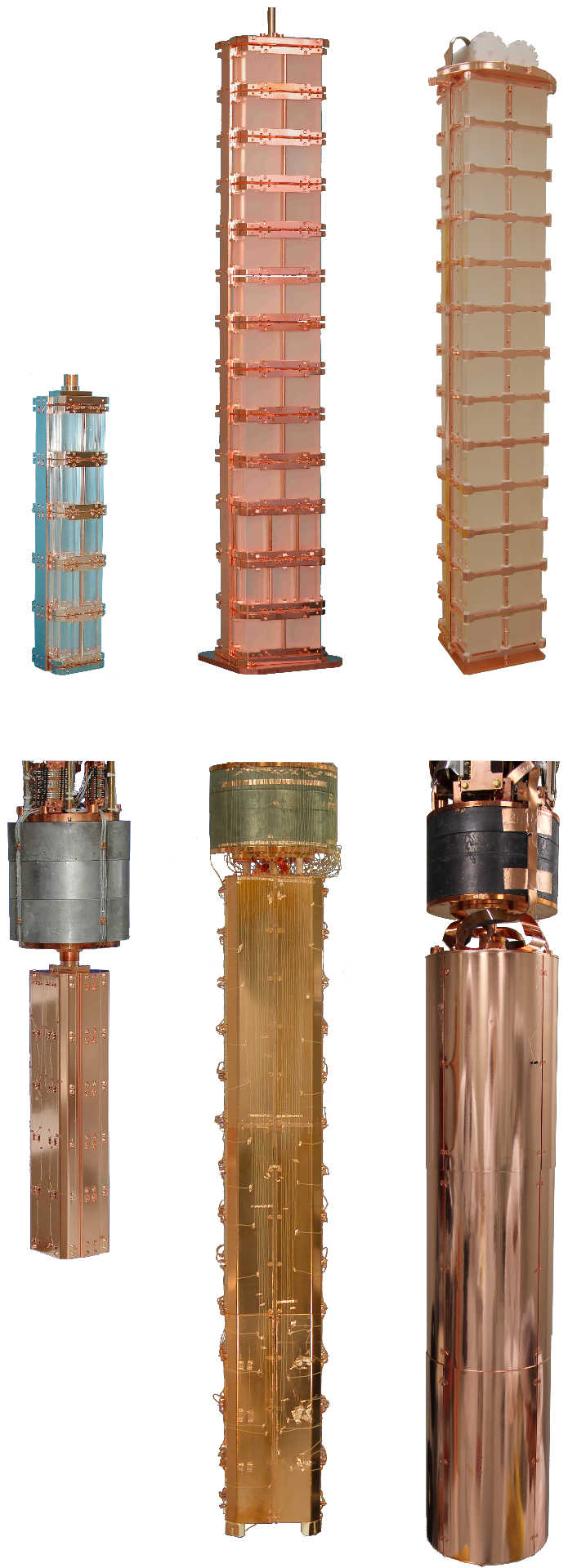}
		\caption{From left to right: the MiDBD-II, Cuoricino and CUORE-0 towers in relative scale.
			(Top) The completed towers. (Bottom) The towers attached to the Hall A DR.
			Adapted from Ref.~\cite{Alduino:2016vjd} and by courtesy of S.\ Pirro.}
		\label{fig:MiDBD-Qino-Q0}
	\end{figure}

	Already during the early preparation phase of the MiDBD experiment (summer 1997), a very ambitious project was proposed 
	by Fiorini and collaborators, i.\,e.\ to build ``an array of $1000$ cryogenic detectors of a mass between $0.5$ and $1$\,kg 
	each''~\cite{Fiorini:1998gj}. The main goal remained the search for \bb, but this experiment would have allowed 
	also studies on the interaction of Weakly-Interacting Massive Particles (WIMPS) and solar axions and on rare decays.
	The name chosen for the new detector was CUORE, acronym for Cryogenic Underground Observatory for Rare Events
	(Italian for `heart').
	The realization of such a complex cryogenic setup would have required the support of a numerous (international) 
	collaboration and many years of preparation.
	As a first step towards CUORE, a simpler (and much less expensive) experiment was also proposed. The original idea 
	foresaw an array made of $100$ $5\times5\times5\,\cm^3$ crystals for a total mass of about $75$\,kg, hence the name 
	Cuoricino (Italian for ``small CUORE''). Cuoricino had not to be purely considered a proof of concept for CUORE, 
	but an independent experiment as well. 
	Apart from becoming by far the new largest cryogenic detector, 
	Cuoricino would have contained about $20$\,kg of \ce{^{130}Te}, a mass three times larger than the whole MiDBD and, 
	more in general, larger than the $\beta\beta$-isotope mass of any other \bb~experiment running at the time.
	
	On the technological side, the change in the growth axis of the crystals 
	from $\langle 1\,0\,0 \rangle$, used until MiDBD, to $\langle 1\,1\,0 \rangle$
	allowed the increase of the final dimension of the specimens.
	Therefore, it was possible to obtain a first $5\times5\times5\,\cm^3$ $760$\,g \ce{TeO_2} bolometer, which
	was operated for $\alpha$- and $\gamma$-ray spectroscopy in the Hall~C DR~\cite{Alessandrello:2000ywa}.
	The achieved FWHM resolution throughout the $\gamma$ region was comparable to that of a Ge detector and the value of 
	$4.2\,\keV$ at the $5407\,\keV$ line of \ce{^{210}Po} was the best ever obtained with any type of detector.
	The fundamental brick for the construction of the new tower was therefore available.

	In its final configuration, Cuoricino consisted of $44$ $5\times5\times5\,\cm^3$ cubic crystals of 
	about $790$\,g mass and $18$ $3\times3\times6\,\cm^3$ crystals coming from MiDBD (including the $4$ enriched ones).
	These were disposed in $13$ floors, $11$ $4$-crystal modules housing the large crystals and 
	$2$ $9$-crystal modules housing the small ones (Fig.~\ref{fig:MiDBD-Qino-Q0}).
	The total mass of \ce{TeO_2} was $40.7$\,kg. It was resized with respect to the proposed one to be housed in the 
	existing Hall A DR, but still large enough to make Cuoricino a competitive experiment for the \bb~search.
	
	A strict protocol was adopted in order to guarantee the radio-purity of the detector.
	All the crystals were grown with pre-tested low radioactivity material by the Shanghai Institute of Ceramics - 
	Chinese Academy of Sciences (SICCAS, China) and shipped to Italy by sea to minimize the activation by cosmic rays.
	These were then lapped with specially selected low-contamination polishing compound.
	All the operations, including the assembly of the tower, were carried out in a clean room environemnt inside sealed 
	glove boxes constantly flushed with \ce{N_2} gas.
	
	Significant improvements interested the whole data production chain, with the implementation of 
	a PID controller for the temperature stabilization~\cite{Arnaboldi:2005xu}, a new front-end 
	electronics~\cite{Arnaboldi:2004jj} and a software trigger for the data acquisition~\cite{Arnaboldi:2008ds}.%
	\footnote{The software trigger was actually introduced in Run-II (see further in text).}

	Cuoricino~\cite{Andreotti:2010vj} took the place of MiDBD in the Hall A cryostat.
	The cool down occurred at the beginning of 2003 and the start 
	of the data taking shortly after (Run\,I).	
	In November 2003, the tower was warmed up to perform maintenance operations and to recover some lost connections.
	The data taking restarted in May 2004 (Run\,II) and lasted until June 2008, with $50/52$ channels operational.
	The total collected exposure was $71.45$\,kg\,yr.

	The best performance, both in terms of background counts and resolution, was reached with the $5$\,cm-side 
	bolometers.
	The harmonic mean FWHM weighted on the physics exposure at the \ce{^{208}Tl} line 
	was $(5.8 \pm 2.1)\,\keV$~\cite{Alduino:2016vjd} (Fig.~\ref{fig:CUORE-0_res}).
	Regarding the background, the flat continuum in the ROI relative to Run-II 
	was $(0.153 \pm 0.006)$ for the large crystals and $(0.17 \pm 0.02)$\,\ckky for the small crystals, respectively
	(anyway compatible).
	
	Cuoricino demonstrated the feasibility of running a very massive and complex bolometric detector array for almost five 
	years with the best performance obtained so far.
	However, although very good, the results in terms of resolution and, especially, of background rate were not yet 
	compliant with the tight limits set for the CUORE detector.

\subsection{Background studies}
\label{sec:bkg_study}

	\begin{figure}[tb]
		\centering
		\includegraphics[width=.6\columnwidth]{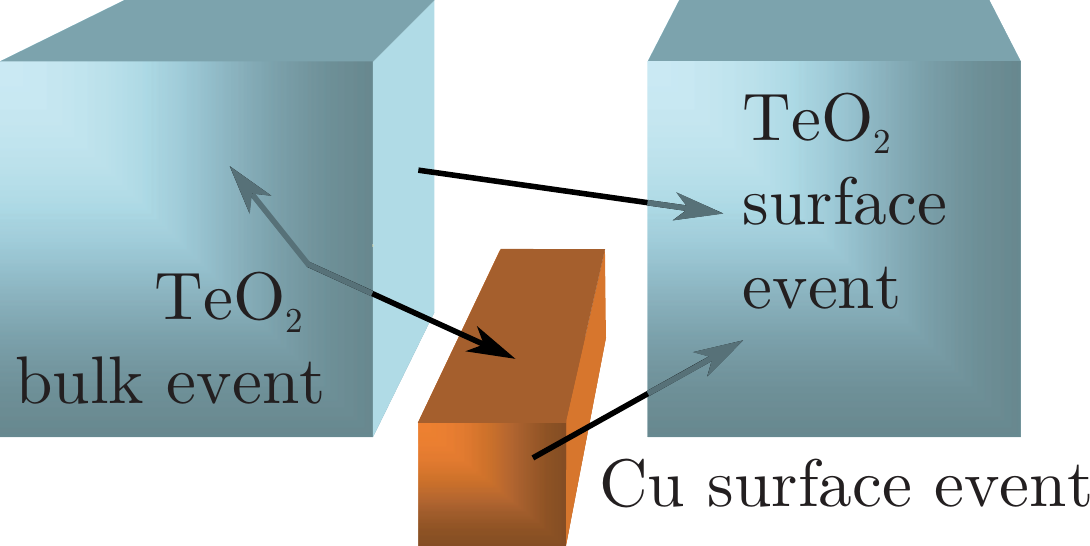}
		\caption{Sketch of events in which the energy release is only partially deposited in one crystal.
			Degraded $\alpha$ particles due to surface contamination of the crystals or of the closest 
			inert material (copper and PTFE) represent a dangerous background for the \bb~search.}
		\label{fig:surf_contams}
	\end{figure}

	With MiDBD, a systematic study of the radioactive contaminations dangerous for the \bb~search was started.
	A background model able to describe the observed spectra in terms of contaminations from the materials 
	directly facing the detector, the whole cryogenic setup and the environmental 
	radioactivity was developed and supported by detailed Monte Carlo simulations of the detector 
	geometry~\cite{Capelli_PhD-thesis:2005}.
	This model allowed to account for the background, both peaks and continuum, measured in the two MiDBD runs 
	and was used to disentangle the origin of that measured with Cuoricino.
	
	The major sources of the background observed in the \bb~region of interest 
	were identified with multi-Compton events from the 
	$2615\,\keV$ $\gamma$ rays of \ce{^{208}Tl} and with degraded $\alpha$s.
	The former type of events was identified to be originated from the \ce{^{232}Th} contamination of the cryostat,
	while the latter was attributed to surface contaminations of the \ce{TeO_2} crystals and of 
	the materials facing the detector, such as copper and PTFE frames.
	In fact, despite the kinetic energy of $\alpha$s (from U and Th daughters) 
	being always far above the ROI, these particles can contribute to the background in the ROI 
	if they deposit only a limited fraction of their energy inside
	a single bolometer (without the possibility of reconstructing the total energy via coincidence analysis).
	This can happen when the contamination is localized on the surface either of the crystal 
	or of the inert material closest to the detector (Fig.~\ref{fig:surf_contams}). 

	Above $2.6\,\MeV$, a flat background continued up to $3.9\,\MeV$ (see the spectra in Figs.~\ref{fig:MiDBD_spectra}, 
	\ref{fig:RAD1_spectrum}, \ref{fig:CAW_spectra}, \ref{fig:TTT_spectra} and \ref{fig:CUORE-0_spec}).
	The only exception was the peak at $3.2\,\MeV$ of \ce{^{190}Pt}, most likely due to an internal contamination of the 
	crystals, since these were grown in platinum crucibles.
	At higher energies, above 4\,MeV, the contribution of the various $\alpha$ lines from the U and Th decay 
	chains was found as expected.
	In particular, the region between 5 and 5.5\,MeV was dominated by the \ce{^{210}Po} contamination.
	
	It is worth to notice that the position and the shape of the $\alpha$ peaks in the spectrum can provide a strong
	indication on the location of the contamination.
	In the case of Gaussian peaks, if the position corresponds to the transition energy of the decay ($\alpha$ + 
	nuclear recoil), this indicates that the contamination is internal to the crystal (bulk contamination).
	If the peak position corresponds to the $\alpha$ energy, instead, this indicates that the contamination 
	is within a thin layer on the surface of the crystal or of the inert material directly facing it
	(surface contamination).
	Long-tailed asymmetric peaks (on the low energy side) are due to deep-surface contaminations in the crystals,
	whereas a flat continuum (without peaks) is expected by either a bulk or a deep-surface contamination of an inert material facing the detector.
			
	\medskip
	In order to investigate the possible contributions to the background, especially the rather flat continuum between 3 and 4\,MeV, in view of the forthcoming CUORE, 
	an intense R\&D activity was carried out both in the material selection and in the surface cleaning procedure optimization. 
	Therefore, in parallel with the Cuoricino running, many tests were performed in the Hall C facility.

\subsubsection{RAD detectors}
	
	\begin{figure}[tb]
		\centering
		\includegraphics[width=.9\columnwidth]{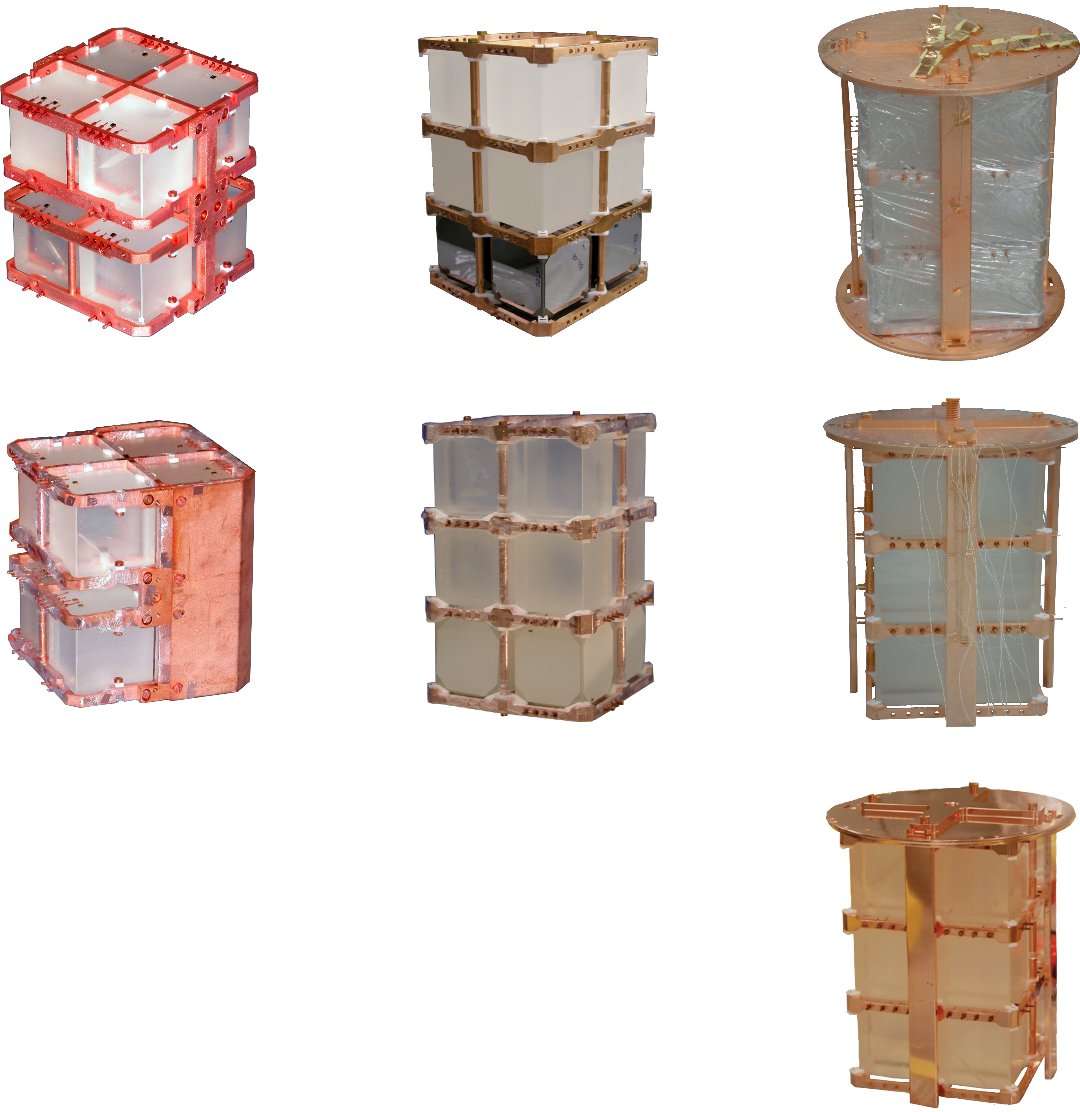}
		\caption{Background study experimental setups. (Left) The RAD\,1 (top) and RAD\,3 (bottom) detectors.
			(Center) The CAW\,1 (top) and CAW\,3 (bottom) detectors. The composite bolometers for the rejection of surface 
			radioactive background are visible at the bottom floors.
			(Right) The TTT detector top, middle and bottom towers.
			Adapted from Refs.~\cite{Gorla_PhD-thesis:2006,Giachero_PhD-thesis:2008,Alessandria:2012zp}.}
		\label{fig:bkg_towers}
	\end{figure}

	A measurement campaign was performed with a dedicated setup, the Radioactivity Array Detector~\cite{Gorla_PhD-thesis:2006,Giachero_PhD-thesis:2008} 
	(RAD), 
	with the aim of better understanding the contribution of the individual detector components.
	The RAD detector consisted of a 2 4-crystal floor array with a structure almost identical to that of Cuoricino 
	(2 Cuoricino single modules, Fig.~\ref{fig:bkg_towers}).
	The measurements, six in total, were carried out between summer 2004 and early 2007.

\begin{figure}[tb]
		\centering
		\includegraphics[width=1.\columnwidth]{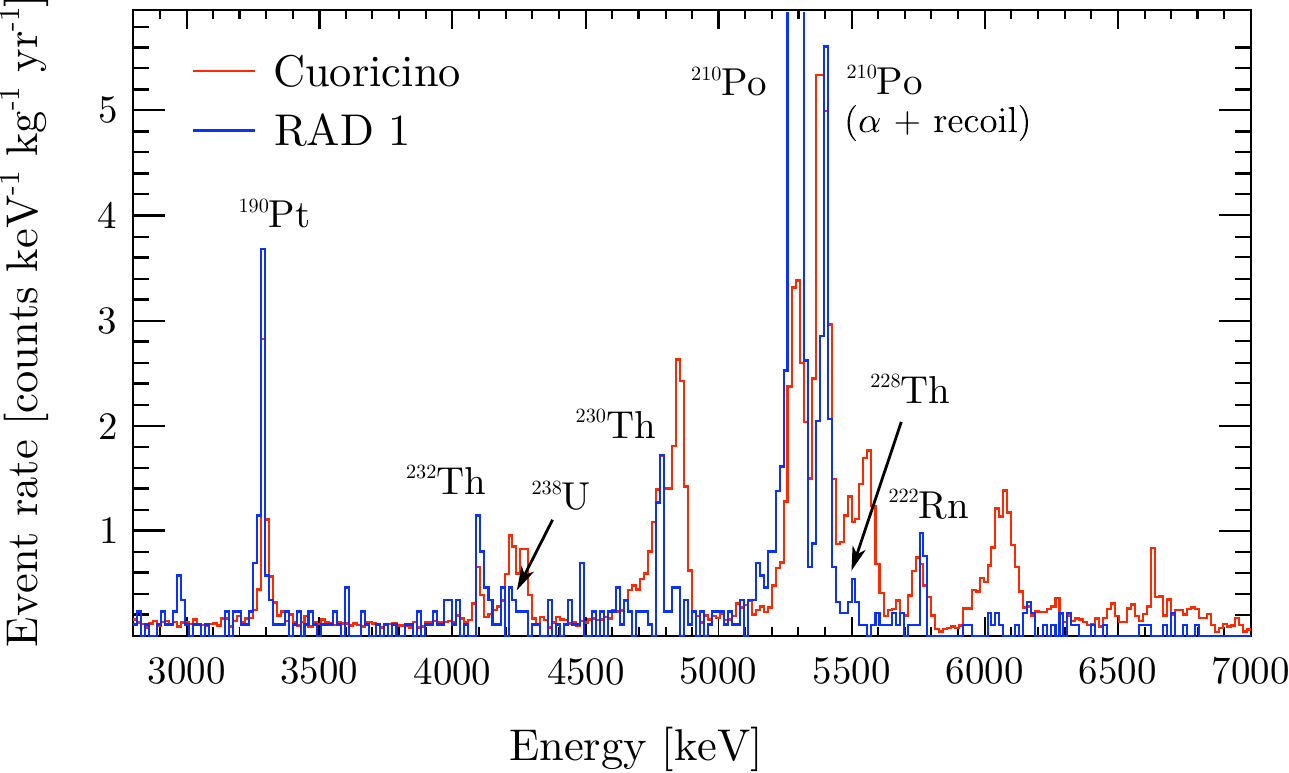}
		\caption{Background comparison between Cuoricino (red) and RAD\,1 (blue).
			The reduction of crystal surface contamination, still visible in the Cuoricino spectrum as large 
			asymmetric peaks, is evident.
			Figure from Ref.~\cite{Giachero_PhD-thesis:2008}.}
		\label{fig:RAD1_spectrum}
	\end{figure}

	The purpose of the RAD\,1 run was to define a new protocol for the surface cleaning.
	The crystals were etched and polished while the copper structures were etched and treated with electroerosion
	and only a few radioclean materials were allowed.
	The result was quite successful and saw the reduction of the \ce{^{238}U} and \ce{^{232}Th} peaks. 
	In particular, the extremely low background allowed for the first time to disentangle the bulk from the surface 
	contamination of the crystals, with Gaussian peaks due to the internal contamination of \ce{TeO_2}
	eventually emerging from the continuum~(Fig.~\ref{fig:RAD1_spectrum}).
	It was observed that the crystals were contaminated in \ce{Th} isotopes, while no evidence of \ce{U} 
	contamination in secular equilibrium was obtained.
	The only other $\alpha$ peaks were those of \ce{^{210}Po} at 5.3 and 5.4\,MeV, with intensities 3 times larger and 
	comparable to Cuoricino, respectively. Finally, despite the strong reduction of the \ce{^{238}U} and \ce{^{232}Th} 
	crystal contamination, no improvement was observed in the flat background over the $(3-4)\,\MeV$ region,
	thus demonstrating the dominant role of inert material contamination.

	The RAD\,2 run was intended to test the background contribution coming from the ``small parts'' of the detector: the PTFE 
	supports, the NTDs, the heaters and the gold bonding wires, glue and pins (used to attach the sensors to the crystal
	and to make the electrical connections).
	The top and bottom plates of the copper mounting structure were thus covered with samples of these parts
	(a PTFE slab, a set of heaters and a set of gold wires).
	The results of this test were very clear. The $(3-4)\,\MeV$ rates measured proved that these materials together
	provided a maximum contribution to the Cuoricino (and RAD\,1) background of about $20\%$.

	Given these results, the RAD\,3 and 4 runs
	investigated the hypothesis whether degraded $\alpha$s could actually account for the large fraction of the background.
	In order to do so, all the copper structure surfaces facing the crystals were covered with a radio-clean plastic layer.
	This would have stopped the $\alpha$s escaping from copper, while nothing would have changed otherwise.
	At the same time, the contribution to the background from 
	the thin foil could have been easily measured with a germanium detector since in this case bulk and surface 
	contaminations were coinciding.
	
	The obtained results showed that the crystal surface contaminations in \ce{^{238}U} and \ce{^{232}Th} were reduced by 
	a factor $\sim 4$ while the counting rate in the $(3-4)\,\MeV$ region by a factor $\sim 2$.
	Also, the 5.3\,MeV \ce{^{210}Po} peak, quite high in RAD\,1, was ascribed to a contamination of the 
	copper surface. In fact, this was efficiently reduced to a level compatible with Cuoricino thanks to the polyethylene 
	film.
	
	During RAD\,4, a neutron shield was added to the cryostat external shielding.
	This was composed of an external 7\,cm polyethylene layer for the thermalization of the fast neutrons plus 
	an internal thin layer made of boron carbide for the capture of the thermal neutrons.
	Dedicated measurements with an Am-Be source in the two configurations with and without the new shield allowed to 
	test the neutron absorption capability. The results of RAD\,4 were found completely compatible with those of RAD\,3, thus 
	indicating that environmental neutrons were not influencing the background in an appreciable manner.
	
	Finally, the RAD\,5 and 6 explored the possibility of exotic (non particle physics) effects as source of the contaminations, focusing in particular 
	on the PTFE holders, which were therefore replaced with phosphorus-bronze clamps. 
	The hypothesis was that, due to the fast cool down, the PTFE
	could undergo internal adjustments once the cryostat base temperature was reached and it could therefore continue to release 
	heat into the crystal absorber even for a long time, originating pulses that would mimic signal events.
	Unfortunately, a too high contamination of the clamps (probably due to \ce{^{210}Pb}) prevented deducing 
	anything about the exotic sources hypothesis. However, from the technical point of view, these last runs
	proved that the clamps could work in the same way as the PTFE supports.
	
	In conclusion, the RAD tests confirmed the Cuoricino background model and proved the relevant role of the copper 
	surface contamination in the $(3-4)\,\MeV$ background region.
	
\subsubsection{CAW detectors}
	
	\begin{figure}[tb]
		\centering
		\includegraphics[width=.4\columnwidth]{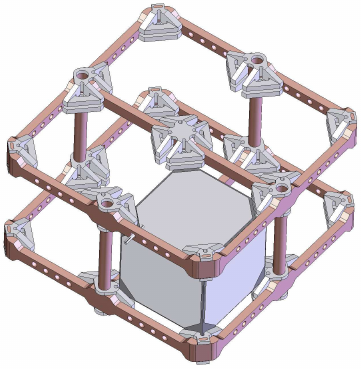}
		\caption{Schematic of the CAW\,2 detector module. The same design for the copper frame and PTFE supports will be 
			the base for the one used in CUORE (compare with Fig.~\ref{fig:Q0_tower}).
			Figure from Ref.~\cite{Giachero_PhD-thesis:2008}.}
		\label{fig:CAW_schematic}
	\end{figure}

	In 2005 and 2006 a new design for the detector structure was tested. 
	The prototype detector, the CUORE Assembling Working group (CAW) detector, 
	consisted of a 3 4-crystal floor array (Fig.~\ref{fig:bkg_towers}). 
	The CAW detector was the result of an intense program of 
	optimization of both the single module structure and its assembly procedure during the tower build up.
	The new structure reduced the amount of copper by a factor $\sim 2$.
	At the same time, the PTFE supports were moved to the vertices of the crystals.
	In the first attempt, used during the CAW\,1 run, too much stress on the vertices actually caused cracks of the crystals. 
	Therefore a further modification of the PTFE parts was implemented and used during the CAW\,2 run (Fig.~\ref{fig:CAW_schematic}).
	The new structure relaxed the tension on the critical points, but introduced a larger amount of material.
	Considering that the hypothesis of thermal stress release from PTFE was still unsolved, in the end it was opted for smaller clamps, 
	following the original idea, but adding small stress-relieving holes on the vertices to avoid the risk of cracks.
	This design became the official choice for CUORE and has been ever since (see Fig.~\ref{fig:Q0_tower}).

\begin{figure}[tb]
		\centering
		\includegraphics[width=1.\columnwidth]{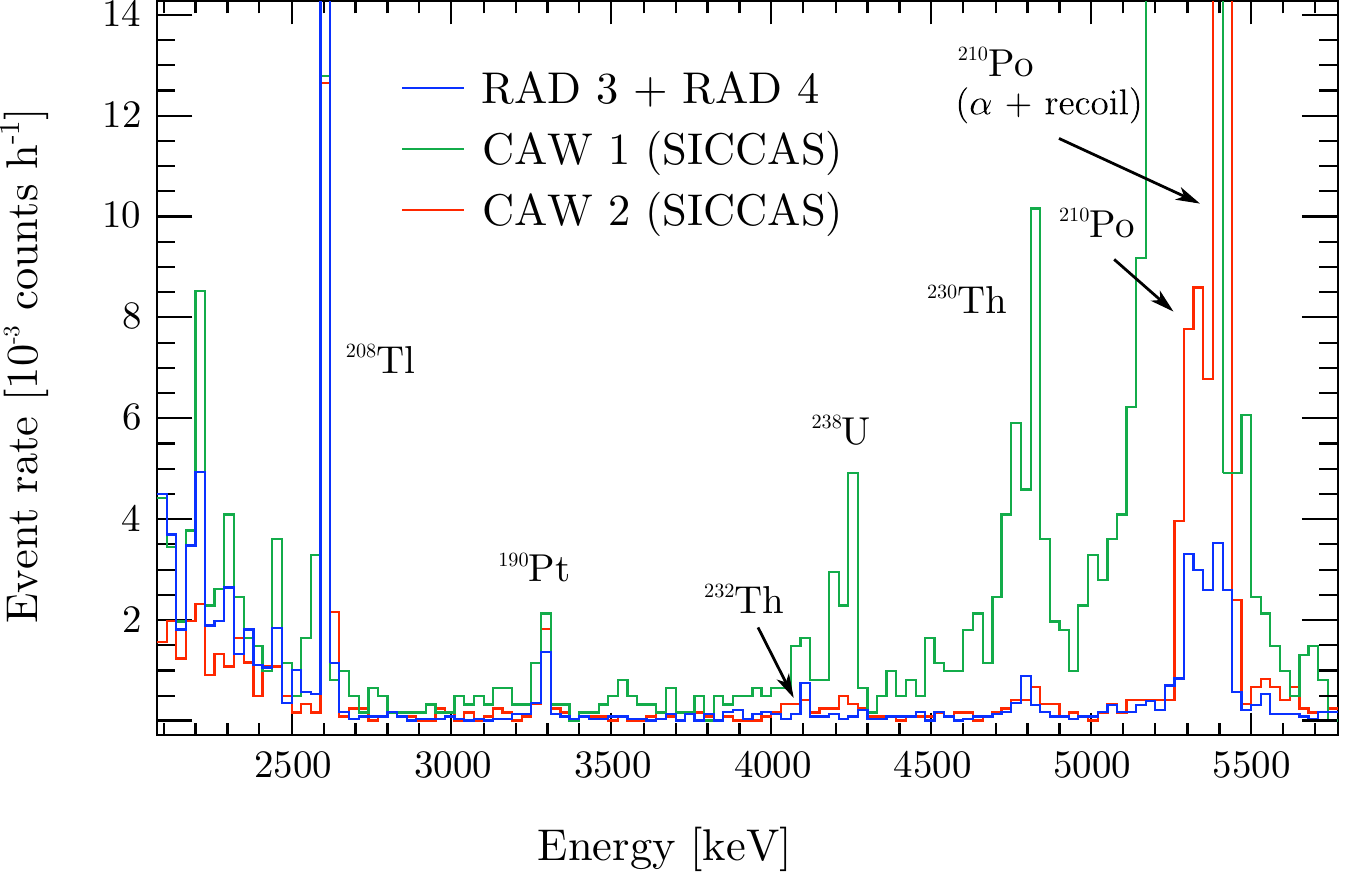}
		\caption{Background comparison between RAD\,3 + RAD\,4 (blue), CAW\,1 (green) and CAW\,2 (red).
			In both the CAW cases, the spectra refer to the crystals produced by SICCAS.
			The comparable spectra of RADs and CAW\,2 proves that the adopted surface treatment is reproducible.
			Figure from Ref.~\cite{Giachero_PhD-thesis:2008}.}
		\label{fig:CAW_spectra}
	\end{figure}

	Concerning radioactivity, the CAW\,1 test was a technical run, aimed at verifying the reproducibility and the effectiveness 
	of the new setup.
	Therefore, no particular cleaning procedure for the setup was adopted, whereas 
	this was done for the CAW\,2 run. 
	In this case, a RAD-like cleaning protocol (crystal surface and copper cleaned as for RAD\,1 + copper covered
	with a polyethylene foil as for RAD\,3) was adopted with the purpose of checking the background achievement 
	obtained with the RAD runs in a different structure.
	With the CAW runs, the RAD results were confirmed, 
	proving that the surface cleaning procedure of crystals and copper were reproducible (Fig.~\ref{fig:CAW_spectra}).
	The new measurements showed an easy, fast and standard assembling procedure. This improved the 
	detector reproducibility and standardization. 

	As it can be seen from the pictures in Fig.~\ref{fig:bkg_towers} (central column), the bottom floors of the CAW detectors
	were covered with a slab of \ce{Ge} and \ce{TeO_2} during the CAW\,1 and 2 runs, respectively.
	This was done in order to study the performance of prototype composite bolometers able to identify events originated at 
	the detector surface, as R\&D projects finalized at the background abatement 
	(see Refs.~\cite{Foggetta:2011nk,Sangiorgio_PhD-thesis:2006} for details).

	During CAW\,2, crystals coming from different producers were also checked. However, these were discarded since they were
	largely contaminated in \ce{^{232}Th} and \ce{^{210}Po}.
	SICCAS was thus confirmed as the producer for the CUORE crystals.
	The CAW detector was then used for the so-called Chinese Crystals Tests (CCTs) in order to validate a new production 
	and treatment protocol for CUORE~\cite{Arnaboldi:2010fj}.

	The CCT runs (two in total) aimed at verifying the crystal quality and the surface treatment on the way to CUORE.
	Up to Cuoricino, all the crystal surface cleaning procedure was performed in a clean room at LNGS.
	Moving to the direction of simplifying the cleaning operations, the CCTs intended to test the performance 
	of the first crystals 
	completely processed directly at SICCAS, where they had been grown. 
	The surface treatment used in China (etching + lapping) was the same successfully optimized at the LNGS and verified 
	with the RAD array.
	
	In terms of bulk contamination, the CCTs showed limits that were comparable with those of Cuoricino,
	already sufficiently good for CUORE.
	Also, the repeated presence of a peak due to \ce{^{190}Pt} consolidated the hypothesis of a possible 
	inclusion of platinum fragments from the crucible used for the crystal growing
	(see the spectra in Figs.~\ref{fig:RAD1_spectrum} and \ref{fig:CAW_spectra}).
	Regarding the surface contamination, all the crystals showed fewer counts than Cuoricino in the flat background region. 
	However, the final sensitivity reached in the CCTs was not enough to establish whether 
	the compliance with the requests for CUORE had been achieved.
	
\subsubsection{TTT detector}
	
	In order to compare the results of different surface treatments of the copper frames, 
	a higher statistics measurement than those of the RAD/CAW detectors was needed. 
	Therefore, at the end of Cuoricino, the Hall A cryostat hosted the Three Tower Test (TTT) detector
	from September 2009 to January 2010~\cite{Alessandria:2012zp}. 
	The TTT detector consisted of 3 CAW-like arrays separated from each other by copper shields, 
	for a total of 36 crystals, all coming from the Cuoricino production series (Fig.~\ref{fig:bkg_towers}).
	To ensure similar contamination levels, history, and treatment, the crystal surfaces were all re-polished and the copper 
	of the towers was taken from one single batch and machined following identical procedures.
	
	The copper for the top tower was treated following the optimal protocol identified with the previous tests.
	This included a soap and chemical cleaning and the wrapping with multiple layers of polyethylene. 
	The middle tower followed a new chemical process, starting from a simple soap cleaning and then adding electroerosion, 
	chemical etching and passivation~\cite{Wojcik:2007zz}.
	These two procedures were still done ``manually'' at LNGS. Instead, an ``industrialized'' cleaning procedure 
	performed at the Laboratori Nazionali di Legnaro of I.\,N.\,F.\,N.\ (Legnaro (PD), Italy) was adopted for the bottom 
	tower. In this case, the copper cleaning procedure consisted of tumbling, electro-polishing, chemical etching and 
	magnetron plasma cleaning.

\begin{figure}[tb]
		\centering
		\includegraphics[width=1.\columnwidth]{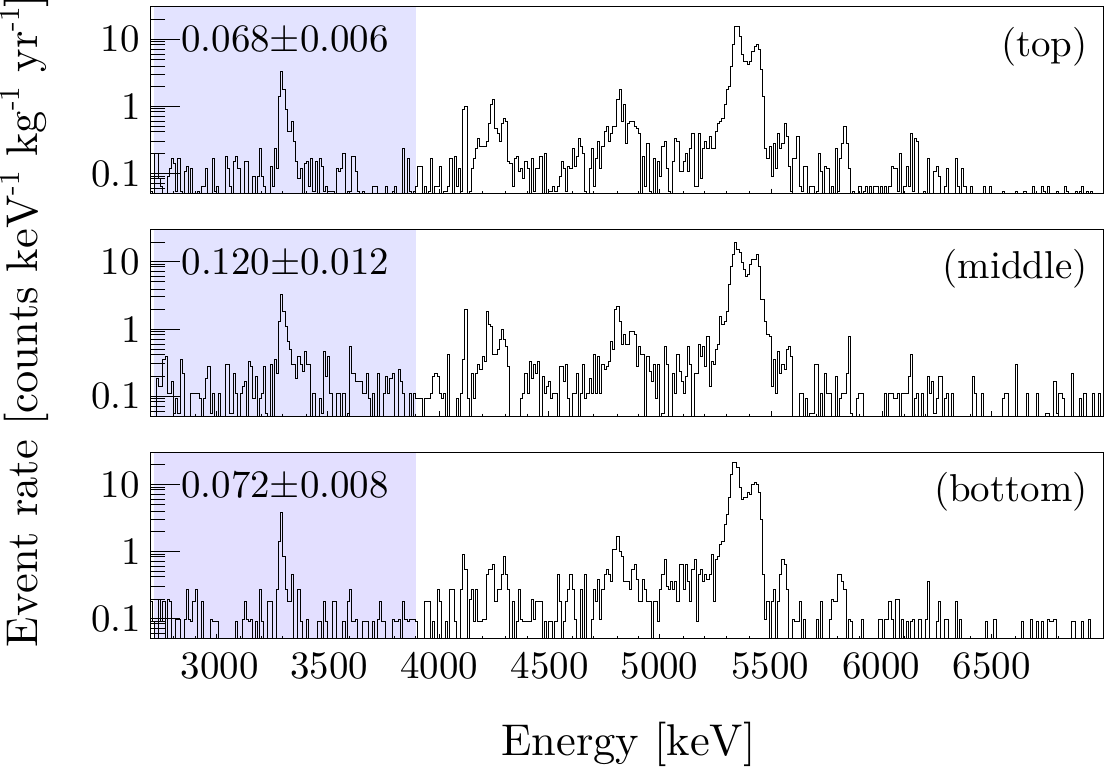}
		\caption{Comparison of the TTT spectra in the $\alpha$ region. The number of counts in the upper part of the plots 
			refer to the $(2700-3900)\,\keV$ region (highlighted). The name in the brackets refer to the tower position 
			in the detector.
			Figure from Ref.~\cite{Alessandria:2012zp}.}
		\label{fig:TTT_spectra}
	\end{figure}

	All these copper cleaning techniques led to significant reduction of the background in the 
	flat background continuum compared to Cuoricino. 
	The improvement was close to a factor $\sim 2$ for the top and bottom towers, which got comparable results, 
	while	the rate of the central tower remained $\sim 70\%$ higher (Fig.~\ref{fig:TTT_spectra}).
	Therefore, looking for the best compromise between cost, reproducibility and background control, the Legnaro protocol 
	was validated for copper parts of the CUORE detector.%
	\footnote{Initially, it was decided to adopt a mixed approach for CUORE, with only the small copper parts to be 
		subjected to the Legnaro protocol, while the large shields facing the external part of the detector
		(MC shield and Tower Support Plate, refer to Fig.~\ref{fig:cryostat_scheme}) 
		had to be covered with polyethylene~\cite{Alessandria:2012zp}. 
		The final option consisted in substituting the polyethylene with copper tiles also cleaned according to the 
		Legnaro protocol.} 
	
\subsubsection{CCVRs}
\label{sec:CCVR}

	After the CCT runs and in parallel with the TTT measurement, 
	the Hall C cryostat hosted the CUORE Crystal Validation Runs (CCVRs)~\cite{Alessandria:2011vj}.
	The CCVRs, ten in total, were carried out between December 2008 and July 2013.
	These were cryogenic measurements designed to check the different batches of crystal production by SICCAS upon their arrival at LNGS. 
	Each time, four crystals randomly chosen from the batch were operated as bolometers 
	in order to test the performance and the compliance to the strict radiopurity requests.

	In general, 
	the results were satisfactory and a reduced background rate with respect to Cuoricino was observed 
	over the whole energy range above $2.7\,\MeV$.
	The upper limits for the crystal bulk contamination in \ce{^{238}U} and \ce{^{232}Th} were set to 
	$0.8\cdot 10^{-6}\,\Bq\,\kg^{-1}$
	while those for the surface contamination from the same isotopes were set to a few $10^{-9}\,\text{Bq\,cm}^{-2}$.
	At the same time, the activities of the bulk and surface contamination of \ce{^{210}Pb} (parent of \ce{^{210}Po}) 
	were estimated to be $3.3\cdot10^{-6}\,\text{Bq\,kg}^{-1}$ and $1.0\cdot10^{-6}\,\text{Bq\,cm}^{-2}$.
	
	Taking these results as input for the Monte Carlo simulations, it was possible to 
	extrapolate the contribution to the CUORE background arising from crystal impurities 
	and to study its impact on the \bb~energy region (see Sec.~\ref{sec:CUORE_bkg}).
	
\subsection{Towards the tonne-scale}

	In Cuoricino, all the operations related to the detector assembly, i.\,e.\ the crystal surface treatments,
	the gluing, the bonding of sensors and heaters, and the tower wiring  were still handcrafted. 
	This reflected in a quite broad distribution for the individual performance of the crystals.
	A standardization and automation of all the construction phases was needed for the handling of a detector 
	almost $20$ times bigger and much more complex, as CUORE is.
	At the same time, the high radiopurity levels of the components had to be conserved during (but also after) the various operations.
	
	First, all the parts necessary to the tower construction, now surface-cleaned either at SICCAS (crystals) 
	or Legnaro (PTFE and copper frames), were vacuum sealed into plastic bags contained in vacuum sealed boxes when shipped to LNGS.

	Then, completely new procedures were designed for the detector assembly:
	the sensor and heater gluing was performed by means of a programmable xyz table with glue dispenser and a robotic arm handling the 
	crystals~\cite{Rusconi_PhD-thesis:2011}, while the CUORE Tower Assembly Line (CTAL~\cite{Buccheri:2014bma}) allowed the transformation of 
	the over 10,000 ultra-clean pieces into 19 ultra-clean towers.
	To avoid recontamination of all the parts due to direct contact with less radiopure materials and exposure to 
	solid (any powder) or gaseous (\ce{Rn}~\cite{Clemenza:2011zz}) contaminants in the atmosphere, the whole gluing and CTAL processes 
	were confined in hermetic volumes constantly flushed with \ce{N_2}, accessible only through isolated glove ports.

	Finally, the bonding (2 wires per pad for redundancy) could be now performed on the sensor and on the 
	corresponding pad at the extremity of the wire tape at one time, thus avoiding any soldering material or mechanical contact. 
	Regarding the detector wiring, the twisted-pair wires that were used up to Cuoricino 
	(see Fig.~\ref{fig:MiDBD-Qino-Q0}) were substituted with Cu-PEN based flat flexible 
	tapes~\cite{Andreotti:2009zza,Brofferio:2013cya}.

	Thanks to the new procedures, the CUORE crystals were never exposed to air from the moment of the polishing to the 
	installation of the detector.
	In order to ultimately validate the ultraclean CTAL procedure and to test the radiopurity of the materials for the upcoming CUORE,
	it was decided to run in the Hall A DR the first tower produced with the new protocol.
	This detector was called CUORE-0.

\subsubsection{CUORE-0}
\label{sec:CUORE-0}

	\begin{figure}[tb]
		\centering
		\includegraphics[width=.9\columnwidth]{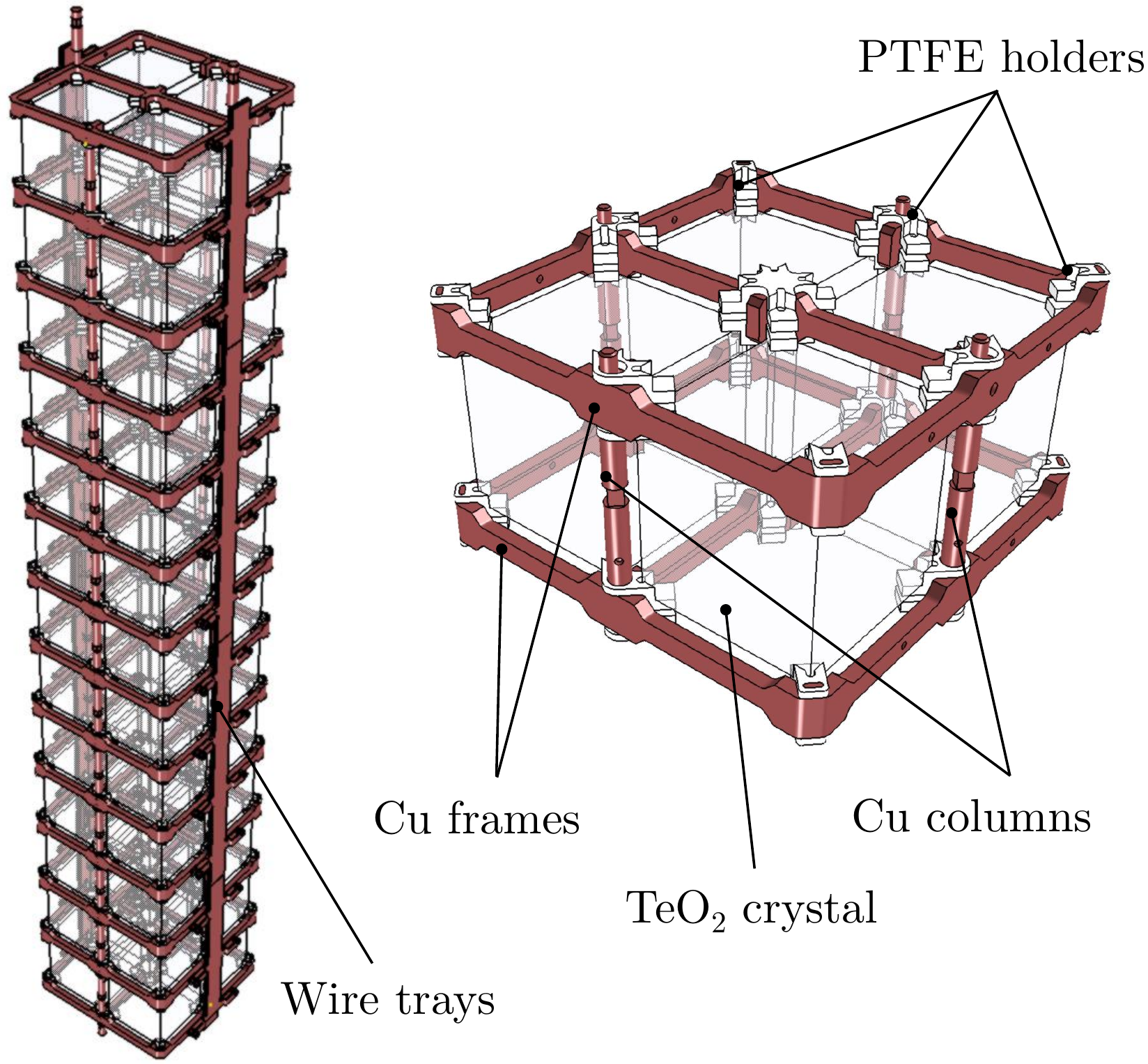}
		\caption{Schematic of the CUORE-0 tower with a detail of a single floor, based on the CAW detector module 
			(4 $5\times5\times5\,\cm^3$ crystals).
			Figure from Ref.~\cite{Alduino:2016vjd}.}
		\label{fig:Q0_tower}
	\end{figure}

	CUORE-0~\cite{Alduino:2016vjd} was a tower made of 52 $50\times50\times50\,\mm^3$ cubic crystal bolometers 
	of about $750$\,g, arranged in 13 4-crystal floors, for a total mass of $39\,\kg$ of \ce{TeO_2} 
	(Figs.~\ref{fig:MiDBD-Qino-Q0} and \ref{fig:Q0_tower}).%
	\footnote{Actually, the number of active bolometers was 51 since one NTD could not be bonded.}
	It collected data from March 2013 until March 2015, with an interruption of about two months in September 2013 due 
	to maintenance operations on the cryogenic system. The total duty cycle of the detector was close to $80\%$, 
	of which about two thirds of physics data was used for the \bb~analysis, with a \ce{^{130}Te} final exposure of $9.8\,\kg\,\yr$. 
	
\begin{figure}[tb]
		\centering
		\includegraphics[width=1.\columnwidth]{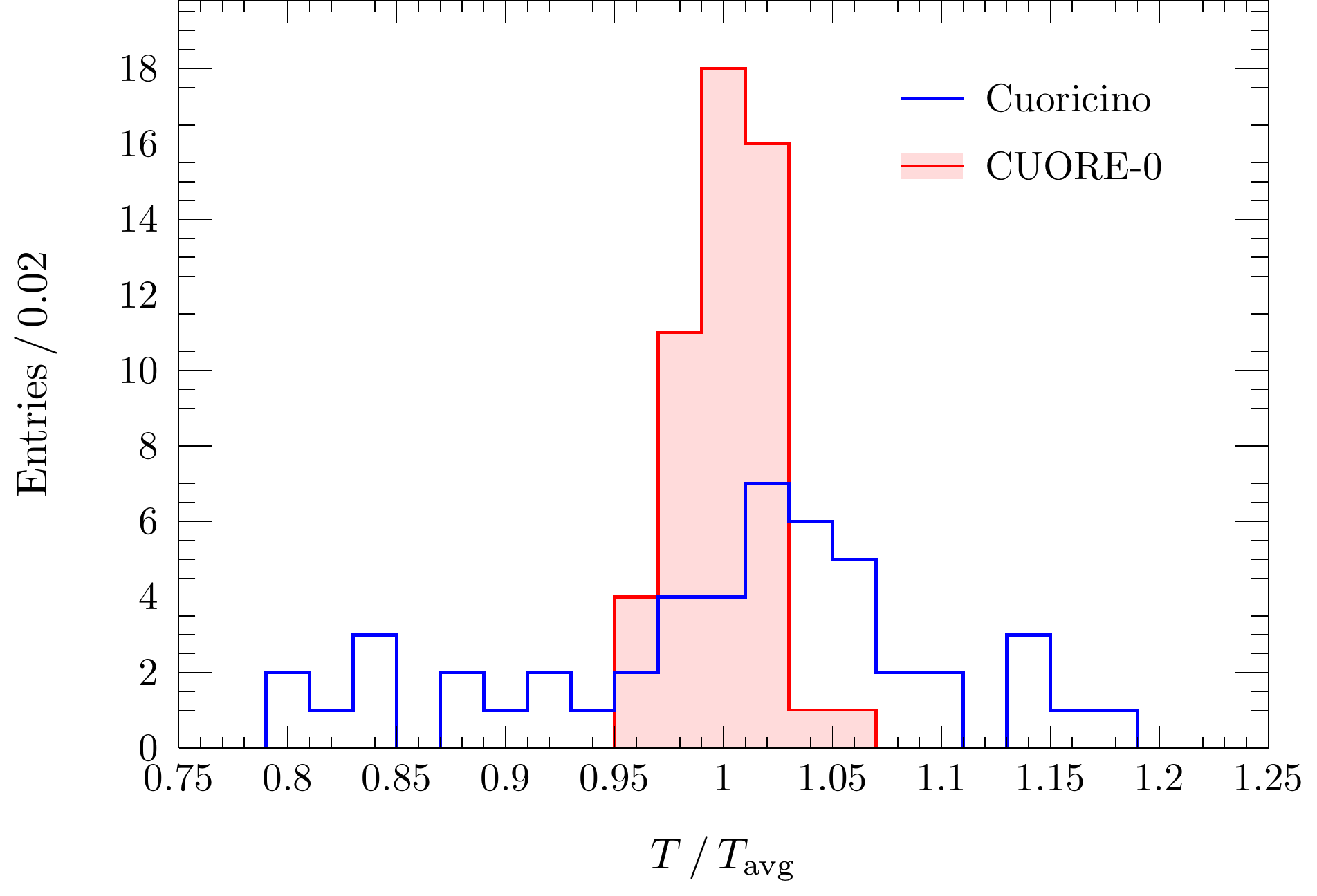}
		\caption{Base temperatures of the individual bolometers normalized to their average temperature for Cuoricino and 
			CUORE-0. The RMS of the distribution decreases from 9\% to 2\% passing from Cuoricino to CUORE-0.
			Figure from Ref.~\cite{Alduino:2016vjd}}
		\label{fig:CUORE-0_base_temp}
		\vspace{20pt}
		\includegraphics[width=1.\columnwidth]{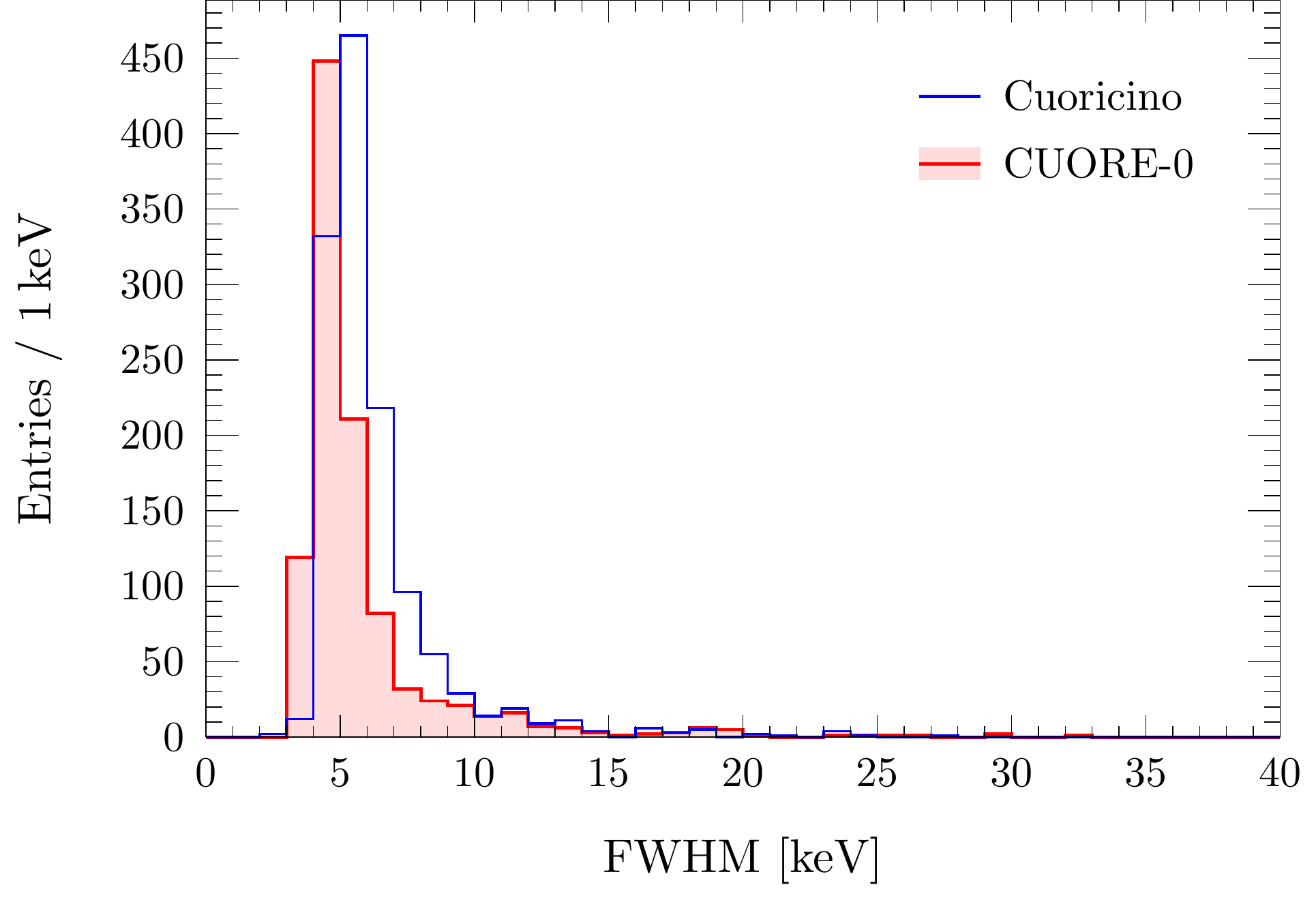}
		\caption{Distribution of the FWHM energy resolution at the \ce{^{208}Tl} line ($2615\,\keV$) for each Cuoricino and 
			CUORE-0 dataset measured during the detector calibrations.
			The effective mean is $4.9\,\keV$ in CUORE-0, while it was $5.8\,\keV$ in Cuoricino.
			Figure from Ref.~\cite{Alduino:2016vjd}}
		\label{fig:CUORE-0_res}
	\end{figure}

	The CUORE-0 data allowed to directly verify the improvements achieved with 
	the new CTAL, gluing and bonding procedures.
	With the new reproducible uniform assembly, the distribution of the single bolometer base temperature,
	which depends on the coupling between the thermistor, the absorber and the heat sink, 
	narrowed the $\Delta T$ RMS from the $9\%$ of Cuoricino to $2\%$ (Fig.~\ref{fig:CUORE-0_base_temp}).
	In terms of bolometric performance, CUORE-0 showed an improvement of the energy resolution with respect to its predecessor.
	The harmonic mean FWHM at $Q_{\beta\beta}$ was $(4.9\pm2.9)\,\keV$ (Fig.~\ref{fig:CUORE-0_res}):
	the CUORE goal, that was set to $5\,\keV$ FWHM in the ROI~\cite{Artusa:2014lgv}, had been reached.

	\begin{figure}[t]
		\includegraphics[width=1.\columnwidth]{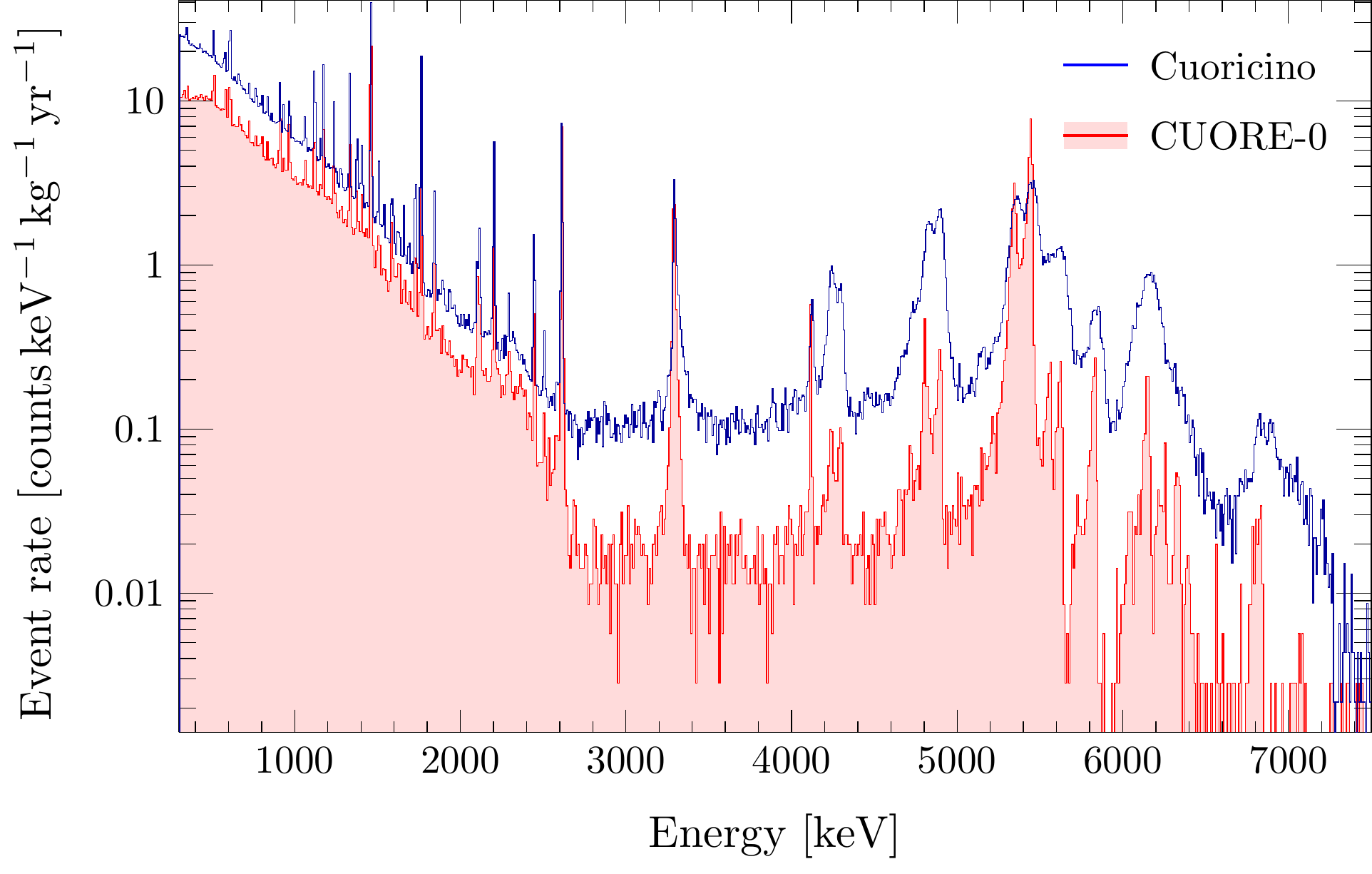}
		\caption{Final spectra of Cuoricino and CUORE-0 in the $(0-7.5)\,\MeV$ region.
			The reduced background rate in CUORE-0 is clearly visible,
			especially in the $\alpha$ region, where it is about a factor $7$ smaller than in Cuoricino.
			Figure from Ref.~\cite{Canonica:2015sva}}
		\label{fig:CUORE-0_spec}
	\end{figure}
	
	Of course, the largest expectations concerned the background.
	CUORE-0 represented in fact the ultimate test on the effectiveness of the long background suppression program.
	A comparison of the CUORE-0 and Cuoricino spectra is shown in Fig.~\ref{fig:CUORE-0_spec}.
	The improvement spans over the whole energy range.
	In \mbox{CUORE-0}, a rate of $(0.016\pm0.001)\,$\ckky~was measured in the flat continuum of the $\alpha$ 
	region~\cite{Alduino:2016zrl}, a value about a factor $7$ smaller than that obtained with Cuoricino,
	$(0.110\pm0.001)$\,\ckky~\cite{Andreotti:2010vj}.
	In particular, the $\alpha$ continuum, that constituted the major contribution to the Cuoricino background in the ROI, 
	in CUORE-0 represented a minor component.
	The background rate in the $Q_{\beta\beta}$ region was now
	($0.058 \pm 0.004\,\mbox{\footnotesize(stat.)} \pm 0.002\,\mbox{\footnotesize(syst.)}$)\,\ckky, 
	a factor of almost $3$ smaller than that obtained by the predecessor.
	As for CUORE, a major improvement in this region was still expected, thanks to the better material 
	selection for its custom-made cryostat and to the scaling effects.
	By using the measured $\alpha$ background index in CUORE-0 as an input to the Monte Carlo simulations of CUORE, 
	it was possible to conclude that the background goal of $0.01$\,\ckky~was within reach~\cite{Alduino:2017qet}.

\subsubsection{CUORE}
\label{sec:CUORE}

\begin{figure}[tb]
		\includegraphics[width=.4\textwidth]{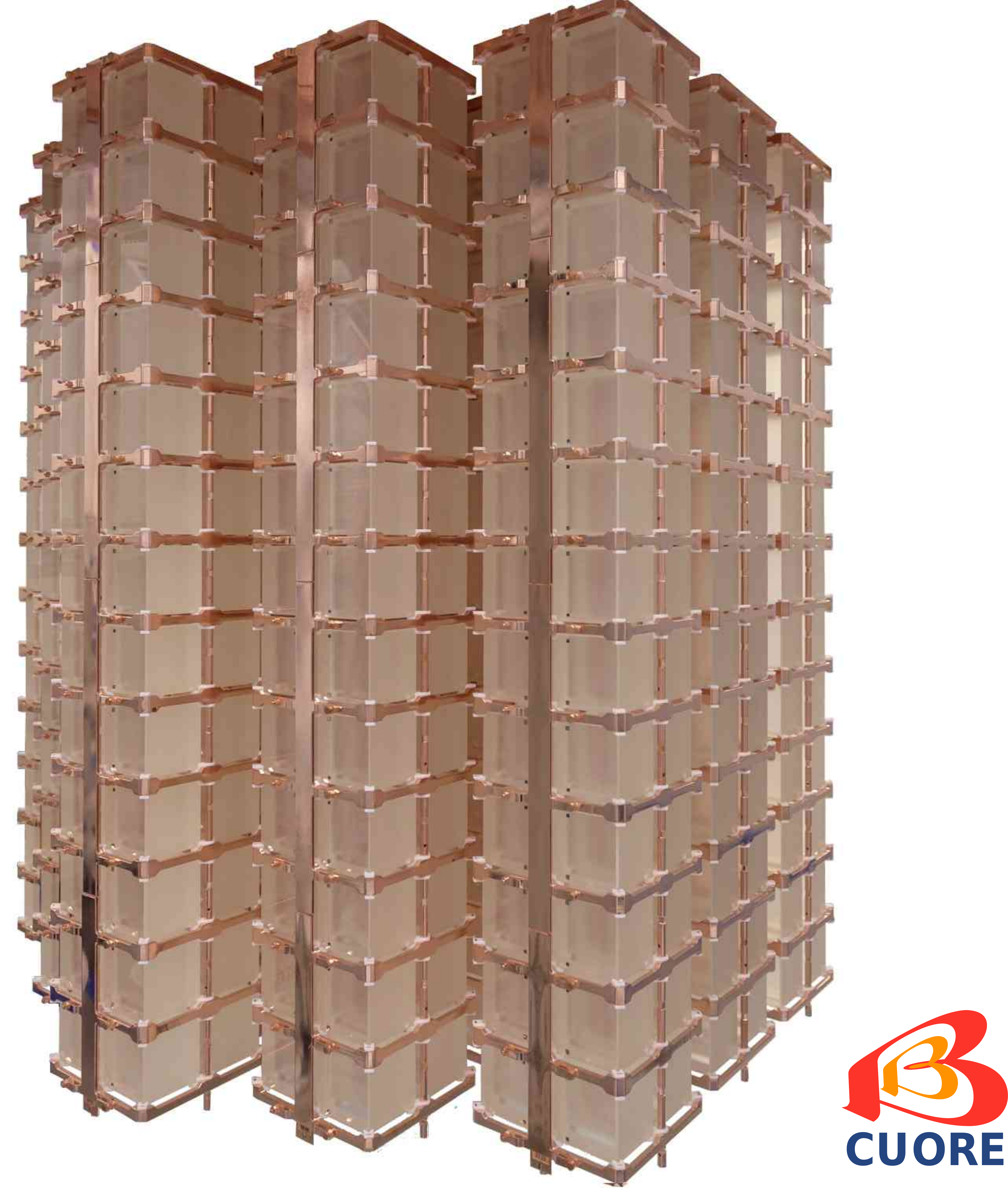}
		\caption{The CUORE detector with the experiment logo~\cite{CUORE_site}.}
		\label{fig:CUORE_detector}
	\end{figure}

	In the long way between the original proposal~\cite{Fiorini:1998gj} and the actual 
	detector construction, the design of 
	CUORE has constantly evolved and improved thanks to the numerous studies and tests performed.
	In the final version~\cite{Artusa:2014lgv}, the array consists of $19$ towers, for a total of 988 \ce{TeO_2} crystals and 
	about $742\,\kg$ of weight, corresponding to $\sim 206\,\kg$ of \ce{^{130}Te} (Fig.~\ref{fig:CUORE_detector}).
	CUORE is by far the largest detector operated as a bolometer.
	
	The full detector assembly took almost two years, going from September 2012 to July 2014. 
	The successful completion of the task definitively proved the effectiveness of the CTAL, gluing and bonding protocols.
	Before being installed inside the CUORE cryostat,
	the towers had to wait for the end of the commissioning of the cryogenic system.
	Therefore, for other two years, these were stored inside the CUORE clean room into sealed containers 
	constantly flushed with clean \ce{N_2} gas to prevent any contamination from \ce{Rn}.
	The tower installation took place during summer 2016.
	The extremely delicate operation was performed in a controlled clean room environment with Rn-free filtered air~\cite{Benato:2017kdf}
	by a specifically trained team. 
	
	CUORE is expected to collect data for a total of $5\,\yr$ live-time.
	The design performance is $5\,\keV$ FWHM at $Q_{\beta\beta}$, while the target background level is
	$0.01$\,\ckky, as a result of the extensive background reduction program and thanks to the new cryostat.
	
\begin{figure}[tb]
		\hspace{17pt}
		\includegraphics[width=1.\columnwidth]{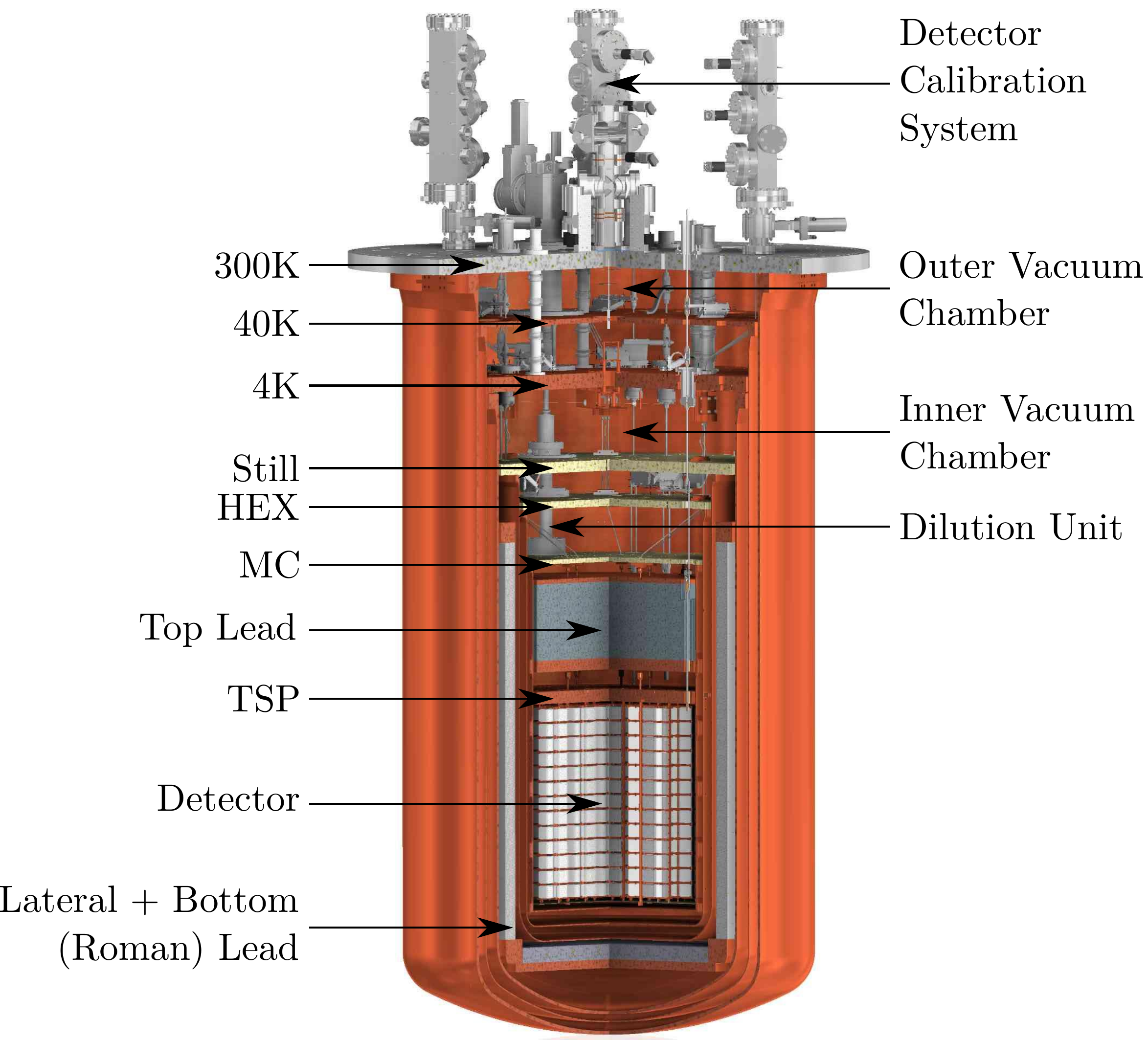}
		\caption{Rendering of the CUORE cryostat. 
			Figure from Ref.~\cite{Cushman:2016cnv}.}
		\label{fig:cryostat_scheme}
	\end{figure}

	In fact, the CUORE detector could not be hosted in the Hall A DR (or in any other standard DR)
	and a custom cryogenic system had to be designed, realized and commissioned~\cite{Delloro_PhD-thesis:2017} in order to guarantee the optimal operation 
	of the detector for a live-time of years, therefore satisfying a set of very stringent experimental requirements in terms of high 
	cooling power, low noise environment and low radioactivity content.
	The adopted solution consisted of a large custom cryogen-free cryostat cooled by 5 Pulse Tube Refrigerators and 
	a high-power \ce{^3He}/\ce{^4He} Dilution Unit~\cite{Alessandria:2013ufa,Alessandria:2018speriamo} (Fig.~\ref{fig:cryostat_scheme}). 
	This cryostat comprises six nested high-purity-copper vessels, the innermost of which encloses an experimental volume of about $1\,\m^3$.
	At its center, the Tower Support Plate holding the detector is attached to a dedicated suspension system in order to reduce the amount of vibrations.
	To avoid radioactive background, only a few construction materials were employed and almost 7\,t of lead had to be cooled below 4\,K. 
	The inner shields include the Top Lead and the Lateral Lead, the latter being made of ancient Roman lead~\cite{ILS:2018speriamo}.
	Outside, the whole cryostat is protected from the environmental radioactivity by the external shield made of 
	a 18\,cm polyethylene + 2\,cm \ce{H_3BO_3} + 25\,cm lead shield on the side and 25\,cm lead + 20\,cm borated polyethylene shield on the bottom.
	In order to calibrate the detector while in operation, the presence of a dedicated system~\cite{Cushman:2016cnv} allows for the insertion and extraction of the sources 
	without perturbing the cryogenic environment.
	
	After the cool down between December 2016 and January 2017 and the identification of suitable working 
	conditions (operating temperature and bias voltages, see Sec.~\ref{sec:det_operation})
	the physics data-taking could finally begin in April 2017~\cite{Cremonesi:2017-TAUP}.
	The first results on the \bb~search, covering the interval between May and September 2017,
	for a total \ce{TeO_2} exposure of $86.3\,\kg\,\yr$,
	have recently been released~\cite{Alduino:2017ehq}.
	These indicate that a background in line with the expectations has been reached, 
	with the value of $(0.014\pm0.002)$\,\ckky observed in the ROI.
	The average energy resolution of the detector at $Q_{\beta\beta}$ is $(7.7\pm0.5)\,\keV$, with an observed improvement
	during the data collection thanks to a data-quality optimization campaign.
	A further improvement down to $\sim 5\,\keV$ is foreseen by further optimizing the experimental operating 
	conditions and through improvements of the analysis.
	
	The successful commissioning and operation of CUORE mark a major step 
	in the application of the bolometric technique and demonstrates the feasibility of future large-mass bolometer arrays 
	for rare event searches.
	
	\bigskip
	Table~\ref{tab:exp_list} summarizes the main features of the chain of bolometric experiments presented in this work.%
	\footnote{To allow the direct comparison of the different parameters, some slight imprecision on the original 
		values could have been introduced. This holds in particular for the older experiments.}
	The series of significant improvements appears here evident. 
	Passing from the 6\,g crystal to CUORE, the detector mass has increased of more than a factor $10^5$. 
	At the same time the background rate has been constantly reduced and the resolution improved,
	reaching values comparable with those of the best performing detectors.
	CUORE is expected to fully exploit the potentiality of the calorimetric technique. The strict protocol 
	designed for the detector construction is the result of an extensive and continuous R\&D program, while 
	the cryogenic system represents a milestone in the dilution refrigeration technology.
	The successful operation of CUORE has opened the way to one of the most sensitive searches for \bb~and to the possibility 
	for tonne-scale bolometric experiments.
	
	\begin{center}
	\begin{table*}[tb]
		\caption{List of the \ce{TeO_2} bolometric experiments searching for the \bb~of \ce{^{130}Te} performed at LNGS. 
			The main characteristics are reported.}
		\begin{ruledtabular}
		\begin{tabular}{l rcr r r r r r}
		Experiment	&\multicolumn{3}{r}{Running period}	&Crystals	&Mass	&Exposure\nota	&BI in ROI	
																																&FWHM @ $Q_{\beta\beta}$\\
						&&&											&				&[kg]	&[kg\,yr]	&[\ckky]				&[keV]\\
		\hline

		6\,g crystal, \cite{Alessandrello:1992aa}		&Jan 1991 &\SC- &\SB Apr 1991		&1		&0.006	&0.0002	&123	&$50$		\\
		
		21\,g crystal, \cite{Giuliani:1991ze}			&May 1991 &\SC- &\SB Jun 1991		&1		&0.021	&0.0004	&88	&$20$		\\	

		34\,g crystal, \cite{Alessandrello:1992jc}	&Jul 1991 &\SC- &\SB Oct 1991		&1		&0.034	&0.0041	&79	&$40$		\\	

		73\,g crystal, \cite{Alessandrello:1993cs}	&Nov 1991 &\SC- &\SB Aug 1992		&1		&0.073	&0.0087		
																	&35 {\scriptsize(Run\,I)} / 17 {\scriptsize(Run\,II)}		&$7$		\\	

		334\,g crystal, \cite{Alessandrello:1994tm,Alessandrello:1994_XX}	
																	&Jan 1993 &\SC- &\SB Aug 1994		&1		&0.334	&0.40		&3.4	&$15$		\\	

		4 crystal array, \cite{Brofferio:1995wx}		&Oct 1994 &\SC- &\SB Oct 1995		&4		&1.32		
																															&0.73\notaB	&4	&$\lesssim 12$\\

		MiDBD, \cite{Arnaboldi:2002te}					&Apr 1998 &\SC- &\SB Dec 2001		&20	&6.8		&4.25
																	&0.6 {\scriptsize(Run\,I)} / 0.3 {\scriptsize(Run\,II)}		&$5-15$	\\

		Cuoricino, \cite{Andreotti:2010vj}
																	&Mar 2003&\SC - &\SB Jun 2008		&62	&40.7		&71.4	
																	&0.20 {\scriptsize(Run\,I)} / 0.15 {\scriptsize(Run\,II)}	&$5.8 \pm 2.1$\notaC\\

		CUORE-0, \cite{Alfonso:2015wka,Alduino:2016vjd}					
																	&Mar 2013&\SC - &\SB Mar 2015		&52	&39.0		&35.2	&$0.058 \pm 0.004$	
																	&$4.9 \pm 2.9$\\

		CUORE  {\scriptsize (first data)}, \cite{Alduino:2017ehq}					
																	&May 2017&\SC - &\SB Sep 2017		&988	&742		&86.3	&$0.014 \pm 0.002$	
																	&$7.7 \pm 0.5$\\[+3pt]

		CUORE {\scriptsize (expected)}, \cite{Artusa:2014lgv}	&May 2017&\SC - & 2023		&988	&742		&3705		&$0.01$	&5	\\
		
		\end{tabular}
		\end{ruledtabular}
		\begin{flushleft}
		\nota {\scriptsize Total exposure of \ce{TeO_2}. To get the corresponding exposure of 
			\ce{^{130}Te}, it is sufficient to multiply the value per $\simeq 0.278$, with the exceptions of MiDBD and Cuoricino, which contained enriched crystals
			} \\
		\notaB {\scriptsize combined with 334\,g crystal.} \\ 
		\notaC {\scriptsize \cite{Alduino:2016vjd}.} 
		\end{flushleft}
		\label{tab:exp_list}
	\end{table*}
	\end{center}

%% file: 4_DBDLimits.tex
\subsection{Limits on the \bb~process}
\label{sec:limits}

	\begin{figure}[p]
		\includegraphics[width=1.\columnwidth]{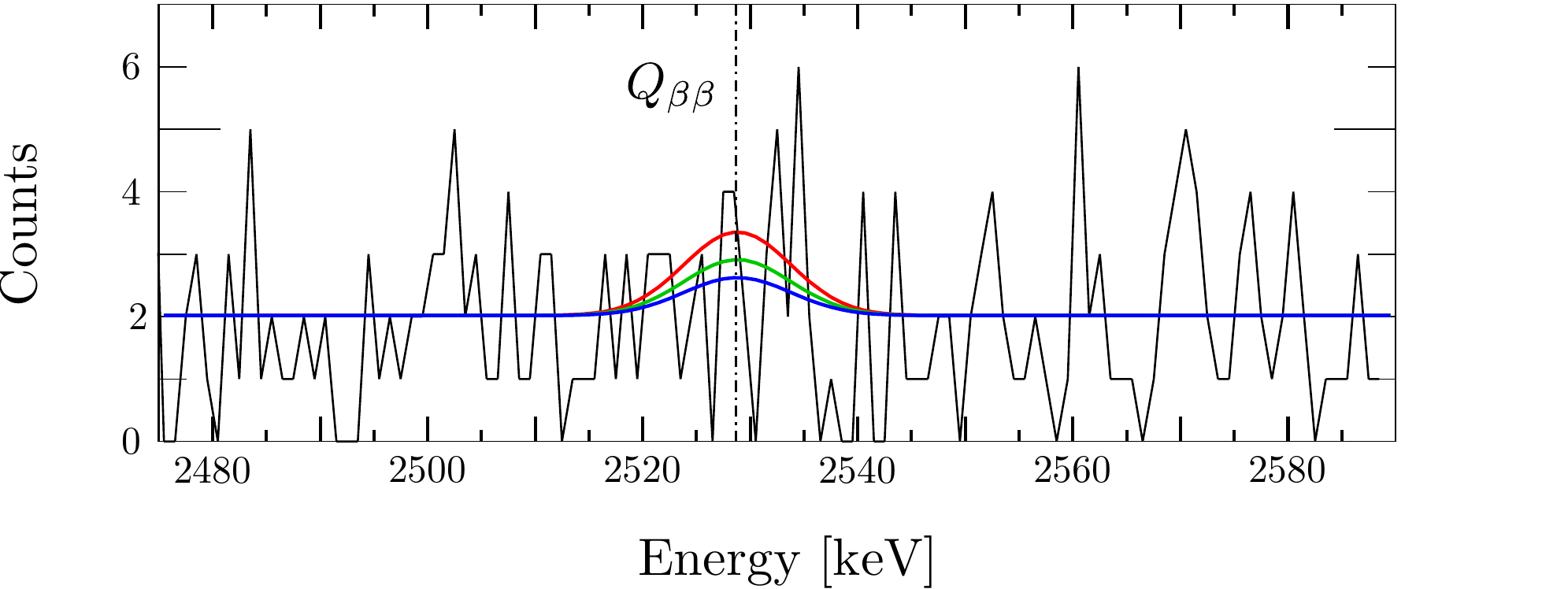} \\[10pt]
		\includegraphics[width=1.\columnwidth]{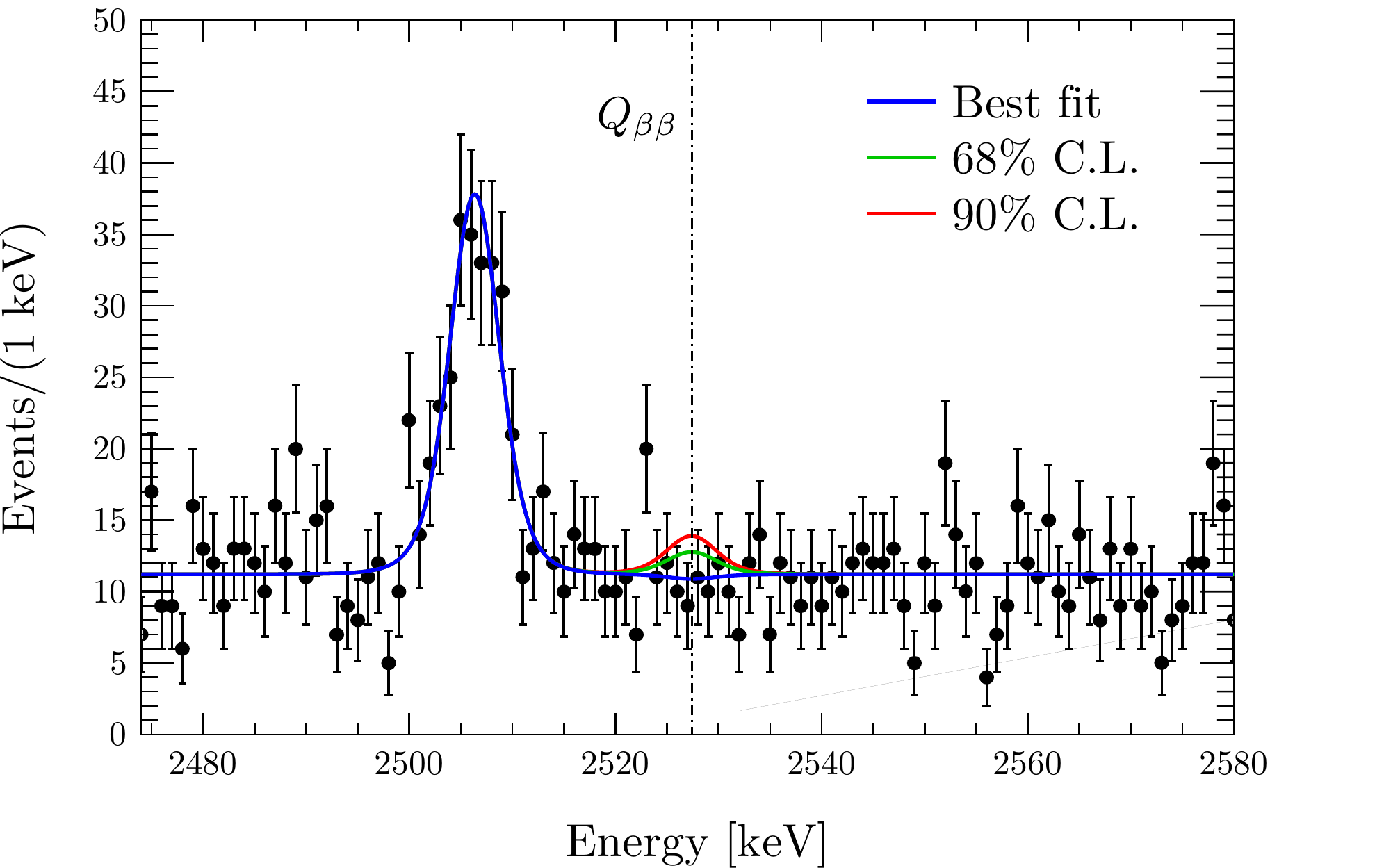} \\[10pt]
		\includegraphics[width=1.\columnwidth]{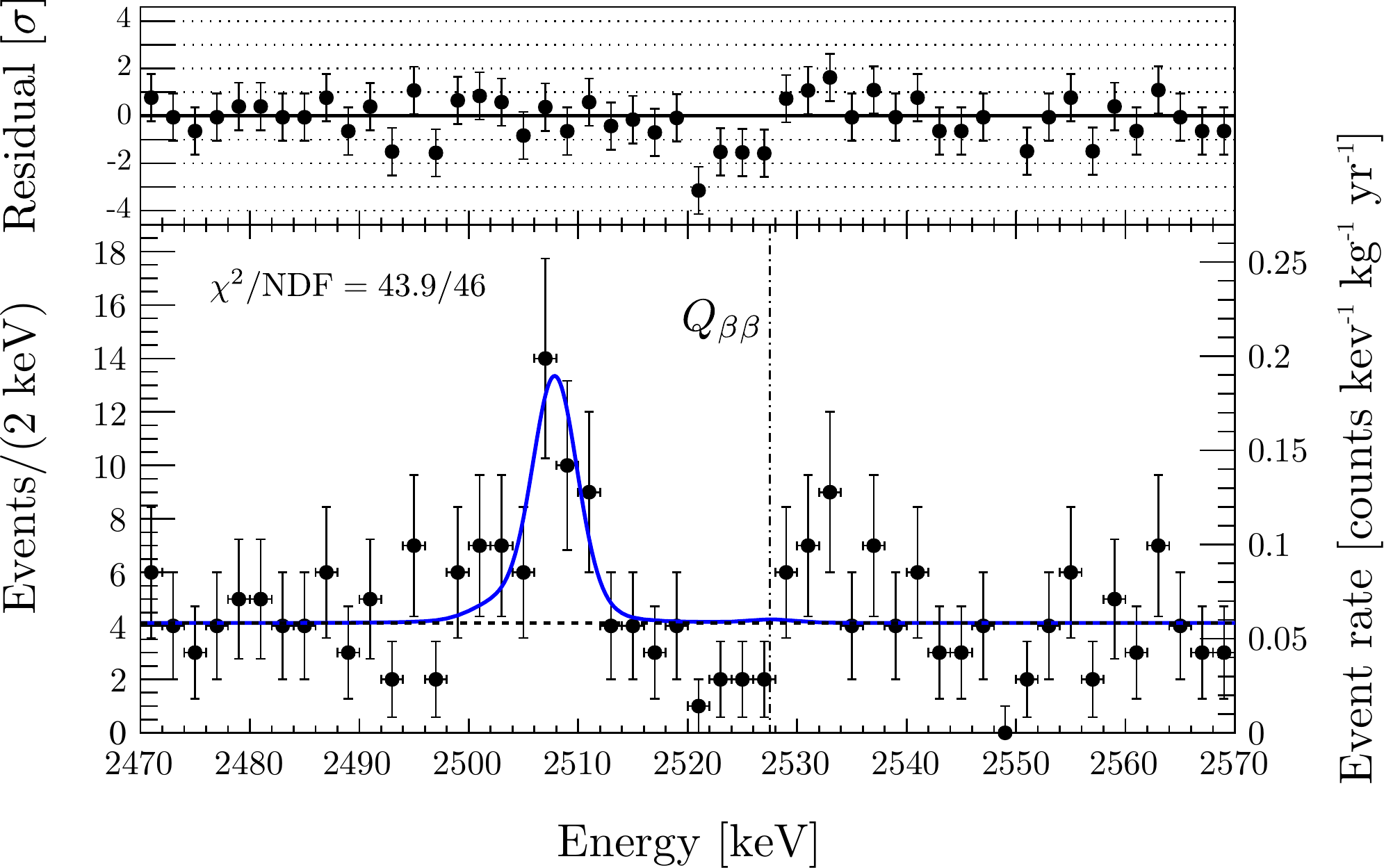} \\[10pt]
		\includegraphics[width=1.\columnwidth]{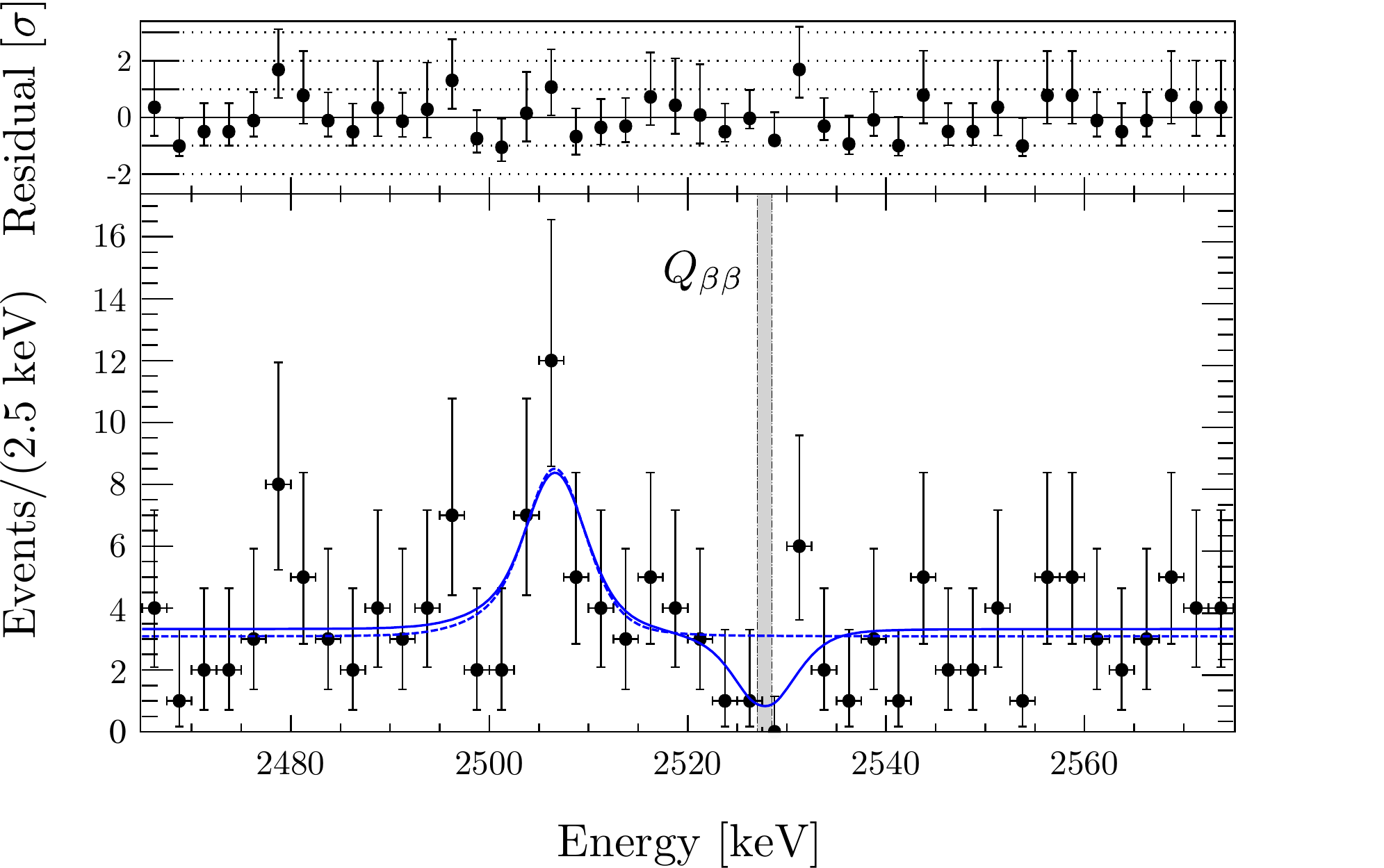}
		\caption{(From Top to Bottom) MiDBD, Cuoricino, \mbox{CUORE-0} and CUORE (first data release) best-fit model 
			in the ROI overlaid on the data points.
			The peak at $\sim~2507\,\keV$ is attributed to \ce{^{60}Co}.
			The vertical dot-dashed line indicates the position of the $Q_{\beta\beta}$ of \ce{^{130}Te}.
		 	Adapted from Refs.~\cite{Arnaboldi:2002te,Andreotti:2010vj,Alfonso:2015wka,Alduino:2017ehq}
		 	(see the references for details).}
		\label{fig:ROI_spectra}
	\end{figure}

	In Fig.~\ref{fig:ROI_spectra}, the zoom on the ROI region of the final MiDBD, Cuoricino and CUORE-0 spectra and 
	of the first CUORE spectrum are shown.
	No evidence for the \bb~of \ce{^{130}Te} has been found so far.
	However, as shown in Table~\ref{tab:half-life}, these and the previous experiments allowed to set more and more stringent limits on the process.
	Starting from the early measurements almost thirty years ago, the evolution of the limit on the decay half-life of the 
	\bb~of \ce{^{130}Te} proceeded in parallel with that of the bolometric measurements.
	The first results presented to the scientific community were immediately competitive%
	\footnote{The most stringent limit on the inclusive double beta decay (\bb~+ \bbvv) of \ce{^{130}Te} at the time of the 6\,g 
		crystal measurement (1991, see Sec.~\ref{sec:6g_crystal}) was $2.6 \cdot 10^{21}\,\yr$ obtained with the 
		geochemical method~\cite{Kirsten:1983jj}.
		The tightest limit on \bb~from a direct measurement was instead $2.8 \cdot 10^{18}\,\yr$ from a 
		measurement with a \ce{CdTe} detector~\cite{Mitchell:1988zz}
		(the result $1.2 \cdot 10^{21}\yr$ from Ref.~\cite{Zdesenko:1980}
		was overestimated since the energy resolution of the detector was not properly taken into 
		account~\cite{Mitchell:1988zz,Alessandrello:1992vk}).}
	and the subsequent improvements translated in an increase of sensitivity of several orders of magnitude.

	Up to date, the limit on the \ce{^{130}Te} \bb~coming from CUORE (combined with CUORE-0 and Cuoricino)
	is $1.5 \cdot 10^{25}$\,yr~\cite{Alduino:2017ehq}.
	It follows the limits on the analogous processes in \ce{^{136}Xe} and 
	\ce{^{76}Ge}, $1.1\cdot10^{26}\,\yr$~\cite{KamLAND-Zen:2016pfg} and 
	$8.0\cdot10^{25}\,\yr$~\cite{Agostini:2018tnm} at $90\%\,$C.\,L., respectively.
	CUORE is eventually expected to raise the half-life limit close to $10^{26}\,\yr$~\cite{Alduino:2017pni}, 
	thus ensuring the central role of 
	the search for \bb~of \ce{^{130}Te} with thermal detectors also in the forthcoming future.

	As illustrated in Sec.~\ref{sec:mbb_limit}, an experimental bound on the decay half-life can be translated into an upper limit on 
	the Majorana effective mass by inverting Eq.~\eqref{eq:tau}.
	In the idea of a relative comparison, the value of $\gA$ has been fixed to $\gAn$ and only one model for the NME has been considered 
	(neither the more favorable nor the more unfavorable for \ce{^{130}Te}) for the results reported in Table~\ref{tab:half-life}.
	
	In order to deal with the uncertainties coming from the theoretical side, more values for the NME and for $\gA$ should be selected.
	This is actually done by the experimental collaborations when presenting their results. However, the set of used NME calculations is arbitrary and the value of the 
	axial coupling constant is commonly fixed to $\gAn$. Also, the error on the PSF is usually not propagated.

\begin{figure}[tb]
		\centering
		\includegraphics[width=1.\columnwidth]{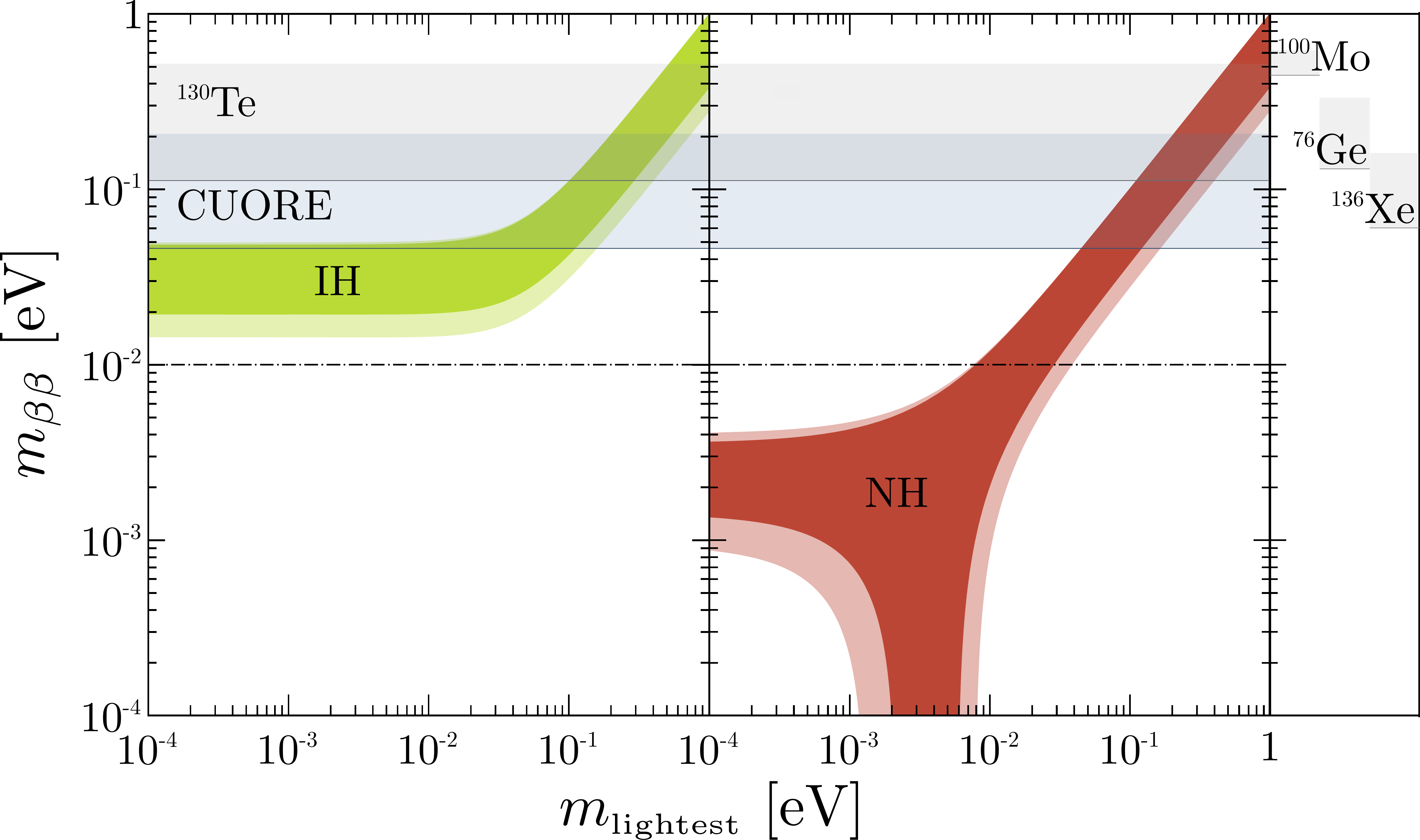}
		\caption{Predictions on $\mbb$ from oscillations with the related $3\sigma$ regions~\cite{Capozzi:2016rtj}
			in the two cases of \NH~and \IH.
			The horizontal bands show the current \bb~limit of \ce{^{130}Te}~\cite{Alduino:2017ehq}
			and the projected CUORE sensitivity~\cite{Alduino:2017pni}.
			The analogous limits on
			\ce{^{76}Ge}~\cite{Agostini:2018tnm}, \ce{^{100}Mo}~\cite{Arnold:2015wpy} and \ce{^{136}Xe}~\cite{KamLAND-Zen:2016pfg}
			are also reported.
			The line $\mbb = 10$\,meV indicates the target value for a future \bb~experiment capable of discriminating
			between \NH~and \IH~(see Sec.~\ref{sec:future}).}
		\label{fig:mbb_ml}
	\end{figure}

	In Fig.~\ref{fig:mbb_ml}, the ``official'' bounds coming from some of the most studied isotopes (see also the discussion in Sec.~\ref{sec:future})
	are plotted adopting the representation $\mbb~vs~m_\mathrm{lightest}$, the latter parameter being the mass of the lightest neutrino, in the two 
	(mutually exclusive) scenarios of Normal and Inverted Hierarchy (\NH~and \IH) of the neutrino mass spectrum. 
	It can be seen that the constraint from \ce{^{130}Te}, i.\,e.\ the current CUORE limit $\mbb < (110-520)\,\meV$~\cite{Alduino:2017ehq},
	is compatible with that from \ce{^{76}Ge}, while the projected CUORE
	sensitiviy is expected to reach (or pass) the most stringent present limit, coming from \ce{^{136}Xe}, approaching the Inverted Hierarchy region.

	\begin{center}
	\begin{table}[tb]
		\caption{Limits on the \bb~half-life of \ce{^{130}Te} and corresponding lower bounds for the Majorana 
			effective mass.
			When comparable, the newer limits are combined with the old ones.
			In order to pass from $\taubb$ to $\mbb$, the PSF from Ref.~\cite{Kotila:2012zza}, the NME from 
			Ref.~\cite{Barea:2015kwa} and $\gAn$ have been used.}
		\begin{ruledtabular}
		\begin{tabular}{l r r}
			Experiment	&$\taubb$ {\scriptsize (90\%\,C.\,L.)}\,[yr]		&$\mbb$\,[eV]	\\
			\hline
			6\,g crystal,~\cite{Alessandrello:1992aa}		&$4.0 \cdot 10^{19}$		&$115$	\\		
			21\,g crystal,~\cite{Giuliani:1991ze}			&$5.3 \cdot 10^{19}$		&$100$	\\	
			34\,g crystal,~\cite{Alessandrello:1992jc}	&$4.4 \cdot 10^{20}$		&$34$		\\	
			73\,g crystal\nota,~\cite{Alessandrello:1993cs}	&$2.7 \cdot 10^{21}$	&$14$		\\	
			334\,g crystal,~\cite{Alessandrello:1994_XX}	&$2.1 \cdot 10^{22}$		&$5$	\\
			4 crystal array,~\cite{Brofferio:1995wx}		&$3.3 \cdot 10^{22}$		&$4$	\\
			MiDBD,~\cite{Arnaboldi:2002te}					&$2.1 \cdot 10^{23}$		&$1.6$	\\
			Cuoricino,~\cite{Andreotti:2010vj}				&$2.8 \cdot 10^{24}$		&$0.43$	\\
			CUORE-0,~\cite{Alfonso:2015wka}					&$4.0 \cdot 10^{24}$		&$0.36$	\\
			CUORE {\scriptsize (first data)},~\cite{Alduino:2017ehq}
																		&$1.5 \cdot 10^{25}$		&$0.19$	\\[+3pt]
			CUORE {\scriptsize (expected)},~\cite{Alduino:2017pni}
																		&$9.0 \cdot 10^{25}$		&$0.076$	\\
		\end{tabular}
		\end{ruledtabular}
		\begin{flushleft}
		\nota {\scriptsize only data from Run\,II were used for the \bb~analysis.} 
		\end{flushleft}
		\label{tab:half-life}
	\end{table}
	\end{center}

%% file: 5_Challenges_Te.tex
\section{Challenges for a next \bb~bolometric detector}
\label{sec:challenges}

	The future of the \bb~search with thermal detectors will not be concluded with CUORE. 
	On the contrary, CUORE itself represents a fundamental step towards the next generation of detectors.
	The knowledge acquired by running the first tonne-scale bolometric array is of crucial importance 
	to the understanding of how to further increase the sensitivity.
	It is thus possible to imagine a future \bb~experiment with improved performance to be placed inside the CUORE 
	cryogenic infrastructure.
	
	Bolometers offer a wide choice of possible absorber materials and thus isotopes different than \ce{^{130}Te} can be 
	in principle selected, thanks to their ``intrinsic'' properties 
	and/or to the more favorable characteristics of the resulting detector.
	This indeed represents a concrete possibility on the path towards a post-CUORE experiment (see Sec.~\ref{sec:CUPID} and 
	Ref.~\cite{Artusa:2014wnl}).
	Anyway, the following discussion will focus on \ce{TeO2}, since this remains a valid choice, with still much room 
	for improvement towards a more powerful search for \bb.
	
\subsection{Improving the sensitivity}

	As shown in Sec.~\ref{sec:limits}, the probing power of the search for \bb~is determined by  
	the experiment live-time, the isotope mass, the energy resolution and the background level.
	These are thus the quantities that need to be addressed in order to try to improve the sensitivity.
	In particular, in the case of a next generation experiment, like the one we are considering, 
	the aim is to reach the zero background condition, i.\,e.\ the relation expressed in Eq.~(\ref{eq:zb}) must hold.
	This sets very stringent requirements on the detector features and performance.

	In order to increase the mass of the isotope under study, it is necessary either to increase the whole detector mass 
	or to enrich the detector material in that isotope (or both).
	In case of a future experiment inside the CUORE cryostat, it can be imagined that an optimization of the detector 
	design will allow to allocate more material than in CUORE. 
	However, this option is ultimately constrained by the experimental volume inside the cryostat.
	On the other side, the use of \ce{TeO_2} crystals enriched in \ce{^{130}Te} seems a viable path to be followed and the first 
	results in this direction are promising.
	The recent run of two \ce{TeO_2} enriched crystal ($\sim 92\%$ of \ce{^{130}Te}) 
	in the Hall C DR~\cite{Artusa:2016mat} proved that performance 
	comparable to those of CUORE-0 are achievable. 
	The crystals (of mass $\sim 435\,\g$ each) showed a FWHM energy resolution at the \ce{^{208}Tl} 
	line of $6.5$ and $4.3\,\keV$, respectively.
	At the same time, the assessment of the crystal internal contamination indicated 
	a relatively small presence of \ce{^{238}U}, although $(10-20)$ times larger than that of the CUORE crystals
	(see Sec.~\ref{sec:CCVR}), while only an upper limit was obtained for \ce{^{232}Th} (within a factor 2 with respect to 
	the CUORE crystals).
	Therefore, it is possible to foresee a potential increase in the isotope mass up to a factor $\sim 3$.

	Since it is unlikely that a \bb~experiment life will be longer than a few years, the optimization of the experiment duty 
	cycle is another important factor in order to maximize the exposure.
	CUORE-0 already showed that values higher than $80\%$ are achievable.%
	\footnote{The duty cycle of CUORE-0 was actually $78.6\%$~\cite{Alduino:2016vjd}. However, this value includes the 
		downtime between the two data taking campaigns (see Sec.~\ref{sec:CUORE-0}).}
	In addition, the use of Pulse Tube Refrigerators in the CUORE cryostat (see Sec.~\ref{sec:CUORE}) further increases the total duty 
	cycle with respect to a LHe bath cryostat (like the Hall A DR), due to the absence of cryogens to refill.
	It is thus reasonable to expect a duty cycle of about $80\%$ for a future experiment.
	
	Regarding the resolution, bolometers fully exploit the potentiality of solid state detectors, showing values 
	of the order of the per mille at the $Q_{\beta\beta}$, close to those of HPGe detectors. 
	As discussed in Sec.~\ref{sec:resolution}, the effective resolution is due to various noise sources. 
	In principle the intrinsic value for CUORE-like bolometers is of the order of some tens of eV. 
	However, in order to try to improve the single detector performance, a better understanding of the bolometer
	behavior is crucial. Several studies aiming at building a working thermal model for the CUORE bolometers have 
	been carried out over the years~\cite{Pedretti_PhD-thesis:2004,Vignati_PhD-thesis:2010,Santone_PhD-thesis:2017}
	and soon a large amount of data will be available thanks to CUORE,
	which will provide useful information in this direction.
	Anyway, since at present we do not know the level of improvement we can expect or how much time will be needed in order to 
	obtain practical effects for the experimental search, no further improvement on this parameter will be considered.
	The value of $5\,\keV$ already achieved by CUORE-0 still represents an
	acceptable target for a forthcoming experiment.
	
	The suppression of the background is the target of most of the studies and R\&D projects aiming at 
	identifying the best configuration and technology for a new generation bolometric detector.
	This in fact is the key to reach the zero background condition and will actually represent
	the largest challenge to deal with.
	The good news is that there is still room for a significant improvement with respect to the $0.01\,$\ckky of CUORE, 
	as it will be discussed in the next section.
	
\subsection{Background suppression}

	In order to identify the most effective ways to reduce the background in a future experiment it is necessary 
	to understand which are the actual contributions to the counting rate observed in CUORE.
	The natural starting point when dealing with this issue is represented by the Monte Carlo simulation already 
	developed since MiDBD and continuously improved thereafter. 
	In fact, thanks to the experience acquired in running the 
	subsequent detectors, especially Cuoricino and \mbox{CUORE-0},
	this work allowed the implementation a background model for CUORE~\cite{Alduino:2017qet}.
	
\subsubsection{Contributions to the CUORE background}
\label{sec:CUORE_bkg}

	As discussed in Sec.~\ref{sec:bkg_study}, three dominant sources were identified by MiDBD and 
	Cuoricino as contributors to the Background Index (BI), i.\,e.\ to the event rate in the ROI:%
	\footnote{The ROI can be intended as the 100 keV-wide region around $Q_{\beta\beta}$, namely $(2470-2570)\,\keV$.}
	\footnote{In the same energy region, we also observe the $2505\,\keV$ \ce{^{60}Co} peak (see Fig.~\ref{fig:ROI_spectra}) which anyway does not contribute to the BI.}
	\begin{itemize}
		\item the multi-Compton events from the $2615\,\keV$ $\gamma$s of \ce{^{208}Tl} originating from the \ce{^{232}Th} 
			contamination of the cryogenic system;
		\item the \ce{^{238}U} and \ce{^{232}Th} contaminations (with the related decay products) on the surface of the 
			\ce{TeO_2} crystals, including the \ce{^{210}Pb} surface implantation from the environmental \ce{^{222}Rn};
		\item the \ce{^{238}U} and \ce{^{232}Th} contaminations (with the related decay products) on the surfaces of the 
			inert materials facing the crystals, most likely of the copper holders.
	\end{itemize}
	This analysis guided the following strategies in view of CUORE, in particular the minimization of the amount of material 
	used in the detector support structure, the design of the shields and dedicated cleaning 
	treatments for both crystals and copper.
	Their effectiveness was proved by CUORE-0, which observed an almost 3 times reduced background 
	with respect to Cuoricino (see Sec.~\ref{sec:CUORE-0}).
	In particular, this was found to be mainly imputable to the cryostat shields ($\sim 74.4\%$~\cite{Alduino:2016vtd}),
	which were still the old ones of the Hall A DR.
	
	In CUORE, an extensive screening campaign preceded the selection and procurement of the materials used in the 
	construction, not only of the detector, but also of the cryostat.
	Different material assay techniques allowed a deep investigation of the role of natural contaminants 
	(especially \ce{^{238}U} and \ce{^{232}Th} with the respective progenies) and of cosmogenically activated
	contaminants (\ce{^{60}Co}, \ce{^{110}Ag} and \ce{^{110m}Ag}~\cite{Barghouty:2010kj})
	which are expected to yield a contribution in the ROI.
	The strict selection of the CUORE components took into account both 
	bulk and surface contaminations, 
	the latter possibly occurring during the part machining and cleaning, or during exposure to contaminated air.%
	\footnote{In addition, the break of the secular equilibrium of a radioactive chain could be relevant for surface 
		contaminations as well, such as in the case of \ce{^{210}Pb} from the \ce{^{238}U} chain.}
	
	\begin{figure}[tb]
		\centering
		\includegraphics[width=1.\columnwidth]{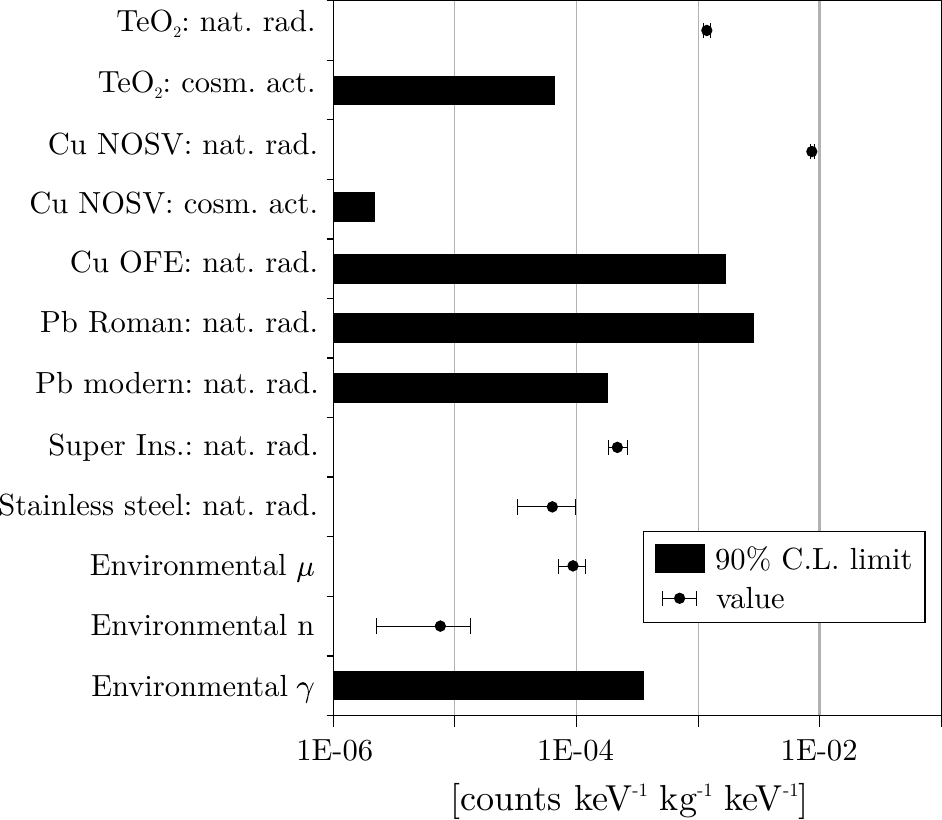}
		\caption{Histogram representing the main BI expected for the various components of CUORE. 
			The bars indicate 90\%\,C.\,L.\ upper limits while the markers derived values 
			(with the relative $1\sigma$ uncertainty).
			The thicker line at $0.01$\,\ckky indicate the goal for the CUORE background.
			Figure from Ref.~\cite{Alduino:2017qet} (see the reference for details).}
		\label{fig:CUORE_bkg}
	\end{figure}

	As a result of this work, the sources that could give sizable contribution to the CUORE BI were identified to be
	(Fig.~\ref{fig:CUORE_bkg}):
	\begin{itemize}
		\item \ce{^{238}U} and \ce{^{232}Th} and their progenies in crystals and holders;
		\item the cosmogenically activated isotopes in crystals and copper parts (both detector frames and Tower Support Plate
			+ MC plate and vessel, see Fig.~\ref{fig:cryostat_scheme});
		\item \ce{^{238}U} and \ce{^{232}Th} in the other vessels and in the lead shields.
	\end{itemize}
	The total projected BI was found to be equal to 
	\mbox{$(1.02 \pm 0.03\,\text{(stat)}^{+0.23}_{-0.10}\,\text{(syst)})\cdot10^{-2}$\,\ckky,}
	by far dominated by the degraded $\alpha$s from surface contaminants of the detector frames.
	This value is perfectly compatible with the observed one of $(0.014\pm0.002)$\,\ckky~\cite{Alduino:2017ehq}.

	\medskip
	As a remark, it is important to notice that the definitive validation or rejection of this projected 
	background model for CUORE will come from the CUORE data themselves (with sufficiently high statistics).
	These will finally allow disentangling the different contributions to the effective counting rate in the ROI.

\subsubsection{Background active rejection}
\label{sec:bkg_active_rej}

	In CUORE, $\alpha$ events and multi-Compton scatters of $\gamma$s are discarded mainly via anti-coincidence analysis, 
	i.\,e.\ by rejecting simultaneous events (within 10\,ms) generated in nearby crystals~\cite{Alduino:2017ehq}.
	This method has a high efficiency, but it only allows to identify decays that deposit energy in multiple crystals.
	
\begin{figure}[tb]
		\includegraphics[width=1.\columnwidth]{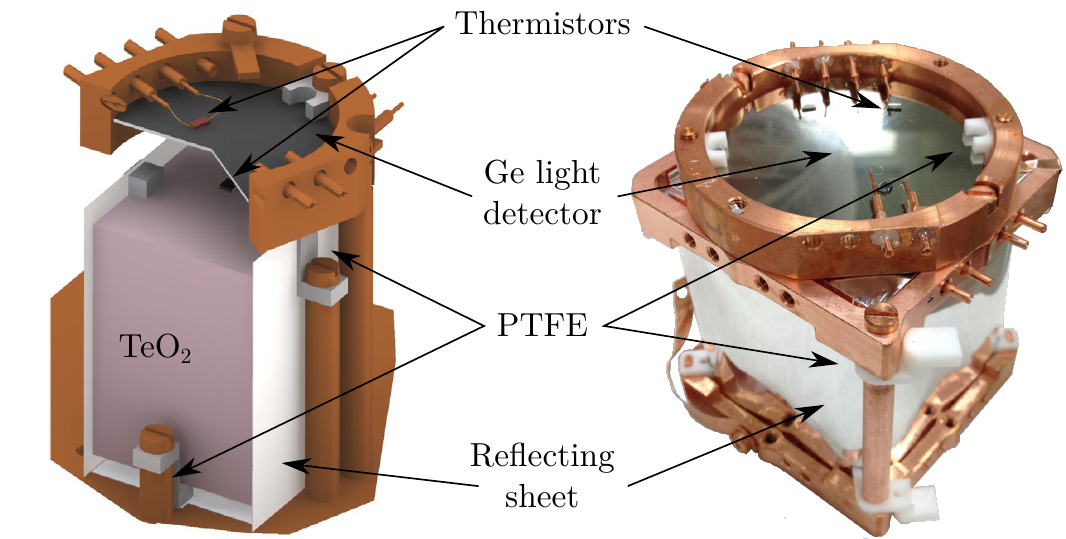}
		\caption{\ce{TeO_2} crystals coupled to a light detector for a simultaneous measurement of heat and light
			signals. The crystals are wrapped with a reflecting sheet in order to enhance the light collection.
			Adapted from Refs.~\cite{Artusa:2016mat,Casali:2014vvt}.}
		\label{fig:TeO2_light}
	\end{figure}

	Therefore, in view of a forthcoming detector, more sensitive approaches for active background reduction 
	against $\alpha$ events have been proposed.
	These foresee the use of a dual readout approach based on the simultaneous measurement of heat and light signals,
	the latter being collected by a thin \ce{Ge} or \ce{Si} slab coupled to the crystal and also employed 
	as a bolometer (Fig.~\ref{fig:TeO2_light}).
	
	Scintillation light due to heavy particles is strongly quenched, thus the measurement of the ratio of the emitted light 
	over the energy dissipated into heat allows to discriminate between $\alpha$ and $\beta/\gamma$ events with the 
	same energy.
	Very good results in this direction have already been obtained
	(see e.\,g.\ Refs.~\cite{Artusa:2016maw,Kim:2015pua,Armengaud:2017hit}), 
	however this approach is limited to materials that exhibit scintillation properties. 
	Therefore, the use of this techniques with \ce{TeO_2} crystals is frustrated by the low luminescence of paratellurite.%
	\footnote{Although in the end standard \ce{TeO_2} was used, dedicated R\&D studies, aiming at increasing the 
		light emission in the crystals by mean of suitable dopants, had been performed in view of 
		CUORE~\cite{Dafinei:2005xy,Dafinei:2007hhc}.}
	
	Nonetheless, it was shown in Ref.~\cite{TabarellideFatis:2009zz} that \ce{TeO_2} crystals have suitable
	optical properties to act as excellent \Cerenkov~radiators, with the threshold for \Cerenkov~emission of about $50\,\keV$ 
	for electrons and of about $400\,\MeV$ for $\alpha$ particles. 
	Therefore, as in the case of scintillation, no light emission is expected from the $\alpha$ background, allowing for 
	a full background rejection, in principle even by detecting one single \Cerenkov~photon.
	A total of about 125 photons (in the range $(350-600)$\,nm) is expected for the two electrons emitted 
	in the \bb~process and a light yield of about 52 \Cerenkov~photons per deposited MeV was actually 
	measured in a CUORE-like crystal~\cite{Bellini:2014yoa}.
	
	The \Cerenkov~emission from \ce{TeO_2} was observed with dedicated measurements, first in a 
	small crystal~\cite{Beeman:2011yc,Bellini:2012rc}
	and then in a full CUORE-size one~\cite{Casali:2014vvt}.
	The results showed that the discrimination of $\alpha$ particles in CUORE is feasible. However, the collected light 
	signal was too small. A higher sensitivity than that provided by standard bolometers is required in order to 
	effectively apply an active discrimination in a future \bb~experiment, thus allowing for a 
	drastic reduction of the background.

\subsubsection{CUPID}
\label{sec:CUPID}

	Starting from the experience, the expertise, and the lessons learned in CUORE, the CUPID project
	(CUORE Upgrade with Particle IDentification~\cite{Wang:2015raa}) aims at developing a future bolometric \bb~experiment
	with sensitivity on $\taubb$ of the order of $(10^{27}-10^{28})\,\yr$.%
	\footnote{As introduced at the beginning of this section, the choice of an isotope different than \ce{^{130}Te}
		is a possibility, but we will only consider the \ce{TeO2} case.}
		
\begin{figure}[tb]
		\centering
		\includegraphics[width=.9\columnwidth]{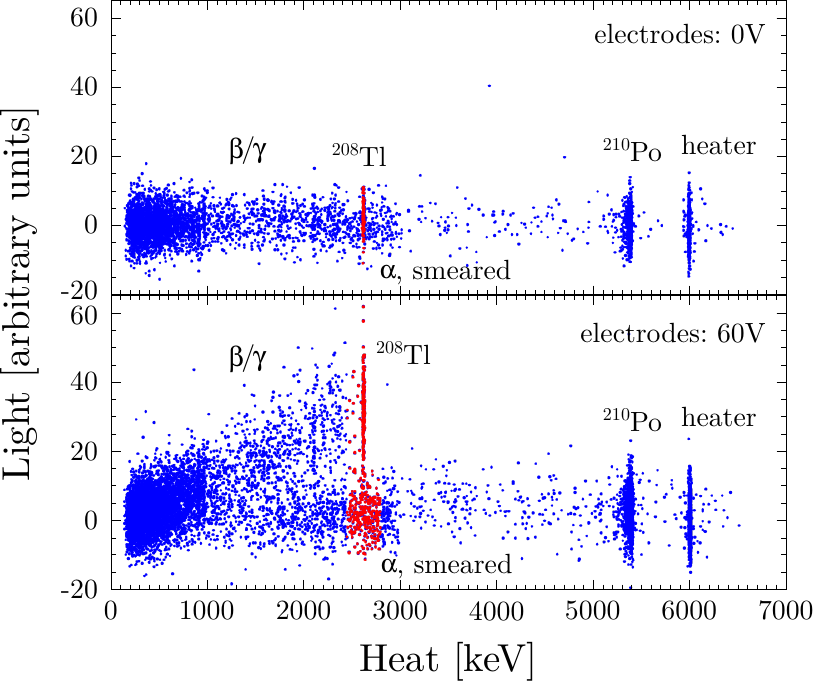}
		\caption{Scatter plots of the heat $vs.$ light signals in the measurements with a CUORE bolometer 
			in coincidence with a NTL \ce{Ge} light detector operated respectively as a standard light 
			detector (0\,V electrode bias, Top) and in a signal amplification regime (60\,V electrode bias, Bottom).
			Figure from Ref.~\cite{Berge:2017nys} (see the reference for details).}
		\label{fig:NL_Berge}
	\end{figure}

	One of the main efforts of CUPID is devoted to the identification of effective strategies to reduce the 
	background in the ROI, especially by developing high-resolution light detectors capable of clearly identify 
	\Cerenkov~emission.
	Several technologies are under investigation in this direction and many R\&D programs are in 
	different stages of evolution~\cite{Wang:2015taa}.
	Among these, very promising solutions foresee the use of 
	microwave kinetic inductance detectors~\cite{Battistelli:2015vha,Bellini:2016lgg},
	transition edge sensors \cite{Schaffner:2014caa,Willers:2014eoa}
	or Neganov-Trofimov-Luke (NTL) amplified light 
	detectors~\cite{Willers:2014eoa,Casali:2015gya,Biassoni:2015eij,Artusa:2016mat,Gironi:2016nae,Berge:2017nys}.
	In particular, the very recent results in Ref.~\cite{Berge:2017nys} demonstrated a complete event-by-event 
	$\alpha~vs.~\beta/\gamma$ separation in a full \ce{TeO_2} CUORE bolometer 
	thanks to a NTL-assisted \ce{Ge} bolometer for the light detection (Fig.~\ref{fig:NL_Berge}).
	
	Complementary studies within CUPID focus on the tagging of surface events. In fact, due to the intrinsic working 
	principle, bolometers do not have a dead layer and are fully sensitive up to the surface.
	While this property is responsible for the excellent energy resolution and high detection efficiency, 
	it represents a problem when surface events constitute a dominant component of the background, 
	since the crystal surface is as sensitive as the bulk.
	In order to address this issue, the addition of passive elements on the bolometer surface has been proposed.
	The deposition of superconducting \ce{Al} films should affect the shape of the events originating a few mm from the 
	surface, thus one expects different signal shapes for surface and bulk events~\cite{Nones:2012tsm}.
	Alternatively, surface energy events could be identified thanks to an external plastic scintillator,
	by encapsulating the crystal with a scintillating foil and by adding a light detector to measure 
	(in addition to the \Cerenkov~light) the light emitted in the interaction of surface events with the 
	foil itself~\cite{Canonica:2013jvz}.
	
	The proposed solutions, together with dedicated strategies for the reduction of the environmental 
	radioactivity~\cite{Wang:2015taa}, should translate into a reduction of the BI of 2 orders of magnitude 
	with respect to CUORE, bringing it to the level of $0.1$\,\ckty and thus allowing the 
	construction of a zero background (see Sec.~\ref{sec:limits}) tonne-size bolometric experiment.

%% file: 6_Future.tex
\section{$\text{Te}$ and the bolometric technique in the future of the \bb~search}
\label{sec:future}

	\begin{center}
	\begin{table*}[tb]
		\caption{Sensitivity and mass of the isotope of a future experiment able to discriminate between \NH~and \IH. 
			The target is to probe $\mbb = 10\,\meV$, assuming small uncertainties on the NMEs, 
			and three possible values for $\gA$ (see the text for details).
			The calculations are performed assuming experiments with 100\% detection efficiency, no fiducial volume cuts and zero background
			with the required $(B\cdot \Delta)$ product to fulfill this condition.}
		\begin{ruledtabular}
		\begin{tabular}{l rr rrr rrr rrr}
			Isotope	&$G_{0\nu}$		&$\mathcal{M}_{0\nu}$
			&\multicolumn{3}{c}{$S_{1/2,\,0\mathrm{bkg}}^{\,0\nu}~[10^{28}\,\yr]$} 	
			&\multicolumn{3}{c}{M [t]}	
			&\multicolumn{3}{c}{$(B\cdot \Delta)_{0\mathrm{bkg}}$~[t$^{-1}$\,yr$^{-1}$]}\\[+2pt]

			\cline{4-6}	\cline{7-9}	\cline{10-12} \\[-12pt]

								&[f\,yr$^{-1}$]	&			&$\ggAn$	&$\ggAq$	&$\ggAp$		&$\ggAn$	&$\ggAq$	&$\ggAp$		&$\ggAn$	&$\ggAq$	&$\ggAp$		\\[+2pt]
			
			\hline \\[-7pt]

			\ce{^{76}Ge}	&2.36					&4.48		&2.1		&5.5		&48			&0.8		&2.0		&17.5			&0.3		&0.1		&0.01		\\[+2pt]

			\ce{^{100}Mo}	&15.9					&4.51		&0.3		&0.8		&8.6			&0.1		&0.4		&4.1			&1.3		&0.5		&0.05		\\[+2pt]

			\ce{^{130}Te}	&14.2					&3.56		&0.6		&1.5		&18			&0.3		&0.9		&11.6			&0.6		&0.2		&0.02		\\[+2pt]

			\ce{^{136}Xe}	&14.6					&2.58		&1.0		&2.7		&36			&0.7		&1.8		&23.2			&0.3		&0.1		&0.009	\\[+3pt]

		\end{tabular}
		\end{ruledtabular}
		\label{tab:mega-ultimate}
	\end{table*}
	\end{center}
	
	In the future of the \bb~search, it is likely that very few isotopes and techniques will be selected to further continue the challenge.
	Therefore, it is important to understand the chances of \ce{^{130}Te} and bolometers to be picked as suitable candidates.

	The cost of the experiment will become a critical aspect and politics will play a central role in the down-selection.
	In any case, all the collaborations will have first of all to demonstrate their capability to reach a certain sensitivity goal.
	Then, let us try to infer some modest but fair considerations by comparing the case of \ce{^{130}Te} with thermal detectors with 
	those of \ce{^{76}Ge}, \ce{^{100}Mo} and \ce{^{136}Xe}, that are at present setting the most stringent limits on \bb.

	Following a similar argument to that presented in Ref.~\cite{Dell'Oro:2014yca}, let us imagine 
	a (not too close) future experiment with enhanced sensitivity, able to discriminate between 
	the two scenarios of \NH~and \IH. Therefore, we require the sensitivity to be equivalent to $\mbb = 10\,\meV$ (Fig.~\ref{fig:mbb_ml}):
	the corresponding limit on $\taubb$ can be obtained by inverting Eq.~\eqref{eq:tau}.

	As discussed in Sec.~\ref{sec:mbb_limit}, when converting the experimental limit on the 
	decay half-life time into a bound on the Majorana effective mass, it is necessary to pass
	through the theoretical calculations of nuclear physics, which up to date present quite large uncertainties.
	We opted to use a value of $\mathcal{M}_{0\nu}$ (refer to Eq.~(\ref{eq:tau}))
	obtained by averaging the individual ones from Refs.~\cite{Barea:2015kwa,Simkovic:2013qiy,Hyvarinen:2015bda}, and to
	consider more possibilities for the value of the axial coupling constant, i.\,e.\ $\gA$ equal to $\gAn$, $\gAq$ or $\gAp$,
	hoping that at some point the issue of the quenching will be sorted out.
	This choice is somehow different from the ``usual'' one, adopted by the experimental collaborations, of selecting 
	multiple NME calculations and fixing $\gA$ (see also Sec.~\ref{sec:limits}).
	However, it is a conservative approach that explores different potential scenarios.
	In fact, assuming less favorable values for $\gA$ is equivalent to require a higher sensitivity 
	on $\mbb$ with $\gA = \gAn$, namely the case $\gA = \gAq$ would corresponds to a goal of $\mbb \simeq 6\,\meV$, while the case $\gA = \gAp$
	to $\mbb \simeq 2\,\meV$, more or less at the center of the \NH~band for the lightest neutrino mass approaching zero.
	
	In this ideal detector, we assume the whole mass as entirely made of the isotope of interest, 100\% detection efficiency and no fiducial volume cuts.
	Therefore, by fixing the live-time, let us say to 5\,yr, the (isotope) mass needed to reach the target sensitivity 
	can be extracted by inverting Eq.~(\ref{eq:0bkg_sens}).
	Since this equation expects the zero background condition to be satisfied, the 
	constraint expressed in Eq.~(\ref{eq:zb}) must hold.
	Up to date only GERDA (\ce{^{76}Ge}) has achieved a low enough BI 
	to fulfill the requirement at the 1\,t-scale~\cite{Agostini:2017iyd}, while
	an analogous ``effective'' demonstration is still missing for the other techniques.

	Table~\ref{tab:mega-ultimate} summarizes the results: \ce{^{100}Mo} and \ce{^{130}Te} appear to the most convenient isotopes in terms of half-life time sensitivity, 
	smaller detector mass needed and, consequently, looser constraint on the background level,
	with a factor $\sim 2.5$ in favor of the former.
	This advantage though decreases to $\sim 1.5$ in the realistic comparison between \ce{TeO_2} and \ce{Li_2MoO_4}/\ce{ZnMoO_4}~\cite{Armengaud:2017hit},
	given that the most promising use of \ce{^{100}Mo} is with a bolometric experiment too.
	Instead, the situation seems less favorable for \ce{^{76}Ge} and \ce{^{136}Xe} approximately at the same level.

	As an additional consideration, quite borderline for a scientific discussion such as the current one, something can be said 
	about the cost of the isotope, which will represent a large (if not the major) fraction of the whole experiment's cost.
	There is large room for speculation and it is possible that the situation will significantly change in the future.
	However, neglecting the procurement of the natural element and its purification, it is realistic to assume that the minimum cost 
	for the enrichment in the isotope of interest will be represented by that for the extraction of \ce{^{136}Xe}, 
	the easiest and least expensive isotope to separate since, 
	being it in the gas form, it can directly undergo centrifugation.
	The current value amounts to 20\,\$ per gram of isotope~\cite{Biller:2013wua}.%
	\footnote{Actually, the cost of enrichment is proportional the ratio between the output and input isotopic abundances
		and the process almost never reaches 100\% extraction. 
		Anyway, for the sake of the present discussion, it is acceptable to approximate 
		the enrichment to a ``complete'' process, fully separating the isotope of interest.}
	A similar cost can be assumed for \ce{^{130}Te}~\cite{Avignone_chat}, while larger ones in a ratio 3 to 5 have to be considered for
	\ce{^{76}Ge} and \ce{^{100}Mo}~\cite{Bucci_WN2014}.
	Although in a rather qualitative way, this information plays for \ce{^{130}Te}.
	
	Anyway, it is important to stress again that, independent from any comparison, each experiment will be asked to prove its
	ability to achieve the corresponding half-life time sensitivity with zero background.
	Therefore, the amount of material and the background level required to \ce{^{130}Te}
	in the most optimistic scenario ($\gA=\gAn$) will be likely fulfilled by CUPID (see Sec.~\ref{sec:CUPID}), 
	while in the case of $\gA=\gAq$ a larger effort would be needed. Still, this could be seen as an open challenge.
	The third and most pessimistic scenario appears unaccessible to all experiments, but a larger understanding of nuclear physics on the theoretical side 
	is needed in order to extract more quantitative information.

	The present discussion provides rather generic indications and makes assumptions whose validity in some cases has yet to be proven. 
	Also, the estimate of the costs only takes into account the isotope procurement and not those of the whole experimental infrastructures and of the feasibility 
	of constructing and running such experiments.
	Anyway, despite their simplicity, these considerations address the main issues concerning a discussion on the future searches for \bb~and provide some useful evaluation elements.

%% file: 7_Summary.tex
\section{Summary}
\label{sec:summary}

	Neutrinoless double beta decay is a key tool to address some of the major outstanding issues in neutrino physics, 
	such as the lepton number conservation and the Majorana nature of the neutrino.

\begin{figure}[b]
		\includegraphics[width=1.\columnwidth]{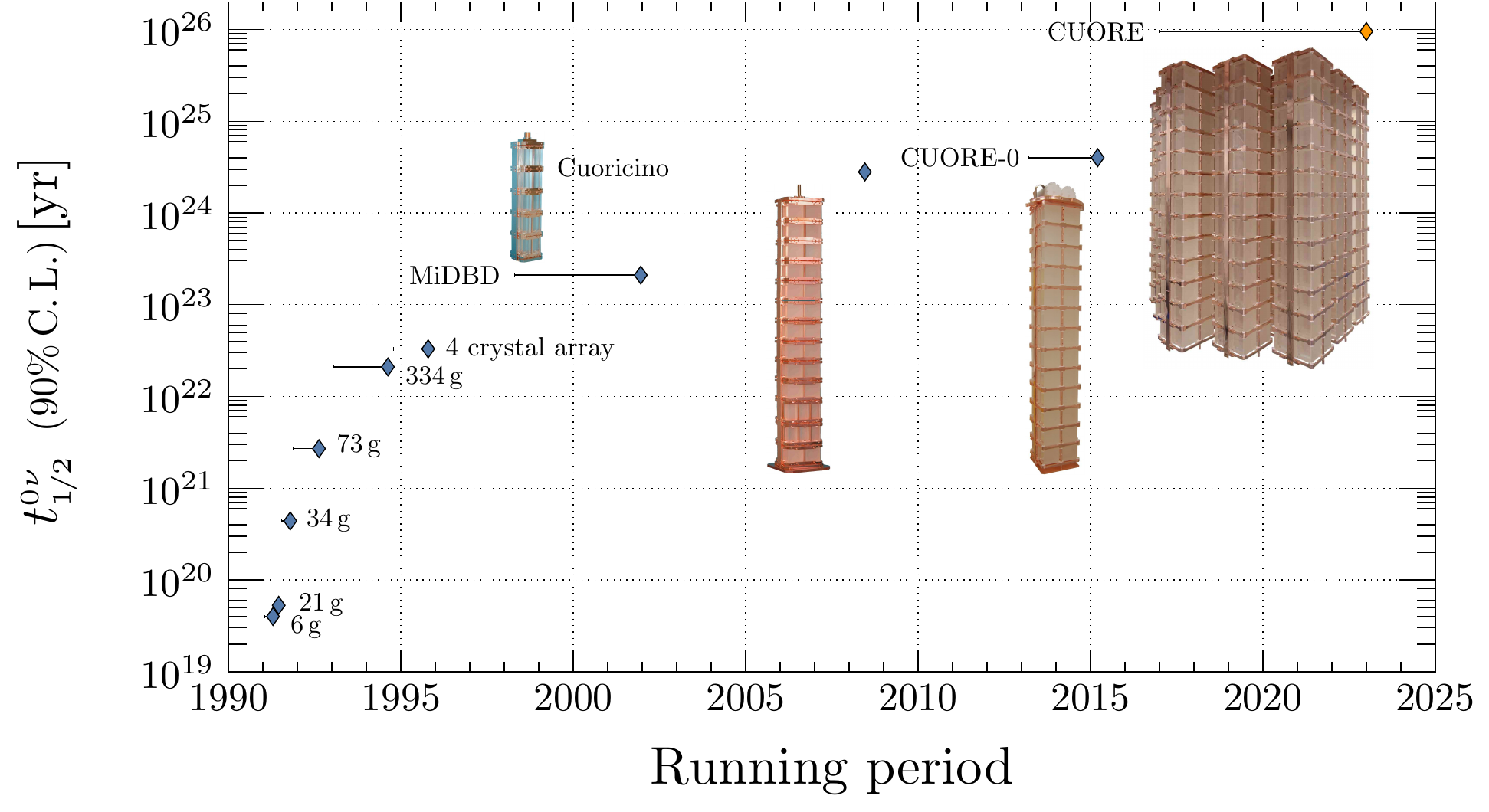}
		\caption{Limits on the \ce{^{130}Te} \bb~half-life achieved by the described experiments. 
			See Table~\ref{tab:half-life} for the references.}
		\label{fig:exposure}
	\end{figure}

	Thermal detectors based on \ce{TeO_2} fulfill the requirements for a competitive experiment searching for this rare process.
	In their almost thirty-year-old history at LNGS, these devices have achieved very important results, setting more and 
	more stringent constraints on the decay \bb~half-life of \ce{^{130}Te} (Fig.~\ref{fig:exposure}).
	The improvements on the performance cover all aspects, from the energy resolution to the background
	reduction and today the present limit by CUORE, $1.5\cdot 10^{25}\,\yr$ at $90\%$\,C.\,L., is among the most 
	competitive ones in the field.

	Looking towards a next generation bolometric experiment searching for \bb, there is still room for a significant 
	enhancement in the sensitivity. The numerous R\&D projects are showing promising results
	towards a complete particle identification, thus leading to a strong suppression of the background.
	The construction of a zero background experiment seems within the reach.
	It is fair to imagine that such a detector will reach sensitivities of the order of $(10^{27}-10^{28})$\,yr.
	
	Within the worldwide contest of the search for \bb, \ce{^{130}Te} represents one of the most 
	favorable candidates and it is likely that bolometric detectors will continue to play a crucial role in this challenge.

%% file: acknowledgments.tex
\begin{acknowledgments}

	The present review is a re-elaboration of a very long history of research activities.
	The reported scientific material comes from original works published initially by the group of Fiorini and Collaborators 
	and later by the Cuoricino and CUORE Collaborations.
	The achieved results have been made possible by the dedication and efforts by these Physicists and by the Technical Staff 
	of the Laboratori Nazionali del Gran Sasso and of the other Institutions involved.
	
	The authors thank A.~McGuckin for proofing the manuscript.

\end{acknowledgments}

%% file: main.bbl
\begin{thebibliography}{153}%
\makeatletter
\providecommand \@ifxundefined [1]{%
 \@ifx{#1\undefined}
}%
\providecommand \@ifnum [1]{%
 \ifnum #1\expandafter \@firstoftwo
 \else \expandafter \@secondoftwo
 \fi
}%
\providecommand \@ifx [1]{%
 \ifx #1\expandafter \@firstoftwo
 \else \expandafter \@secondoftwo
 \fi
}%
\providecommand \natexlab [1]{#1}%
\providecommand \enquote  [1]{``#1''}%
\providecommand \bibnamefont  [1]{#1}%
\providecommand \bibfnamefont [1]{#1}%
\providecommand \citenamefont [1]{#1}%
\providecommand \href@noop [0]{\@secondoftwo}%
\providecommand \href [0]{\begingroup \@sanitize@url \@href}%
\providecommand \@href[1]{\@@startlink{#1}\@@href}%
\providecommand \@@href[1]{\endgroup#1\@@endlink}%
\providecommand \@sanitize@url [0]{\catcode `\\12\catcode `\$12\catcode
  `\&12\catcode `\#12\catcode `\^12\catcode `\_12\catcode `\%12\relax}%
\providecommand \@@startlink[1]{}%
\providecommand \@@endlink[0]{}%
\providecommand \url  [0]{\begingroup\@sanitize@url \@url }%
\providecommand \@url [1]{\endgroup\@href {#1}{\urlprefix }}%
\providecommand \urlprefix  [0]{URL }%
\providecommand \Eprint [0]{\href }%
\providecommand \doibase [0]{http://dx.doi.org/}%
\providecommand \selectlanguage [0]{\@gobble}%
\providecommand \bibinfo  [0]{\@secondoftwo}%
\providecommand \bibfield  [0]{\@secondoftwo}%
\providecommand \translation [1]{[#1]}%
\providecommand \BibitemOpen [0]{}%
\providecommand \bibitemStop [0]{}%
\providecommand \bibitemNoStop [0]{.\EOS\space}%
\providecommand \EOS [0]{\spacefactor3000\relax}%
\providecommand \BibitemShut  [1]{\csname bibitem#1\endcsname}%
\let\auto@bib@innerbib\@empty
\bibitem [{\citenamefont {Furry}(1939)}]{Furry:1939qr}%
  \BibitemOpen
  \bibfield  {author} {\bibinfo {author} {\bibfnamefont {W.~H.}\ \bibnamefont
  {Furry}},\ }\href {\doibase 10.1103/PhysRev.56.1184} {\bibfield  {journal}
  {\bibinfo  {journal} {Phys.\ Rev.}\ }\textbf {\bibinfo {volume} {56}},\
  \bibinfo {pages} {1184} (\bibinfo {year} {1939})}\BibitemShut {NoStop}%
\bibitem [{\citenamefont {Cremonesi}\ and\ \citenamefont
  {Pavan}(2013)}]{Cremonesi:2013vla}%
  \BibitemOpen
  \bibfield  {author} {\bibinfo {author} {\bibfnamefont {O.}~\bibnamefont
  {Cremonesi}}\ and\ \bibinfo {author} {\bibfnamefont {M.}~\bibnamefont
  {Pavan}},\ }\href {\doibase 10.1155/2014/951432} {\bibfield  {journal}
  {\bibinfo  {journal} {Adv.\ High Energy Phys.}\ }\textbf {\bibinfo {volume}
  {2014}},\ \bibinfo {pages} {951432} (\bibinfo {year} {2013})}\BibitemShut
  {NoStop}%
\bibitem [{\citenamefont {Dell'Oro}\ \emph {et~al.}(2016)\citenamefont
  {Dell'Oro}, \citenamefont {Marcocci},\ and\ \citenamefont
  {Vissani}}]{Dell'Oro:2016dbc}%
  \BibitemOpen
  \bibfield  {author} {\bibinfo {author} {\bibfnamefont {S.}~\bibnamefont
  {Dell'Oro}}, \bibinfo {author} {\bibfnamefont {M.}~\bibnamefont {Marcocci},
  \bibfnamefont {S.and~Viel}}, \ and\ \bibinfo {author} {\bibfnamefont
  {F.}~\bibnamefont {Vissani}},\ }\href {\doibase 10.1155/2016/2162659}
  {\bibfield  {journal} {\bibinfo  {journal} {Adv.\ High Energy Phys.}\
  }\textbf {\bibinfo {volume} {2016}},\ \bibinfo {pages} {2162659} (\bibinfo
  {year} {2016})}\BibitemShut {NoStop}%
\bibitem [{\citenamefont {Kotila}\ and\ \citenamefont
  {Iachello}(2012)}]{Kotila:2012zza}%
  \BibitemOpen
  \bibfield  {author} {\bibinfo {author} {\bibfnamefont {J.}~\bibnamefont
  {Kotila}}\ and\ \bibinfo {author} {\bibfnamefont {F.}~\bibnamefont
  {Iachello}},\ }\href {\doibase 10.1103/PhysRevC.85.034316} {\bibfield
  {journal} {\bibinfo  {journal} {Phys.\ Rev.\ C}\ }\textbf {\bibinfo {volume}
  {85}},\ \bibinfo {pages} {034316} (\bibinfo {year} {2012})}\BibitemShut
  {NoStop}%
\bibitem [{\citenamefont {Barea}\ \emph {et~al.}(2015)\citenamefont {Barea},
  \citenamefont {Kotila},\ and\ \citenamefont {Iachello}}]{Barea:2015kwa}%
  \BibitemOpen
  \bibfield  {author} {\bibinfo {author} {\bibfnamefont {J.}~\bibnamefont
  {Barea}}, \bibinfo {author} {\bibfnamefont {J.}~\bibnamefont {Kotila}}, \
  and\ \bibinfo {author} {\bibfnamefont {F.}~\bibnamefont {Iachello}},\ }\href
  {\doibase 10.1103/PhysRevC.91.034304} {\bibfield  {journal} {\bibinfo
  {journal} {Phys.\ Rev.\ C}\ }\textbf {\bibinfo {volume} {91}},\ \bibinfo
  {pages} {034304} (\bibinfo {year} {2015})}\BibitemShut {NoStop}%
\bibitem [{\citenamefont {\u{S}imkovic}\ \emph {et~al.}(2013)\citenamefont
  {\u{S}imkovic}, \citenamefont {Rodin}, \citenamefont {Faessler},\ and\
  \citenamefont {Vogel}}]{Simkovic:2013qiy}%
  \BibitemOpen
  \bibfield  {author} {\bibinfo {author} {\bibfnamefont {F.}~\bibnamefont
  {\u{S}imkovic}}, \bibinfo {author} {\bibfnamefont {V.}~\bibnamefont {Rodin}},
  \bibinfo {author} {\bibfnamefont {A.}~\bibnamefont {Faessler}}, \ and\
  \bibinfo {author} {\bibfnamefont {P.}~\bibnamefont {Vogel}},\ }\href
  {\doibase 10.1103/PhysRevC.87.045501} {\bibfield  {journal} {\bibinfo
  {journal} {Phys.\ Rev.\ C}\ }\textbf {\bibinfo {volume} {87}},\ \bibinfo
  {pages} {045501} (\bibinfo {year} {2013})}\BibitemShut {NoStop}%
\bibitem [{\citenamefont {Faessler}\ \emph {et~al.}(2008)\citenamefont
  {Faessler}, \citenamefont {Fogli}, \citenamefont {Lisi}, \citenamefont
  {Rodin}, \citenamefont {Rotunno},\ and\ \citenamefont
  {\u{S}imkovic}}]{Faessler:2007hu}%
  \BibitemOpen
  \bibfield  {author} {\bibinfo {author} {\bibfnamefont {A.}~\bibnamefont
  {Faessler}}, \bibinfo {author} {\bibfnamefont {G.~L.}\ \bibnamefont {Fogli}},
  \bibinfo {author} {\bibfnamefont {E.}~\bibnamefont {Lisi}}, \bibinfo {author}
  {\bibfnamefont {V.}~\bibnamefont {Rodin}}, \bibinfo {author} {\bibfnamefont
  {A.~M.}\ \bibnamefont {Rotunno}}, \ and\ \bibinfo {author} {\bibfnamefont
  {F.}~\bibnamefont {\u{S}imkovic}},\ }\href {\doibase
  10.1088/0954-3899/35/7/075104} {\bibfield  {journal} {\bibinfo  {journal}
  {J.\ Phys.\ G}\ }\textbf {\bibinfo {volume} {35}},\ \bibinfo {pages} {075104}
  (\bibinfo {year} {2008})}\BibitemShut {NoStop}%
\bibitem [{\citenamefont {Barea}\ \emph {et~al.}(2013)\citenamefont {Barea},
  \citenamefont {Kotila},\ and\ \citenamefont {Iachello}}]{Barea:2013bz}%
  \BibitemOpen
  \bibfield  {author} {\bibinfo {author} {\bibfnamefont {J.}~\bibnamefont
  {Barea}}, \bibinfo {author} {\bibfnamefont {J.}~\bibnamefont {Kotila}}, \
  and\ \bibinfo {author} {\bibfnamefont {F.}~\bibnamefont {Iachello}},\ }\href
  {\doibase 10.1103/PhysRevC.87.014315} {\bibfield  {journal} {\bibinfo
  {journal} {Phys.\ Rev.\ C}\ }\textbf {\bibinfo {volume} {87}},\ \bibinfo
  {pages} {014315} (\bibinfo {year} {2013})}\BibitemShut {NoStop}%
\bibitem [{\citenamefont {Robertson}(2013)}]{Robertson:2013cy}%
  \BibitemOpen
  \bibfield  {author} {\bibinfo {author} {\bibfnamefont {R.~G.~H.}\
  \bibnamefont {Robertson}},\ }\href {\doibase 10.1142/S0217732313500211}
  {\bibfield  {journal} {\bibinfo  {journal} {Mod.\ Phys.\ Lett.\ A}\ }\textbf
  {\bibinfo {volume} {28}},\ \bibinfo {pages} {1350021} (\bibinfo {year}
  {2013})}\BibitemShut {NoStop}%
\bibitem [{\citenamefont {Dell'Oro}\ \emph {et~al.}(2014)\citenamefont
  {Dell'Oro}, \citenamefont {Marcocci},\ and\ \citenamefont
  {Vissani}}]{Dell'Oro:2014yca}%
  \BibitemOpen
  \bibfield  {author} {\bibinfo {author} {\bibfnamefont {S.}~\bibnamefont
  {Dell'Oro}}, \bibinfo {author} {\bibfnamefont {S.}~\bibnamefont {Marcocci}},
  \ and\ \bibinfo {author} {\bibfnamefont {F.}~\bibnamefont {Vissani}},\ }\href
  {\doibase 10.1103/PhysRevD.90.033005} {\bibfield  {journal} {\bibinfo
  {journal} {Phys.\ Rev.\ D}\ }\textbf {\bibinfo {volume} {90}},\ \bibinfo
  {pages} {033005} (\bibinfo {year} {2014})}\BibitemShut {NoStop}%
\bibitem [{\citenamefont {Engel}\ and\ \citenamefont
  {Men\'endez}(2017)}]{Engel:2016xgb}%
  \BibitemOpen
  \bibfield  {author} {\bibinfo {author} {\bibfnamefont {J.}~\bibnamefont
  {Engel}}\ and\ \bibinfo {author} {\bibfnamefont {J.}~\bibnamefont
  {Men\'endez}},\ }\href {\doibase 10.1088/1361-6633/aa5bc5} {\bibfield
  {journal} {\bibinfo  {journal} {Rep.\ Prog.\ Phys.}\ }\textbf {\bibinfo
  {volume} {80}},\ \bibinfo {pages} {046301} (\bibinfo {year}
  {2017})}\BibitemShut {NoStop}%
\bibitem [{\citenamefont {Lisi}\ \emph {et~al.}(2015)\citenamefont {Lisi},
  \citenamefont {Rotunno},\ and\ \citenamefont {\u{S}imkovic}}]{Lisi:2015yma}%
  \BibitemOpen
  \bibfield  {author} {\bibinfo {author} {\bibfnamefont {E.}~\bibnamefont
  {Lisi}}, \bibinfo {author} {\bibfnamefont {A.}~\bibnamefont {Rotunno}}, \
  and\ \bibinfo {author} {\bibfnamefont {F.}~\bibnamefont {\u{S}imkovic}},\
  }\href {\doibase 10.1103/PhysRevD.92.093004} {\bibfield  {journal} {\bibinfo
  {journal} {Phys.\ Rev.\ D}\ }\textbf {\bibinfo {volume} {92}},\ \bibinfo
  {pages} {093004} (\bibinfo {year} {2015})}\BibitemShut {NoStop}%
\bibitem [{\citenamefont {Suhonen}(2017)}]{Suhonen:2017rjf}%
  \BibitemOpen
  \bibfield  {author} {\bibinfo {author} {\bibfnamefont {J.}~\bibnamefont
  {Suhonen}},\ }\href {\doibase 10.1103/PhysRevC.96.055501} {\bibfield
  {journal} {\bibinfo  {journal} {Phys.\ Rev.\ C}\ }\textbf {\bibinfo {volume}
  {96}},\ \bibinfo {pages} {055501} (\bibinfo {year} {2017})}\BibitemShut
  {NoStop}%
\bibitem [{\citenamefont {Giuliani}\ and\ \citenamefont
  {Poves}(2012)}]{Giuliani:2012zu}%
  \BibitemOpen
  \bibfield  {author} {\bibinfo {author} {\bibfnamefont {A.}~\bibnamefont
  {Giuliani}}\ and\ \bibinfo {author} {\bibfnamefont {A.}~\bibnamefont
  {Poves}},\ }\href {\doibase 10.1155/2012/857016} {\bibfield  {journal}
  {\bibinfo  {journal} {Adv.\ High Energy Phys.}\ }\textbf {\bibinfo {volume}
  {2012}},\ \bibinfo {pages} {857016} (\bibinfo {year} {2012})}\BibitemShut
  {NoStop}%
\bibitem [{\citenamefont {Das}\ \emph {et~al.}(2015)\citenamefont {Das},
  \citenamefont {Ghorui}, \citenamefont {Raina}, \citenamefont {Singh},
  \citenamefont {Rath}, \citenamefont {Cappella}, \citenamefont {Cerulli},
  \citenamefont {Laubenstein}, \citenamefont {Belli},\ and\ \citenamefont
  {Bernabei}}]{Das:2015yba}%
  \BibitemOpen
  \bibfield  {author} {\bibinfo {author} {\bibfnamefont {S.}~\bibnamefont
  {Das}}, \bibinfo {author} {\bibfnamefont {S.~K.}\ \bibnamefont {Ghorui}},
  \bibinfo {author} {\bibfnamefont {P.~K.}\ \bibnamefont {Raina}}, \bibinfo
  {author} {\bibfnamefont {A.~K.}\ \bibnamefont {Singh}}, \bibinfo {author}
  {\bibfnamefont {P.~K.}\ \bibnamefont {Rath}}, \bibinfo {author}
  {\bibfnamefont {F.}~\bibnamefont {Cappella}}, \bibinfo {author}
  {\bibfnamefont {R.}~\bibnamefont {Cerulli}}, \bibinfo {author} {\bibfnamefont
  {M.}~\bibnamefont {Laubenstein}}, \bibinfo {author} {\bibfnamefont
  {P.}~\bibnamefont {Belli}}, \ and\ \bibinfo {author} {\bibfnamefont
  {R.}~\bibnamefont {Bernabei}},\ }\href {\doibase 10.1016/j.nima.2015.06.053}
  {\bibfield  {journal} {\bibinfo  {journal} {Nucl.\ Instrum.\ Meth.\ A}\
  }\textbf {\bibinfo {volume} {797}},\ \bibinfo {pages} {130} (\bibinfo {year}
  {2015})}\BibitemShut {NoStop}%
\bibitem [{\citenamefont {Zdesenko}(1980)}]{Zdesenko:1980}%
  \BibitemOpen
  \bibfield  {author} {\bibinfo {author} {\bibfnamefont {Y.~G.}\ \bibnamefont
  {Zdesenko}},\ }\href
  {http://www.jetpletters.ac.ru/ps/1423/article_21633.shtml} {\bibfield
  {journal} {\bibinfo  {journal} {JETP Lett.}\ }\textbf {\bibinfo {volume}
  {32}},\ \bibinfo {pages} {58} (\bibinfo {year} {1980})}\BibitemShut {NoStop}%
\bibitem [{\citenamefont {Mitchell}\ and\ \citenamefont
  {Fisher}(1988)}]{Mitchell:1988zz}%
  \BibitemOpen
  \bibfield  {author} {\bibinfo {author} {\bibfnamefont {L.~W.}\ \bibnamefont
  {Mitchell}}\ and\ \bibinfo {author} {\bibfnamefont {P.~H.}\ \bibnamefont
  {Fisher}},\ }\href {\doibase 10.1103/PhysRevC.38.895} {\bibfield  {journal}
  {\bibinfo  {journal} {Phys.\ Rev.\ C}\ }\textbf {\bibinfo {volume} {38}},\
  \bibinfo {pages} {895} (\bibinfo {year} {1988})}\BibitemShut {NoStop}%
\bibitem [{\citenamefont {Arnold}\ \emph {et~al.}(2011)\citenamefont {Arnold}
  \emph {et~al.}}]{Arnold:2011gq}%
  \BibitemOpen
  \bibfield  {author} {\bibinfo {author} {\bibfnamefont {R.}~\bibnamefont
  {Arnold}} \emph {et~al.} (\bibinfo {collaboration} {NEMO-3 Collaboration}),\
  }\href {\doibase 10.1103/PhysRevLett.107.062504} {\bibfield  {journal}
  {\bibinfo  {journal} {Phys.\ Rev.\ Lett.}\ }\textbf {\bibinfo {volume}
  {107}},\ \bibinfo {pages} {062504} (\bibinfo {year} {2011})}\BibitemShut
  {NoStop}%
\bibitem [{\citenamefont {Biller}(2013)}]{Biller:2013wua}%
  \BibitemOpen
  \bibfield  {author} {\bibinfo {author} {\bibfnamefont {S.~D.}\ \bibnamefont
  {Biller}},\ }\href {\doibase 10.1103/PhysRevD.87.071301} {\bibfield
  {journal} {\bibinfo  {journal} {Phys.\ Rev. D}\ }\textbf {\bibinfo {volume}
  {87}},\ \bibinfo {pages} {071301} (\bibinfo {year} {2013})}\BibitemShut
  {NoStop}%
\bibitem [{\citenamefont {Rahaman}\ \emph {et~al.}(2011)\citenamefont {Rahaman}
  \emph {et~al.}}]{Rahaman:2011zz}%
  \BibitemOpen
  \bibfield  {author} {\bibinfo {author} {\bibfnamefont {S.}~\bibnamefont
  {Rahaman}} \emph {et~al.},\ }\href {\doibase 10.1016/j.physletb.2011.07.078}
  {\bibfield  {journal} {\bibinfo  {journal} {Phys.\ Lett.\ B}\ }\textbf
  {\bibinfo {volume} {703}},\ \bibinfo {pages} {412} (\bibinfo {year}
  {2011})}\BibitemShut {NoStop}%
\bibitem [{\citenamefont {Redshaw}\ \emph {et~al.}(2009)\citenamefont
  {Redshaw}, \citenamefont {Mount}, \citenamefont {Myers},\ and\ \citenamefont
  {Avignone}}]{Redshaw:2009zz}%
  \BibitemOpen
  \bibfield  {author} {\bibinfo {author} {\bibfnamefont {M.}~\bibnamefont
  {Redshaw}}, \bibinfo {author} {\bibfnamefont {B.~J.}\ \bibnamefont {Mount}},
  \bibinfo {author} {\bibfnamefont {E.}~\bibnamefont {Myers}}, \ and\ \bibinfo
  {author} {\bibfnamefont {F.~T.}\ \bibnamefont {Avignone}},\ }\href {\doibase
  10.1103/PhysRevLett.102.212502} {\bibfield  {journal} {\bibinfo  {journal}
  {Phys.\ Rev.\ Lett.}\ }\textbf {\bibinfo {volume} {102}},\ \bibinfo {pages}
  {212502} (\bibinfo {year} {2009})}\BibitemShut {NoStop}%
\bibitem [{\citenamefont {Scielzo}\ \emph {et~al.}(2009)\citenamefont {Scielzo}
  \emph {et~al.}}]{Scielzo:2009nh}%
  \BibitemOpen
  \bibfield  {author} {\bibinfo {author} {\bibfnamefont {N.~D.}\ \bibnamefont
  {Scielzo}} \emph {et~al.},\ }\href {\doibase 10.1103/PhysRevC.80.025501}
  {\bibfield  {journal} {\bibinfo  {journal} {Phys.\ Rev.\ C}\ }\textbf
  {\bibinfo {volume} {80}},\ \bibinfo {pages} {025501} (\bibinfo {year}
  {2009})}\BibitemShut {NoStop}%
\bibitem [{\citenamefont {Alduino}\ \emph
  {et~al.}(2017{\natexlab{a}})\citenamefont {Alduino} \emph
  {et~al.}}]{Alduino:2016vtd}%
  \BibitemOpen
  \bibfield  {author} {\bibinfo {author} {\bibfnamefont {C.}~\bibnamefont
  {Alduino}} \emph {et~al.} (\bibinfo {collaboration} {CUORE Collaboration}),\
  }\href {\doibase 10.1140/epjc/s10052-016-4498-6} {\bibfield  {journal}
  {\bibinfo  {journal} {Eur.\ Phys.\ J.\ C}\ }\textbf {\bibinfo {volume}
  {77}},\ \bibinfo {pages} {13} (\bibinfo {year}
  {2017}{\natexlab{a}})}\BibitemShut {NoStop}%
\bibitem [{\citenamefont {Fehr}\ \emph {et~al.}(2004)\citenamefont {Fehr},
  \citenamefont {Rehkamper},\ and\ \citenamefont {Halliday}}]{Fehr200483}%
  \BibitemOpen
  \bibfield  {author} {\bibinfo {author} {\bibfnamefont {M.~A.}\ \bibnamefont
  {Fehr}}, \bibinfo {author} {\bibfnamefont {M.}~\bibnamefont {Rehkamper}}, \
  and\ \bibinfo {author} {\bibfnamefont {A.~N.}\ \bibnamefont {Halliday}},\
  }\href {\doibase 10.1016/j.ijms.2003.11.006} {\bibfield  {journal} {\bibinfo
  {journal} {Int.\ J.\ Mass Spectrometry}\ }\textbf {\bibinfo {volume} {232}},\
  \bibinfo {pages} {83} (\bibinfo {year} {2004})}\BibitemShut {NoStop}%
\bibitem [{Te-(2016)}]{Te-price}%
  \BibitemOpen
  \href {https://www.metalprices.com/metal/tellurium/tellurium-99-99-ex-china}
  {\enquote {\bibinfo {title} {{https://www.metalprices.com}},}\ } (\bibinfo
  {year} {2016})\BibitemShut {NoStop}%
\bibitem [{Avi(2017)}]{Avignone_chat}%
  \BibitemOpen
  \href@noop {} {\enquote {\bibinfo {title} {{Private communication with Prof.\
  F.\ Avignone}},}\ } (\bibinfo {year} {2017})\BibitemShut {NoStop}%
\bibitem [{\citenamefont {Langley}(1881)}]{Langley:1881}%
  \BibitemOpen
  \bibfield  {author} {\bibinfo {author} {\bibfnamefont {S.~P.}\ \bibnamefont
  {Langley}},\ }\href@noop {} {\bibfield  {journal} {\bibinfo  {journal}
  {Proc.\ Am.\ Acad.\ Arts Sci.\ XVI}\ }\textbf {\bibinfo {volume} {\!\!}},\
  \bibinfo {pages} {342} (\bibinfo {year} {1881})}\BibitemShut {NoStop}%
\bibitem [{\citenamefont {Curie}\ and\ \citenamefont
  {Laborde}(1903)}]{Curie_Laborde:1903}%
  \BibitemOpen
  \bibfield  {author} {\bibinfo {author} {\bibfnamefont {P.}~\bibnamefont
  {Curie}}\ and\ \bibinfo {author} {\bibfnamefont {A.}~\bibnamefont
  {Laborde}},\ }\href@noop {} {\bibfield  {journal} {\bibinfo  {journal} {C.\
  R.\ Acad.\ Sci.}\ }\textbf {\bibinfo {volume} {136}},\ \bibinfo {pages} {673}
  (\bibinfo {year} {1903})}\BibitemShut {NoStop}%
\bibitem [{\citenamefont {Andrews}\ \emph {et~al.}(1949)\citenamefont
  {Andrews}, \citenamefont {Fowler},\ and\ \citenamefont
  {Williams}}]{Andrews:1949}%
  \BibitemOpen
  \bibfield  {author} {\bibinfo {author} {\bibfnamefont {D.~H.}\ \bibnamefont
  {Andrews}}, \bibinfo {author} {\bibfnamefont {R.~D.}\ \bibnamefont {Fowler}},
  \ and\ \bibinfo {author} {\bibfnamefont {M.~C.}\ \bibnamefont {Williams}},\
  }\href {\doibase 10.1103/PhysRev.76.154.2} {\bibfield  {journal} {\bibinfo
  {journal} {Phys.\ Rev.}\ }\textbf {\bibinfo {volume} {76}},\ \bibinfo {pages}
  {154} (\bibinfo {year} {1949})}\BibitemShut {NoStop}%
\bibitem [{\citenamefont {Simon}(1935)}]{Simon:1935}%
  \BibitemOpen
  \bibfield  {author} {\bibinfo {author} {\bibfnamefont {F.}~\bibnamefont
  {Simon}},\ }\href {\doibase 10.1038/135763a0} {\bibfield  {journal} {\bibinfo
   {journal} {Nature}\ }\textbf {\bibinfo {volume} {135}},\ \bibinfo {pages}
  {763} (\bibinfo {year} {1935})}\BibitemShut {NoStop}%
\bibitem [{\citenamefont {Mitsel'makher}\ \emph {et~al.}(1982)\citenamefont
  {Mitsel'makher}, \citenamefont {Neganov},\ and\ \citenamefont
  {Trofimov}}]{Mitsel'makher:1982bb}%
  \BibitemOpen
  \bibfield  {author} {\bibinfo {author} {\bibfnamefont {G.~V.}\ \bibnamefont
  {Mitsel'makher}}, \bibinfo {author} {\bibfnamefont {B.~S.}\ \bibnamefont
  {Neganov}}, \ and\ \bibinfo {author} {\bibfnamefont {V.~N.}\ \bibnamefont
  {Trofimov}},\ }\href@noop {} {} (\bibinfo {year} {1982}),\ \bibinfo {note}
  {[Communication to the Joint Institute for Nuclear Research, Dubna, Russia,
  P8-32-549]}\BibitemShut {NoStop}%
\bibitem [{\citenamefont {Fiorini}\ and\ \citenamefont
  {Niinikoski}(1984)}]{Fiorini:1983yj}%
  \BibitemOpen
  \bibfield  {author} {\bibinfo {author} {\bibfnamefont {E.}~\bibnamefont
  {Fiorini}}\ and\ \bibinfo {author} {\bibfnamefont {T.~O.}\ \bibnamefont
  {Niinikoski}},\ }\href {\doibase 10.1016/0167-5087(84)90449-6} {\bibfield
  {journal} {\bibinfo  {journal} {Nucl.\ Instrum.\ Meth.\ A}\ }\textbf
  {\bibinfo {volume} {224}},\ \bibinfo {pages} {83} (\bibinfo {year}
  {1984})}\BibitemShut {NoStop}%
\bibitem [{\citenamefont {Drukier}\ and\ \citenamefont
  {Stodolsky}(1984)}]{Drukier:1983gj}%
  \BibitemOpen
  \bibfield  {author} {\bibinfo {author} {\bibfnamefont {A.}~\bibnamefont
  {Drukier}}\ and\ \bibinfo {author} {\bibfnamefont {L.}~\bibnamefont
  {Stodolsky}},\ }\href {\doibase 10.1103/PhysRevD.30.2295} {\bibfield
  {journal} {\bibinfo  {journal} {Phys.\ Rev.\ D}\ }\textbf {\bibinfo {volume}
  {30}},\ \bibinfo {pages} {2295} (\bibinfo {year} {1984})}\BibitemShut
  {NoStop}%
\bibitem [{\citenamefont {Moseley}\ \emph {et~al.}(1984)\citenamefont
  {Moseley}, \citenamefont {Mather},\ and\ \citenamefont
  {McCammon}}]{Moseley:1984}%
  \BibitemOpen
  \bibfield  {author} {\bibinfo {author} {\bibfnamefont {S.~H.}\ \bibnamefont
  {Moseley}}, \bibinfo {author} {\bibfnamefont {J.~C.}\ \bibnamefont {Mather}},
  \ and\ \bibinfo {author} {\bibfnamefont {D.}~\bibnamefont {McCammon}},\
  }\href {\doibase 10.1063/1.334129} {\bibfield  {journal} {\bibinfo  {journal}
  {J.\ Appl.\ Phys.}\ }\textbf {\bibinfo {volume} {56}},\ \bibinfo {pages}
  {1257} (\bibinfo {year} {1984})}\BibitemShut {NoStop}%
\bibitem [{LNG()}]{LNGS_site}%
  \BibitemOpen
  \href {https://www.lngs.infn.it/en} {\enquote {\bibinfo {title}
  {{https://www.lngs.infn.it/en}},}\ }\BibitemShut {NoStop}%
\bibitem [{\citenamefont {Catalano}\ \emph {et~al.}(1986)\citenamefont
  {Catalano}, \citenamefont {Cavinato}, \citenamefont {Salvini},\ and\
  \citenamefont {Tozzi}}]{Catalano:1986M}%
  \BibitemOpen
  \bibfield  {author} {\bibinfo {author} {\bibfnamefont {P.~G.}\ \bibnamefont
  {Catalano}}, \bibinfo {author} {\bibfnamefont {G.}~\bibnamefont {Cavinato}},
  \bibinfo {author} {\bibfnamefont {F.}~\bibnamefont {Salvini}}, \ and\
  \bibinfo {author} {\bibfnamefont {M.}~\bibnamefont {Tozzi}},\ }\href@noop {}
  {\bibfield  {journal} {\bibinfo  {journal} {Mem.\ Soc.\ Geol.\ It.}\ }\textbf
  {\bibinfo {volume} {35}},\ \bibinfo {pages} {647} (\bibinfo {year}
  {1986})}\BibitemShut {NoStop}%
\bibitem [{\citenamefont {Ambrosio}\ \emph {et~al.}(1995)\citenamefont
  {Ambrosio} \emph {et~al.}}]{Ambrosio:1995cx}%
  \BibitemOpen
  \bibfield  {author} {\bibinfo {author} {\bibfnamefont {M.}~\bibnamefont
  {Ambrosio}} \emph {et~al.} (\bibinfo {collaboration} {MACRO Collaboration}),\
  }\href {\doibase 10.1103/PhysRevD.52.3793} {\bibfield  {journal} {\bibinfo
  {journal} {Phys.\ Rev.\ D}\ }\textbf {\bibinfo {volume} {52}},\ \bibinfo
  {pages} {3793} (\bibinfo {year} {1995})}\BibitemShut {NoStop}%
\bibitem [{\citenamefont {Best}\ \emph {et~al.}(2016)\citenamefont {Best},
  \citenamefont {G$\ddot{\text{o}}$rres}, \citenamefont {Junker}, \citenamefont
  {Kratz}, \citenamefont {Laubenstein}, \citenamefont {Long}, \citenamefont
  {Nisi}, \citenamefont {Smith},\ and\ \citenamefont
  {Wiescher}}]{Best:2015yma}%
  \BibitemOpen
  \bibfield  {author} {\bibinfo {author} {\bibfnamefont {A.}~\bibnamefont
  {Best}}, \bibinfo {author} {\bibfnamefont {J.}~\bibnamefont
  {G$\ddot{\text{o}}$rres}}, \bibinfo {author} {\bibfnamefont {M.}~\bibnamefont
  {Junker}}, \bibinfo {author} {\bibfnamefont {K.~L.}\ \bibnamefont {Kratz}},
  \bibinfo {author} {\bibfnamefont {M.}~\bibnamefont {Laubenstein}}, \bibinfo
  {author} {\bibfnamefont {A.}~\bibnamefont {Long}}, \bibinfo {author}
  {\bibfnamefont {S.}~\bibnamefont {Nisi}}, \bibinfo {author} {\bibfnamefont
  {K.}~\bibnamefont {Smith}}, \ and\ \bibinfo {author} {\bibfnamefont
  {M.}~\bibnamefont {Wiescher}},\ }\href {\doibase 10.1016/j.nima.2015.12.034}
  {\bibfield  {journal} {\bibinfo  {journal} {Nucl.\ Instrum.\ Meth.\ A}\
  }\textbf {\bibinfo {volume} {812}},\ \bibinfo {pages} {1} (\bibinfo {year}
  {2016})}\BibitemShut {NoStop}%
\bibitem [{CUO()}]{CUORE_site}%
  \BibitemOpen
  \href {https://cuore.lngs.infn.it/en} {\enquote {\bibinfo {title}
  {{https://cuore.lngs.infn.it/en}},}\ }\BibitemShut {NoStop}%
\bibitem [{\citenamefont {Enss}\ and\ \citenamefont
  {Siegfried}(2005)}]{EnssSiegfried:2005}%
  \BibitemOpen
  \bibfield  {author} {\bibinfo {author} {\bibfnamefont {C.}~\bibnamefont
  {Enss}}\ and\ \bibinfo {author} {\bibfnamefont {H.}~\bibnamefont
  {Siegfried}},\ }\href {\doibase 10.1007/b137878} {\emph {\bibinfo {title}
  {{Low-Temperature Physics}}}}\ (\bibinfo  {publisher} {Springer},\ \bibinfo
  {year} {2005})\BibitemShut {NoStop}%
\bibitem [{\citenamefont {Stahl}(2005)}]{Stahl:2005}%
  \BibitemOpen
  \bibfield  {author} {\bibinfo {author} {\bibfnamefont {H.~C.}\ \bibnamefont
  {Stahl}},\ }\href {http://www.springer.com/la/book/9783540201137} {\emph
  {\bibinfo {title} {{Cryogenic Particle Detection}}}}\ (\bibinfo  {publisher}
  {Springer},\ \bibinfo {year} {2005})\BibitemShut {NoStop}%
\bibitem [{\citenamefont {Haller}\ \emph {et~al.}(1984)\citenamefont {Haller},
  \citenamefont {Palaio}, \citenamefont {Rodder}, \citenamefont {Hansen},\ and\
  \citenamefont {Kreysa}}]{Larrabee:1984}%
  \BibitemOpen
  \bibfield  {author} {\bibinfo {author} {\bibfnamefont {E.~E.}\ \bibnamefont
  {Haller}}, \bibinfo {author} {\bibfnamefont {N.~P.}\ \bibnamefont {Palaio}},
  \bibinfo {author} {\bibfnamefont {M.}~\bibnamefont {Rodder}}, \bibinfo
  {author} {\bibfnamefont {W.~L.}\ \bibnamefont {Hansen}}, \ and\ \bibinfo
  {author} {\bibfnamefont {E.}~\bibnamefont {Kreysa}},\ }in\ \href {\doibase
  10.1007/978-1-4613-2695-3_2} {\emph {\bibinfo {booktitle} {{Neutron
  Transmutation Doping of Semiconductor Materials}}}},\ \bibinfo {editor}
  {edited by\ \bibinfo {editor} {\bibfnamefont {R.~D.}\ \bibnamefont
  {Larrabee}}}\ (\bibinfo  {publisher} {Springer US},\ \bibinfo {year} {1984})\
  pp.\ \bibinfo {pages} {21--36}\BibitemShut {NoStop}%
\bibitem [{\citenamefont {Erginsoy}(1950)}]{Erginsoy:1950}%
  \BibitemOpen
  \bibfield  {author} {\bibinfo {author} {\bibfnamefont {C.}~\bibnamefont
  {Erginsoy}},\ }\href {\doibase 10.1103/PhysRev.80.1104} {\bibfield  {journal}
  {\bibinfo  {journal} {Phys.\ Rev.}\ }\textbf {\bibinfo {volume} {80}},\
  \bibinfo {pages} {1104} (\bibinfo {year} {1950})}\BibitemShut {NoStop}%
\bibitem [{\citenamefont {Mott}(1969)}]{Mott:1969nc}%
  \BibitemOpen
  \bibfield  {author} {\bibinfo {author} {\bibfnamefont {N.~F.}\ \bibnamefont
  {Mott}},\ }\href {\doibase 10.1080/14786436908216338} {\bibfield  {journal}
  {\bibinfo  {journal} {Philos.\ Mag.}\ }\textbf {\bibinfo {volume} {19}},\
  \bibinfo {pages} {835} (\bibinfo {year} {1969})}\BibitemShut {NoStop}%
\bibitem [{\citenamefont {Shklovskii}\ and\ \citenamefont
  {Efros}(1984)}]{Shklovskii&Efros:1984}%
  \BibitemOpen
  \bibfield  {author} {\bibinfo {author} {\bibfnamefont {B.~I.}\ \bibnamefont
  {Shklovskii}}\ and\ \bibinfo {author} {\bibfnamefont {A.~L.}\ \bibnamefont
  {Efros}},\ }\href {http://www.springer.com/la/book/9783662024058} {\emph
  {\bibinfo {title} {{Electronic Properties of Doped Semiconductors}}}}\
  (\bibinfo  {publisher} {Springer},\ \bibinfo {year} {1984})\BibitemShut
  {NoStop}%
\bibitem [{\citenamefont {Barucci}\ \emph {et~al.}(2001)\citenamefont
  {Barucci}, \citenamefont {Brofferio}, \citenamefont {Giuliani}, \citenamefont
  {Gottardi}, \citenamefont {Peroni},\ and\ \citenamefont
  {Ventura}}]{Baruccietal:2001}%
  \BibitemOpen
  \bibfield  {author} {\bibinfo {author} {\bibfnamefont {M.}~\bibnamefont
  {Barucci}}, \bibinfo {author} {\bibfnamefont {C.}~\bibnamefont {Brofferio}},
  \bibinfo {author} {\bibfnamefont {A.}~\bibnamefont {Giuliani}}, \bibinfo
  {author} {\bibfnamefont {E.}~\bibnamefont {Gottardi}}, \bibinfo {author}
  {\bibfnamefont {I.}~\bibnamefont {Peroni}}, \ and\ \bibinfo {author}
  {\bibfnamefont {G.}~\bibnamefont {Ventura}},\ }\href {\doibase
  10.1023/A:1017555615150} {\bibfield  {journal} {\bibinfo  {journal} {J.\ Low
  Temp.\ Phys.}\ }\textbf {\bibinfo {volume} {123}},\ \bibinfo {pages} {303}
  (\bibinfo {year} {2001})}\BibitemShut {NoStop}%
\bibitem [{\citenamefont {Mather}(1982)}]{Mather:1982}%
  \BibitemOpen
  \bibfield  {author} {\bibinfo {author} {\bibfnamefont {J.}~\bibnamefont
  {Mather}},\ }\href {\doibase 10.1364/AO.21.001125} {\bibfield  {journal}
  {\bibinfo  {journal} {J.\ Appl.\ Optics}\ }\textbf {\bibinfo {volume} {21}},\
  \bibinfo {pages} {1125} (\bibinfo {year} {1982})}\BibitemShut {NoStop}%
\bibitem [{\citenamefont {Mather}(1984)}]{Mather:1984}%
  \BibitemOpen
  \bibfield  {author} {\bibinfo {author} {\bibfnamefont {J.}~\bibnamefont
  {Mather}},\ }\href {\doibase 10.1364/AO.23.000584} {\bibfield  {journal}
  {\bibinfo  {journal} {J.\ Appl.\ Optics}\ }\textbf {\bibinfo {volume} {23}},\
  \bibinfo {pages} {584} (\bibinfo {year} {1984})}\BibitemShut {NoStop}%
\bibitem [{\citenamefont {Alessandrello}\ \emph
  {et~al.}(1998{\natexlab{a}})\citenamefont {Alessandrello} \emph
  {et~al.}}]{Alessandrello:1998bf}%
  \BibitemOpen
  \bibfield  {author} {\bibinfo {author} {\bibfnamefont {A.}~\bibnamefont
  {Alessandrello}} \emph {et~al.},\ }\href {\doibase
  10.1016/S0168-9002(98)00458-6} {\bibfield  {journal} {\bibinfo  {journal}
  {Nucl.\ Instrum.\ Meth.\ A}\ }\textbf {\bibinfo {volume} {412}},\ \bibinfo
  {pages} {454} (\bibinfo {year} {1998}{\natexlab{a}})}\BibitemShut {NoStop}%
\bibitem [{\citenamefont {Arnaboldi}\ \emph {et~al.}(2011)\citenamefont
  {Arnaboldi}, \citenamefont {Giachero}, \citenamefont {Gotti},\ and\
  \citenamefont {Pessina}}]{Arnaboldi:2011zza}%
  \BibitemOpen
  \bibfield  {author} {\bibinfo {author} {\bibfnamefont {C.}~\bibnamefont
  {Arnaboldi}}, \bibinfo {author} {\bibfnamefont {A.}~\bibnamefont {Giachero}},
  \bibinfo {author} {\bibfnamefont {C.}~\bibnamefont {Gotti}}, \ and\ \bibinfo
  {author} {\bibfnamefont {G.}~\bibnamefont {Pessina}},\ }\href {\doibase
  10.1016/j.nima.2010.09.151} {\bibfield  {journal} {\bibinfo  {journal}
  {Nucl.\ Instrum.\ Meth.\ A}\ }\textbf {\bibinfo {volume} {652}},\ \bibinfo
  {pages} {306} (\bibinfo {year} {2011})}\BibitemShut {NoStop}%
\bibitem [{\citenamefont {Andreotti}\ \emph {et~al.}(2012)\citenamefont
  {Andreotti}, \citenamefont {Brofferio}, \citenamefont {Foggetta},
  \citenamefont {Giuliani}, \citenamefont {Margesin}, \citenamefont {Nones},
  \citenamefont {Pedretti}, \citenamefont {Rusconi}, \citenamefont {Salvioni},\
  and\ \citenamefont {Tenconi}}]{Andreotti:2012zz}%
  \BibitemOpen
  \bibfield  {author} {\bibinfo {author} {\bibfnamefont {E.}~\bibnamefont
  {Andreotti}}, \bibinfo {author} {\bibfnamefont {C.}~\bibnamefont
  {Brofferio}}, \bibinfo {author} {\bibfnamefont {L.}~\bibnamefont {Foggetta}},
  \bibinfo {author} {\bibfnamefont {A.}~\bibnamefont {Giuliani}}, \bibinfo
  {author} {\bibfnamefont {B.}~\bibnamefont {Margesin}}, \bibinfo {author}
  {\bibfnamefont {C.}~\bibnamefont {Nones}}, \bibinfo {author} {\bibfnamefont
  {M.}~\bibnamefont {Pedretti}}, \bibinfo {author} {\bibfnamefont
  {C.}~\bibnamefont {Rusconi}}, \bibinfo {author} {\bibfnamefont
  {C.}~\bibnamefont {Salvioni}}, \ and\ \bibinfo {author} {\bibfnamefont
  {M.}~\bibnamefont {Tenconi}},\ }\href {\doibase 10.1016/j.nima.2011.10.065}
  {\bibfield  {journal} {\bibinfo  {journal} {Nucl.\ Instrum.\ Meth.\ A}\
  }\textbf {\bibinfo {volume} {664}},\ \bibinfo {pages} {161} (\bibinfo {year}
  {2012})}\BibitemShut {NoStop}%
\bibitem [{\citenamefont {Dutton}(1966)}]{Dutton:1966a}%
  \BibitemOpen
  \bibfield  {author} {\bibinfo {author} {\bibfnamefont {W.}~\bibnamefont
  {Dutton}, \bibfnamefont {W.~A. Charles~Cooper}},\ }\href {\doibase
  10.1021/cr60244a003} {\bibfield  {journal} {\bibinfo  {journal} {Chem.\
  Rev.}\ }\textbf {\bibinfo {volume} {66}},\ \bibinfo {pages} {657} (\bibinfo
  {year} {1966})}\BibitemShut {NoStop}%
\bibitem [{\citenamefont {White}\ \emph {et~al.}(1990)\citenamefont {White},
  \citenamefont {Collocott},\ and\ \citenamefont {Collins}}]{White:1990}%
  \BibitemOpen
  \bibfield  {author} {\bibinfo {author} {\bibfnamefont {G.~K.}\ \bibnamefont
  {White}}, \bibinfo {author} {\bibfnamefont {S.}~\bibnamefont {Collocott}}, \
  and\ \bibinfo {author} {\bibfnamefont {J.~G.}\ \bibnamefont {Collins}},\
  }\href {http://stacks.iop.org/0953-8984/2/i=37/a=015} {\bibfield  {journal}
  {\bibinfo  {journal} {J.\ Phys. Condensed Matter}\ }\textbf {\bibinfo
  {volume} {2}},\ \bibinfo {pages} {7715} (\bibinfo {year} {1990})}\BibitemShut
  {NoStop}%
\bibitem [{\citenamefont {Kroeger}\ and\ \citenamefont
  {Swenson}(1977)}]{Kroeger:1977}%
  \BibitemOpen
  \bibfield  {author} {\bibinfo {author} {\bibfnamefont {F.~R.}\ \bibnamefont
  {Kroeger}}\ and\ \bibinfo {author} {\bibfnamefont {C.~A.}\ \bibnamefont
  {Swenson}},\ }\href {http://dx.doi.org/10.1063/1.323746} {\bibfield
  {journal} {\bibinfo  {journal} {J.\ Appl.\ Phys.}\ }\textbf {\bibinfo
  {volume} {48}},\ \bibinfo {pages} {853} (\bibinfo {year} {1977})}\BibitemShut
  {NoStop}%
\bibitem [{\citenamefont {El-Mallawany}(2002)}]{El-Mallawany:2002}%
  \BibitemOpen
  \bibfield  {author} {\bibinfo {author} {\bibfnamefont {R.~A.~H.}\
  \bibnamefont {El-Mallawany}},\ }\href
  {https://www.crcpress.com/Tellurite-Glasses-Handbook-Physical-Properties%
  -and-Data-Second-Edition/ElMallawany/9781439849835} {\emph {\bibinfo {title}
  {{Tellurite Glasses Handbook: Physical Properties and Data}}}}\ (\bibinfo
  {publisher} {CRC Press},\ \bibinfo {year} {2002})\BibitemShut {NoStop}%
\bibitem [{\citenamefont {Chu}\ \emph {et~al.}(2006)\citenamefont {Chu},
  \citenamefont {Li}, \citenamefont {Ge}, \citenamefont {Wu},\ and\
  \citenamefont {Wang}}]{Chu2006158}%
  \BibitemOpen
  \bibfield  {author} {\bibinfo {author} {\bibfnamefont {Y.}~\bibnamefont
  {Chu}}, \bibinfo {author} {\bibfnamefont {Y.}~\bibnamefont {Li}}, \bibinfo
  {author} {\bibfnamefont {Z.}~\bibnamefont {Ge}}, \bibinfo {author}
  {\bibfnamefont {G.}~\bibnamefont {Wu}}, \ and\ \bibinfo {author}
  {\bibfnamefont {H.}~\bibnamefont {Wang}},\ }\href {\doibase
  10.1016/j.jcrysgro.2006.08.009} {\bibfield  {journal} {\bibinfo  {journal}
  {J.\ Cryst.\ Growth}\ }\textbf {\bibinfo {volume} {295}},\ \bibinfo {pages}
  {158} (\bibinfo {year} {2006})}\BibitemShut {NoStop}%
\bibitem [{\citenamefont {Arnaboldi}\ \emph {et~al.}(2010)\citenamefont
  {Arnaboldi} \emph {et~al.}}]{Arnaboldi:2010fj}%
  \BibitemOpen
  \bibfield  {author} {\bibinfo {author} {\bibfnamefont {C.}~\bibnamefont
  {Arnaboldi}} \emph {et~al.},\ }\href {\doibase
  10.1016/j.jcrysgro.2010.06.034} {\bibfield  {journal} {\bibinfo  {journal}
  {J.\ Cryst.\ Growth}\ }\textbf {\bibinfo {volume} {312}},\ \bibinfo {pages}
  {2999} (\bibinfo {year} {2010})}\BibitemShut {NoStop}%
\bibitem [{\citenamefont {Arnaboldi}\ \emph
  {et~al.}(2005{\natexlab{a}})\citenamefont {Arnaboldi}, \citenamefont
  {Brofferio}, \citenamefont {Bucci}, \citenamefont {Gorla}, \citenamefont
  {Pessina},\ and\ \citenamefont {Pirro}}]{Arnaboldi:2005yf}%
  \BibitemOpen
  \bibfield  {author} {\bibinfo {author} {\bibfnamefont {C.}~\bibnamefont
  {Arnaboldi}}, \bibinfo {author} {\bibfnamefont {C.}~\bibnamefont
  {Brofferio}}, \bibinfo {author} {\bibfnamefont {C.}~\bibnamefont {Bucci}},
  \bibinfo {author} {\bibfnamefont {P.}~\bibnamefont {Gorla}}, \bibinfo
  {author} {\bibfnamefont {G.}~\bibnamefont {Pessina}}, \ and\ \bibinfo
  {author} {\bibfnamefont {S.}~\bibnamefont {Pirro}},\ }\href {\doibase
  10.1016/j.nima.2005.07.060} {\bibfield  {journal} {\bibinfo  {journal}
  {Nucl.\ Instrum.\ Meth.\ A}\ }\textbf {\bibinfo {volume} {554}},\ \bibinfo
  {pages} {300} (\bibinfo {year} {2005}{\natexlab{a}})}\BibitemShut {NoStop}%
\bibitem [{\citenamefont {Cardani}\ \emph {et~al.}(2012)\citenamefont
  {Cardani}, \citenamefont {Gironi}, \citenamefont {Beeman}, \citenamefont
  {Dafinei}, \citenamefont {Ge}, \citenamefont {Pessina}, \citenamefont
  {Pirro},\ and\ \citenamefont {Zhu}}]{Cardani:2011vg}%
  \BibitemOpen
  \bibfield  {author} {\bibinfo {author} {\bibfnamefont {L.}~\bibnamefont
  {Cardani}}, \bibinfo {author} {\bibfnamefont {L.}~\bibnamefont {Gironi}},
  \bibinfo {author} {\bibfnamefont {J.~W.}\ \bibnamefont {Beeman}}, \bibinfo
  {author} {\bibfnamefont {I.}~\bibnamefont {Dafinei}}, \bibinfo {author}
  {\bibfnamefont {Z.}~\bibnamefont {Ge}}, \bibinfo {author} {\bibfnamefont
  {G.}~\bibnamefont {Pessina}}, \bibinfo {author} {\bibfnamefont
  {S.}~\bibnamefont {Pirro}}, \ and\ \bibinfo {author} {\bibfnamefont
  {Y.}~\bibnamefont {Zhu}},\ }\href {\doibase 10.1088/1748-0221/7/01/P01020}
  {\bibfield  {journal} {\bibinfo  {journal} {J.\ Instrum.}\ }\textbf {\bibinfo
  {volume} {7}},\ \bibinfo {pages} {P01020} (\bibinfo {year}
  {2012})}\BibitemShut {NoStop}%
\bibitem [{\citenamefont {Alessandria}\ \emph {et~al.}(2012)\citenamefont
  {Alessandria} \emph {et~al.}}]{Alessandria:2011vj}%
  \BibitemOpen
  \bibfield  {author} {\bibinfo {author} {\bibfnamefont {F.}~\bibnamefont
  {Alessandria}} \emph {et~al.},\ }\href {\doibase
  10.1016/j.astropartphys.2012.02.008} {\bibfield  {journal} {\bibinfo
  {journal} {Astropart.\ Phys.}\ }\textbf {\bibinfo {volume} {35}},\ \bibinfo
  {pages} {839} (\bibinfo {year} {2012})}\BibitemShut {NoStop}%
\bibitem [{\citenamefont {Alessandria}\ \emph
  {et~al.}(2013{\natexlab{a}})\citenamefont {Alessandria} \emph
  {et~al.}}]{Alessandria:2012zp}%
  \BibitemOpen
  \bibfield  {author} {\bibinfo {author} {\bibfnamefont {F.}~\bibnamefont
  {Alessandria}} \emph {et~al.},\ }\href {\doibase
  10.1016/j.astropartphys.2013.02.005} {\bibfield  {journal} {\bibinfo
  {journal} {Astropart.\ Phys.}\ }\textbf {\bibinfo {volume} {45}},\ \bibinfo
  {pages} {13} (\bibinfo {year} {2013}{\natexlab{a}})}\BibitemShut {NoStop}%
\bibitem [{\citenamefont {Alessandrello}\ \emph
  {et~al.}(1992{\natexlab{a}})\citenamefont {Alessandrello} \emph
  {et~al.}}]{Alessandrello:1992jc}%
  \BibitemOpen
  \bibfield  {author} {\bibinfo {author} {\bibfnamefont {A.}~\bibnamefont
  {Alessandrello}} \emph {et~al.},\ }\href {\doibase
  10.1016/0920-5632(92)90177-T} {\bibfield  {journal} {\bibinfo  {journal}
  {Nucl.\ Phys.\ B (Proc.\ Suppl.)}\ }\textbf {\bibinfo {volume} {28A}},\
  \bibinfo {pages} {229} (\bibinfo {year} {1992}{\natexlab{a}})}\BibitemShut
  {NoStop}%
\bibitem [{\citenamefont {Alessandrello}\ \emph {et~al.}(1988)\citenamefont
  {Alessandrello}, \citenamefont {Camin}, \citenamefont {Fiorini},
  \citenamefont {Giuliani},\ and\ \citenamefont
  {Pessina}}]{Alessandrello:1988dv}%
  \BibitemOpen
  \bibfield  {author} {\bibinfo {author} {\bibfnamefont {A.}~\bibnamefont
  {Alessandrello}}, \bibinfo {author} {\bibfnamefont {D.~V.}\ \bibnamefont
  {Camin}}, \bibinfo {author} {\bibfnamefont {E.}~\bibnamefont {Fiorini}},
  \bibinfo {author} {\bibfnamefont {A.}~\bibnamefont {Giuliani}}, \ and\
  \bibinfo {author} {\bibfnamefont {G.}~\bibnamefont {Pessina}},\ }\href
  {\doibase 10.1016/0168-9002(88)91108-4} {\bibfield  {journal} {\bibinfo
  {journal} {Nucl.\ Instrum.\ Meth.\ A}\ }\textbf {\bibinfo {volume} {264}},\
  \bibinfo {pages} {93} (\bibinfo {year} {1988})}\BibitemShut {NoStop}%
\bibitem [{\citenamefont {Alessandrello}\ \emph
  {et~al.}(1990{\natexlab{a}})\citenamefont {Alessandrello}, \citenamefont
  {Brofferio}, \citenamefont {Camin}, \citenamefont {Cremonesi}, \citenamefont
  {Fiorini}, \citenamefont {Giuliani},\ and\ \citenamefont
  {Pessina}}]{Alessandrello:1990wd}%
  \BibitemOpen
  \bibfield  {author} {\bibinfo {author} {\bibfnamefont {A.}~\bibnamefont
  {Alessandrello}}, \bibinfo {author} {\bibfnamefont {C.}~\bibnamefont
  {Brofferio}}, \bibinfo {author} {\bibfnamefont {D.~V.}\ \bibnamefont
  {Camin}}, \bibinfo {author} {\bibfnamefont {O.}~\bibnamefont {Cremonesi}},
  \bibinfo {author} {\bibfnamefont {E.}~\bibnamefont {Fiorini}}, \bibinfo
  {author} {\bibfnamefont {A.}~\bibnamefont {Giuliani}}, \ and\ \bibinfo
  {author} {\bibfnamefont {G.}~\bibnamefont {Pessina}},\ }\href {\doibase
  10.1016/0370-2693(90)90923-T} {\bibfield  {journal} {\bibinfo  {journal}
  {Phys.\ Lett.\ B}\ }\textbf {\bibinfo {volume} {247}},\ \bibinfo {pages}
  {442} (\bibinfo {year} {1990}{\natexlab{a}})}\BibitemShut {NoStop}%
\bibitem [{\citenamefont {Alessandrello}\ \emph
  {et~al.}(1990{\natexlab{b}})\citenamefont {Alessandrello}, \citenamefont
  {Brofferio}, \citenamefont {Camin}, \citenamefont {Cremonesi}, \citenamefont
  {Fiorini}, \citenamefont {Giuliani},\ and\ \citenamefont
  {Pessina}}]{Alessandrello:1990te}%
  \BibitemOpen
  \bibfield  {author} {\bibinfo {author} {\bibfnamefont {A.}~\bibnamefont
  {Alessandrello}}, \bibinfo {author} {\bibfnamefont {C.}~\bibnamefont
  {Brofferio}}, \bibinfo {author} {\bibfnamefont {D.~V.}\ \bibnamefont
  {Camin}}, \bibinfo {author} {\bibfnamefont {O.}~\bibnamefont {Cremonesi}},
  \bibinfo {author} {\bibfnamefont {E.}~\bibnamefont {Fiorini}}, \bibinfo
  {author} {\bibfnamefont {A.}~\bibnamefont {Giuliani}}, \ and\ \bibinfo
  {author} {\bibfnamefont {G.}~\bibnamefont {Pessina}},\ }\href {\doibase
  10.1016/S0921-4526(90)80850-I} {\bibfield  {journal} {\bibinfo  {journal}
  {Physica B}\ }\textbf {\bibinfo {volume} {165-166}},\ \bibinfo {pages} {1}
  (\bibinfo {year} {1990}{\natexlab{b}})}\BibitemShut {NoStop}%
\bibitem [{\citenamefont {Alessandrello}\ \emph
  {et~al.}(1992{\natexlab{b}})\citenamefont {Alessandrello} \emph
  {et~al.}}]{Alessandrello:1992aa}%
  \BibitemOpen
  \bibfield  {author} {\bibinfo {author} {\bibfnamefont {A.}~\bibnamefont
  {Alessandrello}} \emph {et~al.},\ }\href {\doibase
  10.1016/0168-9002(92)90713-E} {\bibfield  {journal} {\bibinfo  {journal}
  {Nucl.\ Instrum.\ Meth.\ A}\ }\textbf {\bibinfo {volume} {315}},\ \bibinfo
  {pages} {263} (\bibinfo {year} {1992}{\natexlab{b}})}\BibitemShut {NoStop}%
\bibitem [{\citenamefont {Giuliani}\ \emph {et~al.}(1991)\citenamefont
  {Giuliani} \emph {et~al.}}]{Giuliani:1991ze}%
  \BibitemOpen
  \bibfield  {author} {\bibinfo {author} {\bibfnamefont {A.}~\bibnamefont
  {Giuliani}} \emph {et~al.},\ }\href@noop {} {\bibfield  {journal} {\bibinfo
  {journal} {Proc.\ Joint Int.\ Lepton Photon Symposium at High Energies (15th)
  and European Phys.\ Soc.\ Conf.\ on High-energy Phys.}\ }\textbf {\bibinfo
  {volume} {\!\!}},\ \bibinfo {pages} {670} (\bibinfo {year}
  {1991})}\BibitemShut {NoStop}%
\bibitem [{\citenamefont {Alessandrello}\ \emph {et~al.}(1989)\citenamefont
  {Alessandrello}, \citenamefont {Brofferio}, \citenamefont {Camin},
  \citenamefont {Cremonesi}, \citenamefont {Fiorini}, \citenamefont {Giuliani},
  \citenamefont {Pessina},\ and\ \citenamefont
  {Previtali}}]{Alessandrello:1989bu}%
  \BibitemOpen
  \bibfield  {author} {\bibinfo {author} {\bibfnamefont {A.}~\bibnamefont
  {Alessandrello}}, \bibinfo {author} {\bibfnamefont {C.}~\bibnamefont
  {Brofferio}}, \bibinfo {author} {\bibfnamefont {D.~V.}\ \bibnamefont
  {Camin}}, \bibinfo {author} {\bibfnamefont {O.}~\bibnamefont {Cremonesi}},
  \bibinfo {author} {\bibfnamefont {E.}~\bibnamefont {Fiorini}}, \bibinfo
  {author} {\bibfnamefont {A.}~\bibnamefont {Giuliani}}, \bibinfo {author}
  {\bibfnamefont {G.}~\bibnamefont {Pessina}}, \ and\ \bibinfo {author}
  {\bibfnamefont {E.}~\bibnamefont {Previtali}},\ }\href@noop {} {\bibfield
  {journal} {\bibinfo  {journal} {Proc.\ 3rd Int.\ Workshop on Low Temp.\ Det.\
  for Neutrinos and Dark Matter}\ }\textbf {\bibinfo {volume} {\!\!}},\
  \bibinfo {pages} {253} (\bibinfo {year} {1989})}\BibitemShut {NoStop}%
\bibitem [{\citenamefont {Alessandrello}\ \emph {et~al.}(1991)\citenamefont
  {Alessandrello} \emph {et~al.}}]{Alessandrello:1991fig}%
  \BibitemOpen
  \bibfield  {author} {\bibinfo {author} {\bibfnamefont {A.}~\bibnamefont
  {Alessandrello}} \emph {et~al.},\ }\href@noop {} {\bibfield  {journal}
  {\bibinfo  {journal} {Proc.\ 4th Int.\ Workshop on Low Temp.\ Det.\ for
  Neutrinos and Dark Matter}\ }\textbf {\bibinfo {volume} {\!\!}},\ \bibinfo
  {pages} {99} (\bibinfo {year} {1991})}\BibitemShut {NoStop}%
\bibitem [{\citenamefont {Alessandrello}\ \emph
  {et~al.}(1998{\natexlab{b}})\citenamefont {Alessandrello} \emph
  {et~al.}}]{roman_lead:1998}%
  \BibitemOpen
  \bibfield  {author} {\bibinfo {author} {\bibfnamefont {A.}~\bibnamefont
  {Alessandrello}} \emph {et~al.},\ }\href {\doibase
  10.1016/S0168-583X(98)00279-1} {\bibfield  {journal} {\bibinfo  {journal}
  {Nucl.\ Instrum.\ Meth.\ B}\ }\textbf {\bibinfo {volume} {142}},\ \bibinfo
  {pages} {163} (\bibinfo {year} {1998}{\natexlab{b}})}\BibitemShut {NoStop}%
\bibitem [{\citenamefont {Alessandrello}\ \emph {et~al.}(1993)\citenamefont
  {Alessandrello} \emph {et~al.}}]{Alessandrello:1993cs}%
  \BibitemOpen
  \bibfield  {author} {\bibinfo {author} {\bibfnamefont {A.}~\bibnamefont
  {Alessandrello}} \emph {et~al.},\ }\href {\doibase
  10.1016/0920-5632(93)90118-P} {\bibfield  {journal} {\bibinfo  {journal}
  {Nucl.\ Phys.\ B (Proc.\ Suppl.)}\ }\textbf {\bibinfo {volume} {31}},\
  \bibinfo {pages} {83} (\bibinfo {year} {1993})}\BibitemShut {NoStop}%
\bibitem [{\citenamefont {Alessandrello}\ \emph
  {et~al.}(1992{\natexlab{c}})\citenamefont {Alessandrello} \emph
  {et~al.}}]{Alessandrello:1992fig}%
  \BibitemOpen
  \bibfield  {author} {\bibinfo {author} {\bibfnamefont {A.}~\bibnamefont
  {Alessandrello}} \emph {et~al.},\ }\href {\doibase 10.1117/12.138587}
  {\bibfield  {journal} {\bibinfo  {journal} {Proc.\ SPIE, Gamma-Ray
  Detectors}\ }\textbf {\bibinfo {volume} {1734}},\ \bibinfo {pages} {177}
  (\bibinfo {year} {1992}{\natexlab{c}})}\BibitemShut {NoStop}%
\bibitem [{\citenamefont {Alessandrello}\ \emph {et~al.}(1994)\citenamefont
  {Alessandrello} \emph {et~al.}}]{Alessandrello:1994tm}%
  \BibitemOpen
  \bibfield  {author} {\bibinfo {author} {\bibfnamefont {A.}~\bibnamefont
  {Alessandrello}} \emph {et~al.},\ }\href {\doibase
  10.1016/0370-2693(94)90388-3} {\bibfield  {journal} {\bibinfo  {journal}
  {Phys.\ Lett.\ B}\ }\textbf {\bibinfo {volume} {335}},\ \bibinfo {pages}
  {519} (\bibinfo {year} {1994})}\BibitemShut {NoStop}%
\bibitem [{\citenamefont {Giuliani}\ \emph {et~al.}(1994)\citenamefont
  {Giuliani} \emph {et~al.}}]{Alessandrello:1994_XX}%
  \BibitemOpen
  \bibfield  {author} {\bibinfo {author} {\bibfnamefont {A.}~\bibnamefont
  {Giuliani}} \emph {et~al.},\ }\href@noop {} {\bibfield  {journal} {\bibinfo
  {journal} {Proc.\ 27th Int.\ Conf.\ on High-energy Phys.}\ }\textbf {\bibinfo
  {volume} {\!\!}},\ \bibinfo {pages} {943} (\bibinfo {year}
  {1994})}\BibitemShut {NoStop}%
\bibitem [{\citenamefont {Alessandrello}\ \emph
  {et~al.}(1996{\natexlab{a}})\citenamefont {Alessandrello} \emph
  {et~al.}}]{Brofferio:1995wx}%
  \BibitemOpen
  \bibfield  {author} {\bibinfo {author} {\bibfnamefont {A.}~\bibnamefont
  {Alessandrello}} \emph {et~al.},\ }\href {\doibase
  10.1016/0920-5632(96)00249-6} {\bibfield  {journal} {\bibinfo  {journal}
  {Nucl.\ Phys.\ B (Proc.\ Suppl.)}\ }\textbf {\bibinfo {volume} {48}},\
  \bibinfo {pages} {238} (\bibinfo {year} {1996}{\natexlab{a}})}\BibitemShut
  {NoStop}%
\bibitem [{\citenamefont {Alessandrello}\ \emph {et~al.}(1995)\citenamefont
  {Alessandrello} \emph {et~al.}}]{Alessandrello:1995hq}%
  \BibitemOpen
  \bibfield  {author} {\bibinfo {author} {\bibfnamefont {A.}~\bibnamefont
  {Alessandrello}} \emph {et~al.},\ }\href {\doibase
  10.1016/0168-9002(94)01723-9} {\bibfield  {journal} {\bibinfo  {journal}
  {Nucl.\ Instrum.\ Meth.\ A}\ }\textbf {\bibinfo {volume} {360}},\ \bibinfo
  {pages} {363} (\bibinfo {year} {1995})}\BibitemShut {NoStop}%
\bibitem [{\citenamefont {Pirro}\ \emph {et~al.}(2000)\citenamefont {Pirro}
  \emph {et~al.}}]{Pirro:2000vib}%
  \BibitemOpen
  \bibfield  {author} {\bibinfo {author} {\bibfnamefont {S.}~\bibnamefont
  {Pirro}} \emph {et~al.},\ }\href {\doibase 10.1016/S0168-9002(99)01376-5}
  {\bibfield  {journal} {\bibinfo  {journal} {Nucl.\ Instrum.\ Meth.\ A}\
  }\textbf {\bibinfo {volume} {444}},\ \bibinfo {pages} {331} (\bibinfo {year}
  {2000})}\BibitemShut {NoStop}%
\bibitem [{\citenamefont {Pirro}(2006)}]{Pirro:2006mu}%
  \BibitemOpen
  \bibfield  {author} {\bibinfo {author} {\bibfnamefont {S.}~\bibnamefont
  {Pirro}},\ }\href {\doibase 10.1016/j.nima.2005.12.197} {\bibfield  {journal}
  {\bibinfo  {journal} {Nucl.\ Instrum.\ Meth.\ A}\ }\textbf {\bibinfo {volume}
  {559}},\ \bibinfo {pages} {672} (\bibinfo {year} {2006})}\BibitemShut
  {NoStop}%
\bibitem [{\citenamefont {Alessandrello}\ \emph
  {et~al.}(1996{\natexlab{b}})\citenamefont {Alessandrello} \emph
  {et~al.}}]{Alessandrello:1996dd}%
  \BibitemOpen
  \bibfield  {author} {\bibinfo {author} {\bibfnamefont {A.}~\bibnamefont
  {Alessandrello}} \emph {et~al.},\ }\href {\doibase
  10.1016/0168-9002(95)01125-0} {\bibfield  {journal} {\bibinfo  {journal}
  {Nucl.\ Instrum.\ Meth.\ A}\ }\textbf {\bibinfo {volume} {370}},\ \bibinfo
  {pages} {269} (\bibinfo {year} {1996}{\natexlab{b}})}\BibitemShut {NoStop}%
\bibitem [{\citenamefont {Arnaboldi}\ \emph {et~al.}(2003)\citenamefont
  {Arnaboldi} \emph {et~al.}}]{Arnaboldi:2002te}%
  \BibitemOpen
  \bibfield  {author} {\bibinfo {author} {\bibfnamefont {C.}~\bibnamefont
  {Arnaboldi}} \emph {et~al.},\ }\href {\doibase 10.1016/S0370-2693(03)00212-0}
  {\bibfield  {journal} {\bibinfo  {journal} {Phys.\ Lett.\ B}\ }\textbf
  {\bibinfo {volume} {557}},\ \bibinfo {pages} {167} (\bibinfo {year}
  {2003})}\BibitemShut {NoStop}%
\bibitem [{\citenamefont {Alessandrello}\ \emph
  {et~al.}(2000{\natexlab{a}})\citenamefont {Alessandrello} \emph
  {et~al.}}]{Alessandrello:2000kt}%
  \BibitemOpen
  \bibfield  {author} {\bibinfo {author} {\bibfnamefont {A.}~\bibnamefont
  {Alessandrello}} \emph {et~al.},\ }\href {\doibase
  10.1016/S0370-2693(00)00747-4} {\bibfield  {journal} {\bibinfo  {journal}
  {Phys.\ Lett.\ B}\ }\textbf {\bibinfo {volume} {486}},\ \bibinfo {pages} {13}
  (\bibinfo {year} {2000}{\natexlab{a}})}\BibitemShut {NoStop}%
\bibitem [{\citenamefont {Alessandrello}\ \emph
  {et~al.}(2000{\natexlab{b}})\citenamefont {Alessandrello} \emph
  {et~al.}}]{Alessandrello:2000fs}%
  \BibitemOpen
  \bibfield  {author} {\bibinfo {author} {\bibfnamefont {A.}~\bibnamefont
  {Alessandrello}} \emph {et~al.},\ }\href {\doibase
  10.1016/S0168-9002(99)01340-6} {\bibfield  {journal} {\bibinfo  {journal}
  {Nucl.\ Instrum.\ Meth.\ A}\ }\textbf {\bibinfo {volume} {444}},\ \bibinfo
  {pages} {111} (\bibinfo {year} {2000}{\natexlab{b}})}\BibitemShut {NoStop}%
\bibitem [{\citenamefont {Pessina}(2000)}]{Pessina:2000fu}%
  \BibitemOpen
  \bibfield  {author} {\bibinfo {author} {\bibfnamefont {G.}~\bibnamefont
  {Pessina}},\ }\href {\doibase 10.1016/S0168-9002(99)01345-5} {\bibfield
  {journal} {\bibinfo  {journal} {Nucl.\ Instrum.\ Meth.\ A}\ }\textbf
  {\bibinfo {volume} {444}},\ \bibinfo {pages} {132} (\bibinfo {year}
  {2000})}\BibitemShut {NoStop}%
\bibitem [{\citenamefont {Alessandrello}\ \emph
  {et~al.}(1998{\natexlab{c}})\citenamefont {Alessandrello} \emph
  {et~al.}}]{Alessandrello:1998ey}%
  \BibitemOpen
  \bibfield  {author} {\bibinfo {author} {\bibfnamefont {A.}~\bibnamefont
  {Alessandrello}} \emph {et~al.},\ }\href {\doibase
  10.1016/S0370-2693(98)00645-5} {\bibfield  {journal} {\bibinfo  {journal}
  {Phys.\ Lett.\ B}\ }\textbf {\bibinfo {volume} {433}},\ \bibinfo {pages}
  {156} (\bibinfo {year} {1998}{\natexlab{c}})}\BibitemShut {NoStop}%
\bibitem [{\citenamefont {Capelli}(2005)}]{Capelli_PhD-thesis:2005}%
  \BibitemOpen
  \bibfield  {author} {\bibinfo {author} {\bibfnamefont {S.}~\bibnamefont
  {Capelli}},\ }\href {https://cuore.lngs.infn.it/en/publications/theses} {\
  (\bibinfo {year} {2005})},\ \bibinfo {note} {[Ph.\,D.\ thesis, Universit\`a
  di Milano]}\BibitemShut {NoStop}%
\bibitem [{\citenamefont {Alduino}\ \emph
  {et~al.}(2016{\natexlab{a}})\citenamefont {Alduino} \emph
  {et~al.}}]{Alduino:2016vjd}%
  \BibitemOpen
  \bibfield  {author} {\bibinfo {author} {\bibfnamefont {C.}~\bibnamefont
  {Alduino}} \emph {et~al.} (\bibinfo {collaboration} {CUORE Collaboration}),\
  }\href {\doibase 10.1088/1748-0221/11/07/P07009} {\bibfield  {journal}
  {\bibinfo  {journal} {J.\ Instrum.}\ }\textbf {\bibinfo {volume} {11}},\
  \bibinfo {pages} {P07009} (\bibinfo {year} {2016}{\natexlab{a}})}\BibitemShut
  {NoStop}%
\bibitem [{\citenamefont {Fiorini}(1998)}]{Fiorini:1998gj}%
  \BibitemOpen
  \bibfield  {author} {\bibinfo {author} {\bibfnamefont {E.}~\bibnamefont
  {Fiorini}},\ }\href {\doibase 10.1016/S0370-1573(98)00060-X} {\bibfield
  {journal} {\bibinfo  {journal} {Phys.\ Rep.}\ }\textbf {\bibinfo {volume}
  {307}},\ \bibinfo {pages} {309} (\bibinfo {year} {1998})}\BibitemShut
  {NoStop}%
\bibitem [{\citenamefont {Alessandrello}\ \emph
  {et~al.}(2000{\natexlab{c}})\citenamefont {Alessandrello} \emph
  {et~al.}}]{Alessandrello:2000ywa}%
  \BibitemOpen
  \bibfield  {author} {\bibinfo {author} {\bibfnamefont {A.}~\bibnamefont
  {Alessandrello}} \emph {et~al.},\ }\href {\doibase
  10.1016/S0168-9002(99)00928-6} {\bibfield  {journal} {\bibinfo  {journal}
  {Nucl.\ Instrum.\ Meth.\ A}\ }\textbf {\bibinfo {volume} {440}},\ \bibinfo
  {pages} {397} (\bibinfo {year} {2000}{\natexlab{c}})}\BibitemShut {NoStop}%
\bibitem [{\citenamefont {Arnaboldi}\ \emph
  {et~al.}(2005{\natexlab{b}})\citenamefont {Arnaboldi}, \citenamefont {Bucci},
  \citenamefont {Capelli}, \citenamefont {Gorla}, \citenamefont {Guardincerri},
  \citenamefont {Nucciotti}, \citenamefont {Pessina}, \citenamefont {Pirro},\
  and\ \citenamefont {Sisti}}]{Arnaboldi:2005xu}%
  \BibitemOpen
  \bibfield  {author} {\bibinfo {author} {\bibfnamefont {C.}~\bibnamefont
  {Arnaboldi}}, \bibinfo {author} {\bibfnamefont {C.}~\bibnamefont {Bucci}},
  \bibinfo {author} {\bibfnamefont {S.}~\bibnamefont {Capelli}}, \bibinfo
  {author} {\bibfnamefont {P.}~\bibnamefont {Gorla}}, \bibinfo {author}
  {\bibfnamefont {E.}~\bibnamefont {Guardincerri}}, \bibinfo {author}
  {\bibfnamefont {A.}~\bibnamefont {Nucciotti}}, \bibinfo {author}
  {\bibfnamefont {G.}~\bibnamefont {Pessina}}, \bibinfo {author} {\bibfnamefont
  {S.}~\bibnamefont {Pirro}}, \ and\ \bibinfo {author} {\bibfnamefont
  {M.}~\bibnamefont {Sisti}},\ }\href {\doibase 10.1109/TNS.2005.856758}
  {\bibfield  {journal} {\bibinfo  {journal} {IEEE Trans.\ Nucl.\ Sci.}\
  }\textbf {\bibinfo {volume} {52}},\ \bibinfo {pages} {1630} (\bibinfo {year}
  {2005}{\natexlab{b}})}\BibitemShut {NoStop}%
\bibitem [{\citenamefont {Arnaboldi}\ \emph {et~al.}(2004)\citenamefont
  {Arnaboldi} \emph {et~al.}}]{Arnaboldi:2004jj}%
  \BibitemOpen
  \bibfield  {author} {\bibinfo {author} {\bibfnamefont {C.}~\bibnamefont
  {Arnaboldi}} \emph {et~al.},\ }\href {\doibase 10.1016/j.nima.2003.11.319}
  {\bibfield  {journal} {\bibinfo  {journal} {Nucl.\ Instrum.\ Meth.\ A}\
  }\textbf {\bibinfo {volume} {520}},\ \bibinfo {pages} {578} (\bibinfo {year}
  {2004})}\BibitemShut {NoStop}%
\bibitem [{\citenamefont {Arnaboldi}\ \emph {et~al.}(2008)\citenamefont
  {Arnaboldi} \emph {et~al.}}]{Arnaboldi:2008ds}%
  \BibitemOpen
  \bibfield  {author} {\bibinfo {author} {\bibfnamefont {C.}~\bibnamefont
  {Arnaboldi}} \emph {et~al.} (\bibinfo {collaboration} {Cuoricino
  Collaboration}),\ }\href {\doibase 10.1103/PhysRevC.78.035502} {\bibfield
  {journal} {\bibinfo  {journal} {Phys.\ Rev.\ C}\ }\textbf {\bibinfo {volume}
  {78}},\ \bibinfo {pages} {035502} (\bibinfo {year} {2008})}\BibitemShut
  {NoStop}%
\bibitem [{\citenamefont {Andreotti}\ \emph {et~al.}(2011)\citenamefont
  {Andreotti} \emph {et~al.}}]{Andreotti:2010vj}%
  \BibitemOpen
  \bibfield  {author} {\bibinfo {author} {\bibfnamefont {E.}~\bibnamefont
  {Andreotti}} \emph {et~al.},\ }\href {\doibase
  10.1016/j.astropartphys.2011.02.002} {\bibfield  {journal} {\bibinfo
  {journal} {Astropart.\ Phys.}\ }\textbf {\bibinfo {volume} {34}},\ \bibinfo
  {pages} {822} (\bibinfo {year} {2011})}\BibitemShut {NoStop}%
\bibitem [{\citenamefont {Gorla}(2006)}]{Gorla_PhD-thesis:2006}%
  \BibitemOpen
  \bibfield  {author} {\bibinfo {author} {\bibfnamefont {P.}~\bibnamefont
  {Gorla}},\ }\href {https://cuore.lngs.infn.it/en/publications/theses} {\
  (\bibinfo {year} {2006})},\ \bibinfo {note} {[Ph.\,D.\ thesis, Universit\`a
  di Milano-Bicocca]}\BibitemShut {NoStop}%
\bibitem [{\citenamefont {Giachero}(2008)}]{Giachero_PhD-thesis:2008}%
  \BibitemOpen
  \bibfield  {author} {\bibinfo {author} {\bibfnamefont {A.}~\bibnamefont
  {Giachero}},\ }\href {http://giachero.mib.infn.it/documents/} {\  (\bibinfo
  {year} {2008})},\ \bibinfo {note} {[Ph.\,D.\ thesis, Universit\`a di
  Genova]}\BibitemShut {NoStop}%
\bibitem [{\citenamefont {Foggetta}\ \emph {et~al.}(2011)\citenamefont
  {Foggetta}, \citenamefont {Giuliani}, \citenamefont {Nones}, \citenamefont
  {Pedretti}, \citenamefont {Salvioni},\ and\ \citenamefont
  {Sangiorgio}}]{Foggetta:2011nk}%
  \BibitemOpen
  \bibfield  {author} {\bibinfo {author} {\bibfnamefont {L.}~\bibnamefont
  {Foggetta}}, \bibinfo {author} {\bibfnamefont {A.}~\bibnamefont {Giuliani}},
  \bibinfo {author} {\bibfnamefont {C.}~\bibnamefont {Nones}}, \bibinfo
  {author} {\bibfnamefont {M.}~\bibnamefont {Pedretti}}, \bibinfo {author}
  {\bibfnamefont {C.}~\bibnamefont {Salvioni}}, \ and\ \bibinfo {author}
  {\bibfnamefont {S.}~\bibnamefont {Sangiorgio}},\ }\href {\doibase
  10.1016/j.astropartphys.2011.02.004} {\bibfield  {journal} {\bibinfo
  {journal} {Astropart.\ Phys.}\ }\textbf {\bibinfo {volume} {34}},\ \bibinfo
  {pages} {809} (\bibinfo {year} {2011})}\BibitemShut {NoStop}%
\bibitem [{\citenamefont {Sangiorgio}(2006)}]{Sangiorgio_PhD-thesis:2006}%
  \BibitemOpen
  \bibfield  {author} {\bibinfo {author} {\bibfnamefont {S.}~\bibnamefont
  {Sangiorgio}},\ }\href {https://cuore.lngs.infn.it/en/publications/theses} {\
   (\bibinfo {year} {2006})},\ \bibinfo {note} {[Ph.\,D.\ thesis, Universit\`a
  dell'Insubria]}\BibitemShut {NoStop}%
\bibitem [{\citenamefont {W\'ojcik}\ and\ \citenamefont
  {Zuzel}(2007)}]{Wojcik:2007zz}%
  \BibitemOpen
  \bibfield  {author} {\bibinfo {author} {\bibfnamefont {M.}~\bibnamefont
  {W\'ojcik}}\ and\ \bibinfo {author} {\bibfnamefont {G.}~\bibnamefont
  {Zuzel}},\ }\href {\doibase 10.1063/1.2722068} {\bibfield  {journal}
  {\bibinfo  {journal} {AIP Conf.\ Proc.}\ }\textbf {\bibinfo {volume} {897}},\
  \bibinfo {pages} {53} (\bibinfo {year} {2007})}\BibitemShut {NoStop}%
\bibitem [{\citenamefont {Rusconi}(2011)}]{Rusconi_PhD-thesis:2011}%
  \BibitemOpen
  \bibfield  {author} {\bibinfo {author} {\bibfnamefont {C.}~\bibnamefont
  {Rusconi}},\ }\href {https://cuore.lngs.infn.it/en/publications/theses} {\
  (\bibinfo {year} {2011})},\ \bibinfo {note} {[Ph.\,D.\ thesis, Universit\`a
  dell'Insubria]}\BibitemShut {NoStop}%
\bibitem [{\citenamefont {Buccheri}\ \emph {et~al.}(2014)\citenamefont
  {Buccheri}, \citenamefont {Capodiferro}, \citenamefont {Morganti},
  \citenamefont {Orio}, \citenamefont {Pelosi},\ and\ \citenamefont
  {Pettinacci}}]{Buccheri:2014bma}%
  \BibitemOpen
  \bibfield  {author} {\bibinfo {author} {\bibfnamefont {E.}~\bibnamefont
  {Buccheri}}, \bibinfo {author} {\bibfnamefont {M.}~\bibnamefont
  {Capodiferro}}, \bibinfo {author} {\bibfnamefont {S.}~\bibnamefont
  {Morganti}}, \bibinfo {author} {\bibfnamefont {F.}~\bibnamefont {Orio}},
  \bibinfo {author} {\bibfnamefont {A.}~\bibnamefont {Pelosi}}, \ and\ \bibinfo
  {author} {\bibfnamefont {V.}~\bibnamefont {Pettinacci}},\ }\href {\doibase
  10.1016/j.nima.2014.09.046} {\bibfield  {journal} {\bibinfo  {journal}
  {Nucl.\ Instrum.\ Meth.\ A}\ }\textbf {\bibinfo {volume} {768}},\ \bibinfo
  {pages} {130} (\bibinfo {year} {2014})}\BibitemShut {NoStop}%
\bibitem [{\citenamefont {Clemenza}\ \emph {et~al.}(2011)\citenamefont
  {Clemenza}, \citenamefont {Maiano}, \citenamefont {Pattavina},\ and\
  \citenamefont {Previtali}}]{Clemenza:2011zz}%
  \BibitemOpen
  \bibfield  {author} {\bibinfo {author} {\bibfnamefont {M.}~\bibnamefont
  {Clemenza}}, \bibinfo {author} {\bibfnamefont {C.}~\bibnamefont {Maiano}},
  \bibinfo {author} {\bibfnamefont {L.}~\bibnamefont {Pattavina}}, \ and\
  \bibinfo {author} {\bibfnamefont {E.}~\bibnamefont {Previtali}},\ }\href
  {\doibase 10.1140/epjc/s10052-011-1805-0} {\bibfield  {journal} {\bibinfo
  {journal} {Eur.\ Phys.\ J.\ C}\ }\textbf {\bibinfo {volume} {71}},\ \bibinfo
  {pages} {1805} (\bibinfo {year} {2011})}\BibitemShut {NoStop}%
\bibitem [{\citenamefont {Andreotti}\ \emph {et~al.}(2009)\citenamefont
  {Andreotti} \emph {et~al.}}]{Andreotti:2009zza}%
  \BibitemOpen
  \bibfield  {author} {\bibinfo {author} {\bibfnamefont {E.}~\bibnamefont
  {Andreotti}} \emph {et~al.},\ }\href {\doibase 10.1088/1748-0221/4/09/P09003}
  {\bibfield  {journal} {\bibinfo  {journal} {J.\ Instrum.}\ }\textbf {\bibinfo
  {volume} {4}},\ \bibinfo {pages} {P09003} (\bibinfo {year}
  {2009})}\BibitemShut {NoStop}%
\bibitem [{\citenamefont {Brofferio}\ \emph {et~al.}(2013)\citenamefont
  {Brofferio}, \citenamefont {Canonica}, \citenamefont {Giachero},
  \citenamefont {Gotti}, \citenamefont {Maino},\ and\ \citenamefont
  {Pessina}}]{Brofferio:2013cya}%
  \BibitemOpen
  \bibfield  {author} {\bibinfo {author} {\bibfnamefont {C.}~\bibnamefont
  {Brofferio}}, \bibinfo {author} {\bibfnamefont {L.}~\bibnamefont {Canonica}},
  \bibinfo {author} {\bibfnamefont {A.}~\bibnamefont {Giachero}}, \bibinfo
  {author} {\bibfnamefont {C.}~\bibnamefont {Gotti}}, \bibinfo {author}
  {\bibfnamefont {M.}~\bibnamefont {Maino}}, \ and\ \bibinfo {author}
  {\bibfnamefont {G.}~\bibnamefont {Pessina}},\ }\href {\doibase
  10.1016/j.nima.2012.08.047} {\bibfield  {journal} {\bibinfo  {journal}
  {Nucl.\ Instrum.\ Meth.\ A}\ }\textbf {\bibinfo {volume} {718}},\ \bibinfo
  {pages} {211} (\bibinfo {year} {2013})}\BibitemShut {NoStop}%
\bibitem [{\citenamefont {Artusa}\ \emph {et~al.}(2015)\citenamefont {Artusa}
  \emph {et~al.}}]{Artusa:2014lgv}%
  \BibitemOpen
  \bibfield  {author} {\bibinfo {author} {\bibfnamefont {D.~R.}\ \bibnamefont
  {Artusa}} \emph {et~al.} (\bibinfo {collaboration} {CUORE Collaboration}),\
  }\href {\doibase 10.1155/2015/879871} {\bibfield  {journal} {\bibinfo
  {journal} {Adv.\ High Energy Phys.}\ }\textbf {\bibinfo {volume} {2015}},\
  \bibinfo {pages} {879871} (\bibinfo {year} {2015})}\BibitemShut {NoStop}%
\bibitem [{\citenamefont {Canonica}\ \emph {et~al.}(2015)\citenamefont
  {Canonica} \emph {et~al.}}]{Canonica:2015sva}%
  \BibitemOpen
  \bibfield  {author} {\bibinfo {author} {\bibfnamefont {L.}~\bibnamefont
  {Canonica}} \emph {et~al.},\ }\href {\doibase
  10.1016/j.nuclphysbps.2015.06.020} {\bibfield  {journal} {\bibinfo  {journal}
  {Nucl.\ Part.\ Phys.\ Proc.}\ }\textbf {\bibinfo {volume} {265-266}},\
  \bibinfo {pages} {73} (\bibinfo {year} {2015})}\BibitemShut {NoStop}%
\bibitem [{\citenamefont {Alduino}\ \emph
  {et~al.}(2016{\natexlab{b}})\citenamefont {Alduino} \emph
  {et~al.}}]{Alduino:2016zrl}%
  \BibitemOpen
  \bibfield  {author} {\bibinfo {author} {\bibfnamefont {C.}~\bibnamefont
  {Alduino}} \emph {et~al.} (\bibinfo {collaboration} {CUORE Collaboration}),\
  }\href {\doibase 10.1103/PhysRevC.93.045503} {\bibfield  {journal} {\bibinfo
  {journal} {Phys.\ Rev.\ C}\ }\textbf {\bibinfo {volume} {93}},\ \bibinfo
  {pages} {045503} (\bibinfo {year} {2016}{\natexlab{b}})}\BibitemShut
  {NoStop}%
\bibitem [{\citenamefont {Alduino}\ \emph
  {et~al.}(2017{\natexlab{b}})\citenamefont {Alduino} \emph
  {et~al.}}]{Alduino:2017qet}%
  \BibitemOpen
  \bibfield  {author} {\bibinfo {author} {\bibfnamefont {C.}~\bibnamefont
  {Alduino}} \emph {et~al.} (\bibinfo {collaboration} {CUORE Collaboration}),\
  }\href {\doibase 10.1140/epjc/s10052-017-5080-6} {\bibfield  {journal}
  {\bibinfo  {journal} {Eur.\ Phys.\ J.\ C}\ }\textbf {\bibinfo {volume}
  {77}},\ \bibinfo {pages} {543} (\bibinfo {year}
  {2017}{\natexlab{b}})}\BibitemShut {NoStop}%
\bibitem [{\citenamefont {Benato}\ \emph {et~al.}(2018)\citenamefont {Benato}
  \emph {et~al.}}]{Benato:2017kdf}%
  \BibitemOpen
  \bibfield  {author} {\bibinfo {author} {\bibfnamefont {G.}~\bibnamefont
  {Benato}} \emph {et~al.},\ }\href {\doibase 10.1088/1748-0221/13/01/P01010}
  {\bibfield  {journal} {\bibinfo  {journal} {J.\ Instrum.}\ }\textbf {\bibinfo
  {volume} {13}},\ \bibinfo {pages} {P01010} (\bibinfo {year}
  {2018})}\BibitemShut {NoStop}%
\bibitem [{\citenamefont {Dell'Oro}(2017)}]{Delloro_PhD-thesis:2017}%
  \BibitemOpen
  \bibfield  {author} {\bibinfo {author} {\bibfnamefont {S.}~\bibnamefont
  {Dell'Oro}},\ }\href {https://cuore.lngs.infn.it/en/publications/theses} {\
  (\bibinfo {year} {2017})},\ \bibinfo {note} {[Ph.\,D.\ thesis, INFN - Gran
  Sasso Science Institute]}\BibitemShut {NoStop}%
\bibitem [{\citenamefont {Alessandria}\ \emph
  {et~al.}(2013{\natexlab{b}})\citenamefont {Alessandria} \emph
  {et~al.}}]{Alessandria:2013ufa}%
  \BibitemOpen
  \bibfield  {author} {\bibinfo {author} {\bibfnamefont {F.}~\bibnamefont
  {Alessandria}} \emph {et~al.},\ }\href {\doibase 10.1016/j.nima.2013.06.015}
  {\bibfield  {journal} {\bibinfo  {journal} {Nucl.\ Instrum.\ Meth.}\ }\textbf
  {\bibinfo {volume} {A\,727}},\ \bibinfo {pages} {65} (\bibinfo {year}
  {2013}{\natexlab{b}})}\BibitemShut {NoStop}%
\bibitem [{\citenamefont {Alessandria}\ \emph {et~al.}(2018)\citenamefont
  {Alessandria} \emph {et~al.}}]{Alessandria:2018speriamo}%
  \BibitemOpen
  \bibfield  {author} {\bibinfo {author} {\bibfnamefont {F.}~\bibnamefont
  {Alessandria}} \emph {et~al.},\ }\href@noop {} {\enquote {\bibinfo {title}
  {{The CUORE cryostat: a 10\,mK infrastructure for large bolometric
  arrays}},}\ } (\bibinfo {year} {2018}),\ \bibinfo {note} {[Paper in
  preparation]}\BibitemShut {NoStop}%
\bibitem [{\citenamefont {Bucci}\ \emph {et~al.}(2018)\citenamefont {Bucci}
  \emph {et~al.}}]{ILS:2018speriamo}%
  \BibitemOpen
  \bibfield  {author} {\bibinfo {author} {\bibfnamefont {C.}~\bibnamefont
  {Bucci}} \emph {et~al.},\ }\href@noop {} {\enquote {\bibinfo {title} {{The
  Roman lead shield of the CUORE experiment}},}\ } (\bibinfo {year} {2018}),\
  \bibinfo {note} {[Paper in preparation]}\BibitemShut {NoStop}%
\bibitem [{\citenamefont {Cushman}\ \emph {et~al.}(2017)\citenamefont {Cushman}
  \emph {et~al.}}]{Cushman:2016cnv}%
  \BibitemOpen
  \bibfield  {author} {\bibinfo {author} {\bibfnamefont {J.~S.}\ \bibnamefont
  {Cushman}} \emph {et~al.},\ }\href {\doibase 10.1016/j.nima.2016.11.020}
  {\bibfield  {journal} {\bibinfo  {journal} {Nucl.\ Instrum.\ Meth.\ A}\
  }\textbf {\bibinfo {volume} {844}},\ \bibinfo {pages} {32} (\bibinfo {year}
  {2017})}\BibitemShut {NoStop}%
\bibitem [{\citenamefont {Cremonesi}(2017)}]{Cremonesi:2017-TAUP}%
  \BibitemOpen
  \bibfield  {author} {\bibinfo {author} {\bibfnamefont {O.}~\bibnamefont
  {Cremonesi}},\ }\href@noop {} {\  (\bibinfo {year} {2017})},\ \bibinfo {note}
  {[To appear in the \href{https://taup2017.snolab.ca/proceedings}{proceedings
  of TAUP 2017}]}\BibitemShut {NoStop}%
\bibitem [{\citenamefont {Alduino}\ \emph {et~al.}(2018)\citenamefont {Alduino}
  \emph {et~al.}}]{Alduino:2017ehq}%
  \BibitemOpen
  \bibfield  {author} {\bibinfo {author} {\bibfnamefont {C.}~\bibnamefont
  {Alduino}} \emph {et~al.} (\bibinfo {collaboration} {CUORE Collaboration}),\
  }\href {\doibase 10.1103/PhysRevLett.120.132501} {\bibfield  {journal}
  {\bibinfo  {journal} {Phys.\ Rev.\ Lett.}\ }\textbf {\bibinfo {volume}
  {120}},\ \bibinfo {pages} {132501} (\bibinfo {year} {2018})}\BibitemShut
  {NoStop}%
\bibitem [{\citenamefont {Alfonso}\ \emph {et~al.}(2015)\citenamefont {Alfonso}
  \emph {et~al.}}]{Alfonso:2015wka}%
  \BibitemOpen
  \bibfield  {author} {\bibinfo {author} {\bibfnamefont {K.}~\bibnamefont
  {Alfonso}} \emph {et~al.} (\bibinfo {collaboration} {CUORE Collaboration}),\
  }\href {\doibase 10.1103/PhysRevLett.115.102502} {\bibfield  {journal}
  {\bibinfo  {journal} {Phys.\ Rev.\ Lett.}\ }\textbf {\bibinfo {volume}
  {115}},\ \bibinfo {pages} {102502} (\bibinfo {year} {2015})}\BibitemShut
  {NoStop}%
\bibitem [{\citenamefont {Kirsten}\ \emph {et~al.}(1983)\citenamefont
  {Kirsten}, \citenamefont {Richter},\ and\ \citenamefont
  {Jessberger}}]{Kirsten:1983jj}%
  \BibitemOpen
  \bibfield  {author} {\bibinfo {author} {\bibfnamefont {T.}~\bibnamefont
  {Kirsten}}, \bibinfo {author} {\bibfnamefont {H.}~\bibnamefont {Richter}}, \
  and\ \bibinfo {author} {\bibfnamefont {E.}~\bibnamefont {Jessberger}},\
  }\href {\doibase 10.1103/PhysRevLett.50.474} {\bibfield  {journal} {\bibinfo
  {journal} {Phys.\ Rev.\ Lett.}\ }\textbf {\bibinfo {volume} {50}},\ \bibinfo
  {pages} {474} (\bibinfo {year} {1983})}\BibitemShut {NoStop}%
\bibitem [{\citenamefont {Alessandrello}\ \emph
  {et~al.}(1992{\natexlab{d}})\citenamefont {Alessandrello} \emph
  {et~al.}}]{Alessandrello:1992vk}%
  \BibitemOpen
  \bibfield  {author} {\bibinfo {author} {\bibfnamefont {A.}~\bibnamefont
  {Alessandrello}} \emph {et~al.},\ }\href {\doibase
  10.1016/0370-2693(92)91319-5} {\bibfield  {journal} {\bibinfo  {journal}
  {Phys.\ Lett.\ B}\ }\textbf {\bibinfo {volume} {285}},\ \bibinfo {pages}
  {176} (\bibinfo {year} {1992}{\natexlab{d}})}\BibitemShut {NoStop}%
\bibitem [{\citenamefont {Gando}\ \emph {et~al.}(2016)\citenamefont {Gando}
  \emph {et~al.}}]{KamLAND-Zen:2016pfg}%
  \BibitemOpen
  \bibfield  {author} {\bibinfo {author} {\bibfnamefont {A.}~\bibnamefont
  {Gando}} \emph {et~al.} (\bibinfo {collaboration} {KamLAND-Zen
  Collaboration}),\ }\href {\doibase 10.1103/PhysRevLett.117.109903} {\bibfield
   {journal} {\bibinfo  {journal} {Phys.\ Rev.\ Lett.}\ }\textbf {\bibinfo
  {volume} {117}},\ \bibinfo {pages} {082503} (\bibinfo {year}
  {2016})}\BibitemShut {NoStop}%
\bibitem [{\citenamefont {Agostini}\ \emph {et~al.}(2018)\citenamefont
  {Agostini} \emph {et~al.}}]{Agostini:2018tnm}%
  \BibitemOpen
  \bibfield  {author} {\bibinfo {author} {\bibfnamefont {M.}~\bibnamefont
  {Agostini}} \emph {et~al.} (\bibinfo {collaboration} {GERDA Collaboration}),\
  }\href {\doibase 10.1103/PhysRevLett.120.132503} {\bibfield  {journal}
  {\bibinfo  {journal} {Phys.\ Rev.\ Lett.}\ }\textbf {\bibinfo {volume}
  {120}},\ \bibinfo {pages} {132503} (\bibinfo {year} {2018})}\BibitemShut
  {NoStop}%
\bibitem [{\citenamefont {Alduino}\ \emph
  {et~al.}(2017{\natexlab{c}})\citenamefont {Alduino} \emph
  {et~al.}}]{Alduino:2017pni}%
  \BibitemOpen
  \bibfield  {author} {\bibinfo {author} {\bibfnamefont {C.}~\bibnamefont
  {Alduino}} \emph {et~al.} (\bibinfo {collaboration} {CUORE Collaboration}),\
  }\href {\doibase 10.1140/epjc/s10052-017-5098-9} {\bibfield  {journal}
  {\bibinfo  {journal} {Eur.\ Phys.\ J.\ C}\ }\textbf {\bibinfo {volume}
  {77}},\ \bibinfo {pages} {532} (\bibinfo {year}
  {2017}{\natexlab{c}})}\BibitemShut {NoStop}%
\bibitem [{\citenamefont {Capozzi}\ \emph {et~al.}(2016)\citenamefont
  {Capozzi}, \citenamefont {Lisi}, \citenamefont {Marrone}, \citenamefont
  {Montanino},\ and\ \citenamefont {Palazzo}}]{Capozzi:2016rtj}%
  \BibitemOpen
  \bibfield  {author} {\bibinfo {author} {\bibfnamefont {F.}~\bibnamefont
  {Capozzi}}, \bibinfo {author} {\bibfnamefont {E.}~\bibnamefont {Lisi}},
  \bibinfo {author} {\bibfnamefont {A.}~\bibnamefont {Marrone}}, \bibinfo
  {author} {\bibfnamefont {D.}~\bibnamefont {Montanino}}, \ and\ \bibinfo
  {author} {\bibfnamefont {A.}~\bibnamefont {Palazzo}},\ }\href {\doibase
  10.1016/j.nuclphysb.2016.02.016} {\bibfield  {journal} {\bibinfo  {journal}
  {Nucl.\ Phys.\ B}\ }\textbf {\bibinfo {volume} {908}},\ \bibinfo {pages}
  {218} (\bibinfo {year} {2016})}\BibitemShut {NoStop}%
\bibitem [{\citenamefont {Arnold}\ \emph {et~al.}(2015)\citenamefont {Arnold}
  \emph {et~al.}}]{Arnold:2015wpy}%
  \BibitemOpen
  \bibfield  {author} {\bibinfo {author} {\bibfnamefont {R.}~\bibnamefont
  {Arnold}} \emph {et~al.} (\bibinfo {collaboration} {NEMO-3 Collaboration}),\
  }\href {\doibase 10.1103/PhysRevD.92.072011} {\bibfield  {journal} {\bibinfo
  {journal} {Phys.\ Rev.\ D}\ }\textbf {\bibinfo {volume} {92}},\ \bibinfo
  {pages} {072011} (\bibinfo {year} {2015})}\BibitemShut {NoStop}%
\bibitem [{\citenamefont {Artusa}\ \emph {et~al.}(2014)\citenamefont {Artusa}
  \emph {et~al.}}]{Artusa:2014wnl}%
  \BibitemOpen
  \bibfield  {author} {\bibinfo {author} {\bibfnamefont {D.~R.}\ \bibnamefont
  {Artusa}} \emph {et~al.} (\bibinfo {collaboration} {CUORE Collaboration}),\
  }\href {\doibase 10.1140/epjc/s10052-014-3096-8} {\bibfield  {journal}
  {\bibinfo  {journal} {Eur.\ Phys.\ J.\ C}\ }\textbf {\bibinfo {volume}
  {74}},\ \bibinfo {pages} {3096} (\bibinfo {year} {2014})}\BibitemShut
  {NoStop}%
\bibitem [{\citenamefont {Artusa}\ \emph {et~al.}(2017)\citenamefont {Artusa}
  \emph {et~al.}}]{Artusa:2016mat}%
  \BibitemOpen
  \bibfield  {author} {\bibinfo {author} {\bibfnamefont {D.~R.}\ \bibnamefont
  {Artusa}} \emph {et~al.},\ }\href {\doibase 10.1016/j.physletb.2017.02.011}
  {\bibfield  {journal} {\bibinfo  {journal} {Phys.\ Lett.\ B}\ }\textbf
  {\bibinfo {volume} {767}},\ \bibinfo {pages} {321} (\bibinfo {year}
  {2017})}\BibitemShut {NoStop}%
\bibitem [{\citenamefont {Pedretti}(2004)}]{Pedretti_PhD-thesis:2004}%
  \BibitemOpen
  \bibfield  {author} {\bibinfo {author} {\bibfnamefont {M.}~\bibnamefont
  {Pedretti}},\ }\href {https://cuore.lngs.infn.it/en/publications/theses} {\
  (\bibinfo {year} {2004})},\ \bibinfo {note} {[Ph.\,D.\ thesis, Universit\`a
  dell'Insubria]}\BibitemShut {NoStop}%
\bibitem [{\citenamefont {Vignati}(2010)}]{Vignati_PhD-thesis:2010}%
  \BibitemOpen
  \bibfield  {author} {\bibinfo {author} {\bibfnamefont {M.}~\bibnamefont
  {Vignati}},\ }\href {http://www.springer.com/us/book/9789400712317} {\
  (\bibinfo {year} {2010})},\ \bibinfo {note} {[Ph.\,D.\ thesis, Universit\`a
  di Roma-La Sapienza]}\BibitemShut {NoStop}%
\bibitem [{\citenamefont {Santone}(2017)}]{Santone_PhD-thesis:2017}%
  \BibitemOpen
  \bibfield  {author} {\bibinfo {author} {\bibfnamefont {D.}~\bibnamefont
  {Santone}},\ }\href@noop {} {\  (\bibinfo {year} {2017})},\ \bibinfo {note}
  {[Ph.\,D.\ thesis, Universit\`a dell'Aquila]}\BibitemShut {NoStop}%
\bibitem [{\citenamefont {Barghouty}\ \emph {et~al.}(2013)\citenamefont
  {Barghouty} \emph {et~al.}}]{Barghouty:2010kj}%
  \BibitemOpen
  \bibfield  {author} {\bibinfo {author} {\bibfnamefont {A.~F.}\ \bibnamefont
  {Barghouty}} \emph {et~al.},\ }\href {\doibase 10.1016/j.nimb.2012.10.008}
  {\bibfield  {journal} {\bibinfo  {journal} {Nucl.\ Instrum.\ Meth.\ B}\
  }\textbf {\bibinfo {volume} {295}},\ \bibinfo {pages} {16} (\bibinfo {year}
  {2013})}\BibitemShut {NoStop}%
\bibitem [{\citenamefont {Casali}\ \emph {et~al.}(2015)\citenamefont {Casali}
  \emph {et~al.}}]{Casali:2014vvt}%
  \BibitemOpen
  \bibfield  {author} {\bibinfo {author} {\bibfnamefont {N.}~\bibnamefont
  {Casali}} \emph {et~al.},\ }\href {\doibase 10.1140/epjc/s10052-014-3225-4}
  {\bibfield  {journal} {\bibinfo  {journal} {Eur.\ Phys.\ J.\ C}\ }\textbf
  {\bibinfo {volume} {75}},\ \bibinfo {pages} {12} (\bibinfo {year}
  {2015})}\BibitemShut {NoStop}%
\bibitem [{\citenamefont {Artusa}\ \emph {et~al.}(2016)\citenamefont {Artusa}
  \emph {et~al.}}]{Artusa:2016maw}%
  \BibitemOpen
  \bibfield  {author} {\bibinfo {author} {\bibfnamefont {D.~R.}\ \bibnamefont
  {Artusa}} \emph {et~al.},\ }\href {\doibase 10.1140/epjc/s10052-016-4223-5}
  {\bibfield  {journal} {\bibinfo  {journal} {Eur.\ Phys.\ J.\ C}\ }\textbf
  {\bibinfo {volume} {76}},\ \bibinfo {pages} {364} (\bibinfo {year}
  {2016})}\BibitemShut {NoStop}%
\bibitem [{\citenamefont {Kim}\ \emph {et~al.}(2015)\citenamefont {Kim} \emph
  {et~al.}}]{Kim:2015pua}%
  \BibitemOpen
  \bibfield  {author} {\bibinfo {author} {\bibfnamefont {G.~B.}\ \bibnamefont
  {Kim}} \emph {et~al.},\ }\href {\doibase 10.1155/2015/817530} {\bibfield
  {journal} {\bibinfo  {journal} {Adv.\ High Energy Phys.}\ }\textbf {\bibinfo
  {volume} {2015}},\ \bibinfo {pages} {817530} (\bibinfo {year}
  {2015})}\BibitemShut {NoStop}%
\bibitem [{\citenamefont {Armengaud}\ \emph {et~al.}(2017)\citenamefont
  {Armengaud} \emph {et~al.}}]{Armengaud:2017hit}%
  \BibitemOpen
  \bibfield  {author} {\bibinfo {author} {\bibfnamefont {E.}~\bibnamefont
  {Armengaud}} \emph {et~al.},\ }\href {\doibase
  10.1140/epjc/s10052-017-5343-2} {\bibfield  {journal} {\bibinfo  {journal}
  {Eur.\ Phys.\ J.\ C}\ }\textbf {\bibinfo {volume} {77}},\ \bibinfo {pages}
  {785} (\bibinfo {year} {2017})}\BibitemShut {NoStop}%
\bibitem [{\citenamefont {Dafinei}\ \emph {et~al.}(2005)\citenamefont
  {Dafinei}, \citenamefont {Diemoz}, \citenamefont {Longo}, \citenamefont
  {Peter},\ and\ \citenamefont {Foldvari}}]{Dafinei:2005xy}%
  \BibitemOpen
  \bibfield  {author} {\bibinfo {author} {\bibfnamefont {I.}~\bibnamefont
  {Dafinei}}, \bibinfo {author} {\bibfnamefont {M.}~\bibnamefont {Diemoz}},
  \bibinfo {author} {\bibfnamefont {E.}~\bibnamefont {Longo}}, \bibinfo
  {author} {\bibfnamefont {A.}~\bibnamefont {Peter}}, \ and\ \bibinfo {author}
  {\bibfnamefont {I.}~\bibnamefont {Foldvari}},\ }\href {\doibase
  10.1016/j.nima.2005.08.010} {\bibfield  {journal} {\bibinfo  {journal}
  {Nucl.\ Instrum.\ Meth.\ A}\ }\textbf {\bibinfo {volume} {554}},\ \bibinfo
  {pages} {195} (\bibinfo {year} {2005})}\BibitemShut {NoStop}%
\bibitem [{\citenamefont {Dafinei}\ \emph {et~al.}(2007)\citenamefont
  {Dafinei}, \citenamefont {Dujardin}, \citenamefont {Longo},\ and\
  \citenamefont {Vignati}}]{Dafinei:2007hhc}%
  \BibitemOpen
  \bibfield  {author} {\bibinfo {author} {\bibfnamefont {I.}~\bibnamefont
  {Dafinei}}, \bibinfo {author} {\bibfnamefont {C.}~\bibnamefont {Dujardin}},
  \bibinfo {author} {\bibfnamefont {E.}~\bibnamefont {Longo}}, \ and\ \bibinfo
  {author} {\bibfnamefont {M.}~\bibnamefont {Vignati}},\ }\href {\doibase
  10.1002/pssa.200622458} {\bibfield  {journal} {\bibinfo  {journal} {Phys.\
  Status Solidi A}\ }\textbf {\bibinfo {volume} {204}},\ \bibinfo {pages}
  {1567} (\bibinfo {year} {2007})}\BibitemShut {NoStop}%
\bibitem [{\citenamefont {Tabarelli~de Fatis}(2010)}]{TabarellideFatis:2009zz}%
  \BibitemOpen
  \bibfield  {author} {\bibinfo {author} {\bibfnamefont {T.}~\bibnamefont
  {Tabarelli~de Fatis}},\ }\href {\doibase 10.1140/epjc/s10052-009-1207-8}
  {\bibfield  {journal} {\bibinfo  {journal} {Eur.\ Phys.\ J.\ C}\ }\textbf
  {\bibinfo {volume} {65}},\ \bibinfo {pages} {359} (\bibinfo {year}
  {2010})}\BibitemShut {NoStop}%
\bibitem [{\citenamefont {Bellini}\ \emph {et~al.}(2014)\citenamefont {Bellini}
  \emph {et~al.}}]{Bellini:2014yoa}%
  \BibitemOpen
  \bibfield  {author} {\bibinfo {author} {\bibfnamefont {F.}~\bibnamefont
  {Bellini}} \emph {et~al.},\ }\href {\doibase 10.1088/1748-0221/9/10/P10014}
  {\bibfield  {journal} {\bibinfo  {journal} {J.\ Instrum.}\ }\textbf {\bibinfo
  {volume} {9}},\ \bibinfo {pages} {P10014} (\bibinfo {year}
  {2014})}\BibitemShut {NoStop}%
\bibitem [{\citenamefont {Beeman}\ \emph {et~al.}(2012)\citenamefont {Beeman}
  \emph {et~al.}}]{Beeman:2011yc}%
  \BibitemOpen
  \bibfield  {author} {\bibinfo {author} {\bibfnamefont {J.~W.}\ \bibnamefont
  {Beeman}} \emph {et~al.},\ }\href {\doibase
  10.1016/j.astropartphys.2011.12.004} {\bibfield  {journal} {\bibinfo
  {journal} {Astropart.\ Phys.}\ }\textbf {\bibinfo {volume} {35}},\ \bibinfo
  {pages} {558} (\bibinfo {year} {2012})}\BibitemShut {NoStop}%
\bibitem [{\citenamefont {Bellini}\ \emph {et~al.}(2012)\citenamefont
  {Bellini}, \citenamefont {Casali}, \citenamefont {Dafinei}, \citenamefont
  {Marafini}, \citenamefont {Morganti}, \citenamefont {Orio}, \citenamefont
  {Pinci}, \citenamefont {Vignati},\ and\ \citenamefont
  {Voena}}]{Bellini:2012rc}%
  \BibitemOpen
  \bibfield  {author} {\bibinfo {author} {\bibfnamefont {F.}~\bibnamefont
  {Bellini}}, \bibinfo {author} {\bibfnamefont {N.}~\bibnamefont {Casali}},
  \bibinfo {author} {\bibfnamefont {I.}~\bibnamefont {Dafinei}}, \bibinfo
  {author} {\bibfnamefont {M.}~\bibnamefont {Marafini}}, \bibinfo {author}
  {\bibfnamefont {S.}~\bibnamefont {Morganti}}, \bibinfo {author}
  {\bibfnamefont {F.}~\bibnamefont {Orio}}, \bibinfo {author} {\bibfnamefont
  {D.}~\bibnamefont {Pinci}}, \bibinfo {author} {\bibfnamefont
  {M.}~\bibnamefont {Vignati}}, \ and\ \bibinfo {author} {\bibfnamefont
  {C.}~\bibnamefont {Voena}},\ }\href {\doibase 10.1088/1748-0221/7/11/P11014}
  {\bibfield  {journal} {\bibinfo  {journal} {J.\ Instrum.}\ }\textbf {\bibinfo
  {volume} {7}},\ \bibinfo {pages} {P11014} (\bibinfo {year}
  {2012})}\BibitemShut {NoStop}%
\bibitem [{\citenamefont {Wang}\ \emph
  {et~al.}(2015{\natexlab{a}})\citenamefont {Wang} \emph
  {et~al.}}]{Wang:2015raa}%
  \BibitemOpen
  \bibfield  {author} {\bibinfo {author} {\bibfnamefont {G.}~\bibnamefont
  {Wang}} \emph {et~al.} (\bibinfo {collaboration} {CUPID Collaboration}),\
  }\href@noop {} {\  (\bibinfo {year} {2015}{\natexlab{a}})},\ \Eprint
  {http://arxiv.org/abs/1504.03599} {arXiv:1504.03599 [physics.ins-det]}
  \BibitemShut {NoStop}%
\bibitem [{\citenamefont {Wang}\ \emph
  {et~al.}(2015{\natexlab{b}})\citenamefont {Wang} \emph
  {et~al.}}]{Wang:2015taa}%
  \BibitemOpen
  \bibfield  {author} {\bibinfo {author} {\bibfnamefont {G.}~\bibnamefont
  {Wang}} \emph {et~al.} (\bibinfo {collaboration} {CUPID Collaboration}),\
  }\href@noop {} {\  (\bibinfo {year} {2015}{\natexlab{b}})},\ \Eprint
  {http://arxiv.org/abs/1504.03612} {arXiv:1504.03612 [physics.ins-det]}
  \BibitemShut {NoStop}%
\bibitem [{\citenamefont {Battistelli}\ \emph {et~al.}(2015)\citenamefont
  {Battistelli} \emph {et~al.}}]{Battistelli:2015vha}%
  \BibitemOpen
  \bibfield  {author} {\bibinfo {author} {\bibfnamefont {E.~S.}\ \bibnamefont
  {Battistelli}} \emph {et~al.},\ }\href {\doibase
  10.1140/epjc/s10052-015-3575-6} {\bibfield  {journal} {\bibinfo  {journal}
  {Eur.\ Phys.\ J.\ C}\ }\textbf {\bibinfo {volume} {75}},\ \bibinfo {pages}
  {353} (\bibinfo {year} {2015})}\BibitemShut {NoStop}%
\bibitem [{\citenamefont {Bellini}\ \emph {et~al.}(2017)\citenamefont {Bellini}
  \emph {et~al.}}]{Bellini:2016lgg}%
  \BibitemOpen
  \bibfield  {author} {\bibinfo {author} {\bibfnamefont {F.}~\bibnamefont
  {Bellini}} \emph {et~al.},\ }\href {\doibase 10.1063/1.4974082} {\bibfield
  {journal} {\bibinfo  {journal} {Appl.\ Phys.\ Lett.}\ }\textbf {\bibinfo
  {volume} {110}},\ \bibinfo {pages} {033504} (\bibinfo {year}
  {2017})}\BibitemShut {NoStop}%
\bibitem [{\citenamefont {Sch$\ddot{\text a}$ffner}\ \emph
  {et~al.}(2015)\citenamefont {Sch$\ddot{\text a}$ffner} \emph
  {et~al.}}]{Schaffner:2014caa}%
  \BibitemOpen
  \bibfield  {author} {\bibinfo {author} {\bibfnamefont {K.}~\bibnamefont
  {Sch$\ddot{\text a}$ffner}} \emph {et~al.},\ }\href {\doibase
  10.1016/j.astropartphys.2015.03.008} {\bibfield  {journal} {\bibinfo
  {journal} {Astropart.\ Phys.}\ }\textbf {\bibinfo {volume} {69}},\ \bibinfo
  {pages} {30} (\bibinfo {year} {2015})}\BibitemShut {NoStop}%
\bibitem [{\citenamefont {Willers}\ \emph {et~al.}(2015)\citenamefont {Willers}
  \emph {et~al.}}]{Willers:2014eoa}%
  \BibitemOpen
  \bibfield  {author} {\bibinfo {author} {\bibfnamefont {M.}~\bibnamefont
  {Willers}} \emph {et~al.},\ }\href {\doibase 10.1088/1748-0221/10/03/P03003}
  {\bibfield  {journal} {\bibinfo  {journal} {J.\ Instrum.}\ }\textbf {\bibinfo
  {volume} {10}},\ \bibinfo {pages} {P03003} (\bibinfo {year}
  {2015})}\BibitemShut {NoStop}%
\bibitem [{\citenamefont {Pattavina}\ \emph {et~al.}(2016)\citenamefont
  {Pattavina} \emph {et~al.}}]{Casali:2015gya}%
  \BibitemOpen
  \bibfield  {author} {\bibinfo {author} {\bibfnamefont {L.}~\bibnamefont
  {Pattavina}} \emph {et~al.},\ }\href {\doibase 10.1007/s10909-015-1404-9}
  {\bibfield  {journal} {\bibinfo  {journal} {J.\ Low.\ Temp.\ Phys.}\ }\textbf
  {\bibinfo {volume} {184}},\ \bibinfo {pages} {286} (\bibinfo {year}
  {2016})}\BibitemShut {NoStop}%
\bibitem [{\citenamefont {Biassoni}\ \emph {et~al.}(2015)\citenamefont
  {Biassoni} \emph {et~al.}}]{Biassoni:2015eij}%
  \BibitemOpen
  \bibfield  {author} {\bibinfo {author} {\bibfnamefont {M.}~\bibnamefont
  {Biassoni}} \emph {et~al.},\ }\href {\doibase 10.1140/epjc/s10052-015-3712-2}
  {\bibfield  {journal} {\bibinfo  {journal} {Eur.\ Phys.\ J.\ C}\ }\textbf
  {\bibinfo {volume} {75}},\ \bibinfo {pages} {480} (\bibinfo {year}
  {2015})}\BibitemShut {NoStop}%
\bibitem [{\citenamefont {Gironi}\ \emph {et~al.}(2016)\citenamefont {Gironi}
  \emph {et~al.}}]{Gironi:2016nae}%
  \BibitemOpen
  \bibfield  {author} {\bibinfo {author} {\bibfnamefont {L.}~\bibnamefont
  {Gironi}} \emph {et~al.},\ }\href {\doibase 10.1103/PhysRevC.94.054608}
  {\bibfield  {journal} {\bibinfo  {journal} {Phys.\ Rev.\ C}\ }\textbf
  {\bibinfo {volume} {94}},\ \bibinfo {pages} {054608} (\bibinfo {year}
  {2016})}\BibitemShut {NoStop}%
\bibitem [{\citenamefont {Berg\'e}\ \emph {et~al.}(2018)\citenamefont {Berg\'e}
  \emph {et~al.}}]{Berge:2017nys}%
  \BibitemOpen
  \bibfield  {author} {\bibinfo {author} {\bibfnamefont {L.}~\bibnamefont
  {Berg\'e}} \emph {et~al.},\ }\href {\doibase 10.1103/PhysRevC.97.032501}
  {\bibfield  {journal} {\bibinfo  {journal} {Phys.\ Rev.\ C}\ }\textbf
  {\bibinfo {volume} {97}},\ \bibinfo {pages} {032501} (\bibinfo {year}
  {2018})}\BibitemShut {NoStop}%
\bibitem [{\citenamefont {Nones}\ \emph {et~al.}(2012)\citenamefont {Nones},
  \citenamefont {Berg\'e}, \citenamefont {Dumoulin}, \citenamefont
  {Marnieros},\ and\ \citenamefont {Olivieri}}]{Nones:2012tsm}%
  \BibitemOpen
  \bibfield  {author} {\bibinfo {author} {\bibfnamefont {C.}~\bibnamefont
  {Nones}}, \bibinfo {author} {\bibfnamefont {L.}~\bibnamefont {Berg\'e}},
  \bibinfo {author} {\bibfnamefont {L.}~\bibnamefont {Dumoulin}}, \bibinfo
  {author} {\bibfnamefont {S.}~\bibnamefont {Marnieros}}, \ and\ \bibinfo
  {author} {\bibfnamefont {E.}~\bibnamefont {Olivieri}},\ }\href {\doibase
  10.1007/s10909-012-0558-y} {\bibfield  {journal} {\bibinfo  {journal} {J.\
  Low.\ Temp.\ Phys.}\ }\textbf {\bibinfo {volume} {167}},\ \bibinfo {pages}
  {1029} (\bibinfo {year} {2012})}\BibitemShut {NoStop}%
\bibitem [{\citenamefont {Canonica}\ \emph {et~al.}(2013)\citenamefont
  {Canonica}, \citenamefont {Biassoni}, \citenamefont {Brofferio},
  \citenamefont {Bucci}, \citenamefont {Calvano}, \citenamefont {Di~Vacri},
  \citenamefont {Goett}, \citenamefont {Gorla}, \citenamefont {Pavan},\ and\
  \citenamefont {Yeh}}]{Canonica:2013jvz}%
  \BibitemOpen
  \bibfield  {author} {\bibinfo {author} {\bibfnamefont {L.}~\bibnamefont
  {Canonica}}, \bibinfo {author} {\bibfnamefont {M.}~\bibnamefont {Biassoni}},
  \bibinfo {author} {\bibfnamefont {C.}~\bibnamefont {Brofferio}}, \bibinfo
  {author} {\bibfnamefont {C.}~\bibnamefont {Bucci}}, \bibinfo {author}
  {\bibfnamefont {S.}~\bibnamefont {Calvano}}, \bibinfo {author} {\bibfnamefont
  {M.~L.}\ \bibnamefont {Di~Vacri}}, \bibinfo {author} {\bibfnamefont
  {J.}~\bibnamefont {Goett}}, \bibinfo {author} {\bibfnamefont
  {P.}~\bibnamefont {Gorla}}, \bibinfo {author} {\bibfnamefont
  {M.}~\bibnamefont {Pavan}}, \ and\ \bibinfo {author} {\bibfnamefont
  {M.}~\bibnamefont {Yeh}},\ }\href {\doibase 10.1016/j.nima.2013.05.114}
  {\bibfield  {journal} {\bibinfo  {journal} {Nucl.\ Instrum.\ Meth.\ A}\
  }\textbf {\bibinfo {volume} {732}},\ \bibinfo {pages} {286} (\bibinfo {year}
  {2013})}\BibitemShut {NoStop}%
\bibitem [{\citenamefont {Hyvarinen}\ and\ \citenamefont
  {Suhonen}(2015)}]{Hyvarinen:2015bda}%
  \BibitemOpen
  \bibfield  {author} {\bibinfo {author} {\bibfnamefont {J.}~\bibnamefont
  {Hyvarinen}}\ and\ \bibinfo {author} {\bibfnamefont {J.}~\bibnamefont
  {Suhonen}},\ }\href {\doibase 10.1103/PhysRevC.91.024613} {\bibfield
  {journal} {\bibinfo  {journal} {Phys.\ Rev.\ C}\ }\textbf {\bibinfo {volume}
  {91}},\ \bibinfo {pages} {024613} (\bibinfo {year} {2015})}\BibitemShut
  {NoStop}%
\bibitem [{\citenamefont {Agostini}\ \emph {et~al.}(2017)\citenamefont
  {Agostini} \emph {et~al.}}]{Agostini:2017iyd}%
  \BibitemOpen
  \bibfield  {author} {\bibinfo {author} {\bibfnamefont {M.}~\bibnamefont
  {Agostini}} \emph {et~al.} (\bibinfo {collaboration} {GERDA Collaboration}),\
  }\href {\doibase 10.1038/nature21717} {\bibfield  {journal} {\bibinfo
  {journal} {Nature}\ }\textbf {\bibinfo {volume} {544}},\ \bibinfo {pages}
  {47} (\bibinfo {year} {2017})}\BibitemShut {NoStop}%
\bibitem [{\citenamefont {Bucci}(2014)}]{Bucci_WN2014}%
  \BibitemOpen
  \bibfield  {author} {\bibinfo {author} {\bibfnamefont {C.}~\bibnamefont
  {Bucci}},\ }\href
  {https://agenda.infn.it/contributionDisplay.py?contribId=1&confId=8474}
  {\enquote {\bibinfo {title} {{What Next - $0\nu\beta\beta$}},}\ } (\bibinfo
  {year} {2014}),\ \bibinfo {note} {[Public seminar at LNGS]}\BibitemShut
  {NoStop}%
\end{thebibliography}%
